\documentclass[floats,floatfix,showpacs,amssymb,prd,twocolumn,superscriptaddress,nofootinbib]{revtex4-1}
\usepackage{amssymb,amsmath,verbatim,color,mathtools,needspace,enumitem,etoolbox,xcolor,esint}
\usepackage{hyperref}
\definecolor{linkcolor}{rgb}{0.0,0.3,0.5}
\definecolor{urlcolor}{rgb}{0.27,0.55,0.}
\definecolor{funcolor}{rgb}{0.65, 0.16, 0.16}
\hypersetup{colorlinks=true,linkcolor=linkcolor,citecolor=linkcolor,filecolor=linkcolor,urlcolor=linkcolor}

\usepackage{etoolbox}
\makeatletter
\preto{\@verbatim}{\topsep=3.5pt \partopsep=4pt }
\makeatother
\usepackage{graphicx}
\usepackage{capt-of}
\usepackage{colortbl}
\usepackage{pifont}%
\newcommand{\cmark}{\ding{51}}%
\newcommand{\xmark}{\ding{55}}%

\DeclareMathOperator\erf{erf}

\def\araa{\rm{ARA\&A}}
\def\apj{\rm{ApJ}}
\def\apjl{\rm{ApJ}}

\def\aap{\rm{A\&A}}

\def\jcap{\rm{J. Cosmology Astropart. Phys.}}

\def\mnras{\rm{MNRAS}}

\def\prd{\rm{PRD}}

\def\prl{\rm{PRL}}

\def\nat{\rm{Nature}}

\def\physrep{\rm{Phys.~Rep.}}

\def\cqg{\rm{CQG}}

\begin{document}

\title{Are merging black holes born from stellar collapse or previous mergers?}
\author{Davide Gerosa}
\thanks{Einstein Fellow}
\email{dgerosa@caltech.edu}
\affiliation{TAPIR 350-17, California Institute of Technology, 1200 E California
Boulevard, Pasadena, CA 91125, USA}
\author{Emanuele Berti}
\email{eberti@olemiss.edu}
\affiliation{Department of Physics and Astronomy, The University of 
Mississippi, University, MS 38677, USA}
\affiliation{CENTRA, Departamento de F\'isica, Instituto Superior
T\'ecnico, Universidade de Lisboa, Avenida Rovisco Pais 1,
1049 Lisboa, Portugal}


\date{\today}

\begin{abstract}
  Advanced LIGO detectors at Hanford and Livingston made two confirmed
  and one marginal detection of binary black holes during their first
  observing run.  The first event, GW150914, was from the merger of
  two black holes much heavier that those whose masses have been
  estimated so far, indicating a formation scenario that might differ
  from ``ordinary'' stellar evolution. One possibility is that these
  heavy black holes resulted from a previous merger. When the
  progenitors of a black hole binary merger result from previous
  mergers, they should (on average) merge later, be more massive, and
  have spin magnitudes clustered around a dimensionless spin
  $\sim 0.7$.  Here we ask the following question: can
  gravitational-wave observations determine whether merging black
  holes were born from the collapse of massive stars (``first
  generation''), rather than being the end product of earlier mergers
  (``second generation'')? We construct simple, observationally
  motivated populations of black hole binaries, and we use Bayesian
  model selection to show that measurements of the masses, luminosity
  distance (or redshift), and ``effective spin'' of black hole binaries
  can indeed distinguish between these different formation
  scenarios.
\end{abstract}

\maketitle
\section{Introduction}

The observation of gravitational waves (GWs) from merging black hole
(BH) binaries was a milestone in physics and
astronomy~\cite{2016PhRvL.116f1102A,2016arXiv160604855T,2016PhRvX...6d1015A}.
During their first observing run (O1), the Advanced LIGO detectors
detected two GW events (GW150914 and GW151226) and a marginal
candidate LVT151012, which is also likely to be of astrophysical
origin. The second observing run (O2) is currently ongoing, and
Advanced Virgo is expected to join the detector network soon. Dozens
of BH mergers may be detected by the end of O2 or in the third run
(O3), allowing for statistical studies of their populations.

These events can
further our understanding of the formation channels of binary
BHs~\citep{2016ApJ...818L..22A}, because different astrophysical
scenarios predict different binary properties. As the number of
detections grows, a statistical analysis of the observed binary
parameters should eventually allow us to identify or constrain the
physical processes responsible for the formation and merger of compact
binaries. Currently favored scenarios include stellar evolution of
field binaries \citep{2014LRR....17....3P} and the dynamical capture
of BHs in globular clusters \citep{2013LRR....16....4B}. Recent work
showed that both field formation
\citep{2009MNRAS.395L..71M,2010ApJ...715L.138B,2012ApJ...759...52D,2013ApJ...779...72D,2015MNRAS.451.4086S,2015ApJ...806..263D,2016arXiv160204531B,2016ApJ...819..108B}
and cluster formation
\citep{2015PhRvL.115e1101R,2016PhRvD..93h4029R,2016arXiv160300884C,2016ApJ...824L...8R}
are broadly compatible with current Advanced LIGO
observations~\citep{2016ApJ...818L..22A}. 

It is quite likely that both field and cluster formation channels are
at work in nature. The first event, GW150914, was the most surprising,
because the merging BHs are much heavier that those whose masses have
been estimated so far in x-ray
binaries~\cite{2006ARA&A..44...49R,2013arXiv1312.6698N},
indicating a formation scenario that might differ from ``ordinary''
stellar evolution.
{Alternative theoretical scenarios which could explain the unexpected properties of GW150914 include}
formation via hierarchical triples~\citep{
  2003ApJ...598..419W,2014ApJ...781...45A,2016ApJ...816...65A}, a
Population III origin for the binary members
\citep{2014MNRAS.442.2963K,2016MNRAS.460L..74H,2016arXiv161201524B},
chemically homogeneous evolution in short-period
binaries~\citep{2016MNRAS.458.2634M,2016MNRAS.460.3545D,2016A&A...588A..50M},
and a primordial origin for the merging BHs
\citep{2016PhRvL.116t1301B,2016PhRvD..94h4013C}.

\begin{figure*}
\includegraphics[width=\textwidth]{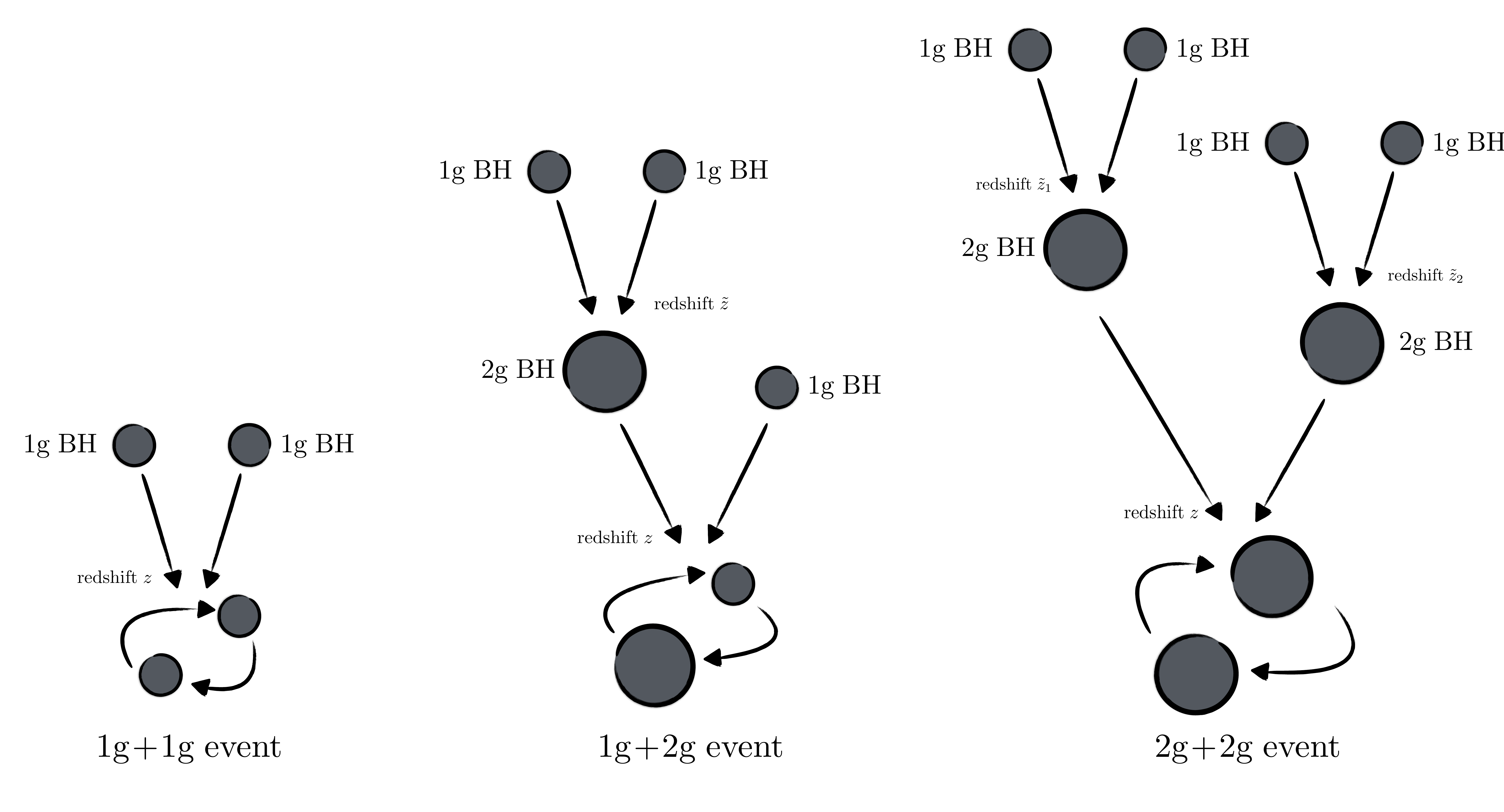}
\caption{Cartoon sketch of the three possible scenarios for the merger
  of two BHs. First generation (1g) BHs resulting from stellar
  collapse can form second generation (2g) BHs via mergers. Imprints
  of these formation channels are left in the statistical distribution
  of masses, spins and redshift of the detected events.}
\label{fig:cartoon}
\end{figure*}
 
One possibility to explain the high mass of the merging BHs in
GW150914 is that these BHs did not form following stellar collapse,
but rather from previous BH mergers. Field formation scenarios
typically predict long delay times between the formation and merger of
a BH binary
\cite{2012ApJ...759...52D}, so repeated mergers seem unlikely. However
gravitational encounters are more common in dense stellar
environments, and some scenarios suggest that repeated mergers may be
possible
\cite{2016ApJ...831..187A,2016MNRAS.459.3432M,2016ApJ...824L..12O}. The
most likely environment to host multiple mergers are nuclear clusters
\cite{2016ApJ...831..187A}, which present larger escape speeds
compared to globular and open clusters and can, therefore, more easily
retains merger remnants with substantial recoils
\cite{2004ApJ...607L...9M}. Stellar-mass BH binaries may also form in
AGN gaseous discs \cite{2017arXiv170207818M}, where migration traps
can be invoked to assemble multiple generations of mergers
\cite{2016ApJ...819L..17B}.
Primordial BHs are also expected to merge very
quickly~\cite{2016PhRvL.116t1301B,2016PhRvD..94h4013C}, so the
possibility of repeated mergers in this scenario should not be
excluded~\cite{2017PDU....15..142C}.

In this paper, we ask the following question: can GW observations
determine whether merging BHs such as those in GW150914 were born
directly from the collapse of massive stars (``first-generation'' BHs,
henceforth 1g) rather than being the end product of previous mergers
(``second-generation'' BHs, henceforth 2g)?

Roughly speaking, one can expect mergers to leave several
statistically observable imprints in 2g BHs, namely,
\pagebreak
\begin{enumerate}
\item[(i)] 2g BHs should be more massive than BHs born from stellar
  collapse;
\item[(ii)] quite independently of the distribution of spin magnitudes
  following core collapse (which is highly uncertain
  \cite{2016MNRAS.462..844K}), the spin magnitudes of 2g BHs should
  cluster (on average) around the dimensionless spin $\sim 0.7$
  resulting from the merger of nonspinning
  BHs~\cite{2008ApJ...684..822B};
\item[(iii)] statistically, the merger of BH binaries including 2g components
  should occur later (i.e., at smaller redshift or luminosity distance
  from GW detectors) because of the delay time between BH formation and
  merger.
\end{enumerate}

In this paper we make these arguments more quantitative and rigorous
by developing a simple but physically motivated model to describe the
bulk theoretical properties of 1g and 2g binary BH mergers
(Sec.~\ref{sec:distributions}).  Then we consider a set of present
and future GW detectors, and we simulate \emph{observable
  distributions} by selecting detectable binaries and estimating the
expected measurement errors on their parameters
(Sec.~\ref{sec:obsdistr}). Finally we set up a Bayesian model
selection framework (Sec.~\ref{sec:modsel}) to address what can be
done with current observations, and to quantify the capabilities of
future detectors to distinguish between different models
(Sec.~\ref{sec:results}). We conclude by summarizing our results
and pointing out possible extensions (Sec.~\ref{sec:conclusions}).

\section{Theoretical distributions}
\label{sec:distributions}

Our goal in this section is to develop a simple prescription to build
populations of binary BHs. Our greatly oversimplified model is not
meant to capture the complexity of binary evolution in an
astrophysical setting, but just the main features distinguishing 1g
and 2g BHs.

As illustrated by the cartoon in Fig.~\ref{fig:cartoon}, we
construct three \emph{theoretical distributions}, labeled by
``1g+1g,'' ``1g+2g'' and ``2g+2g''. In this context, ``1g'' means that
one of the binary components is a first-generation BH produced by
stellar collapse, whereas ``2g'' means that it is a second-generation
BH produced by a previous merger.

\subsection{The 1g+1g population}
\label{1g1gpop}

Following the LIGO-Virgo Scientific
Collaboration~\cite{2016PhRvX...6d1015A}, for the 1g+1g population, we
adopt three different prescriptions for the distribution of
source-frame masses:

\begin{itemize}
\item[(i)] {\bf Model ``flat'':} we assume uniformly distributed
  source-frame masses $m_1$ and $m_2$ in the range
  $m_i\in [5 M_\odot,50 M_\odot]$ ($i=1,\,2$), where hereafter
  $m_1>m_2$. %
\item[(ii)] {\bf Model ``log'':} we take the logarithm of the
  source-frame masses to be uniformly distributed in the same range,
  so that the probability distribution $p(m_1,m_2)\propto 1/m_1 m_2$.
\item[(iii)] {\bf Model ``power law'':} we adopt a power-law
  distribution with spectral index $\alpha=-2.5$ for the primary BH
  (i.e. $p(m_1)\propto m^{\alpha}$), while the secondary mass is
  uniformly distributed in $m_2\in[5 M_\odot, m_1]$. %
\end{itemize}
{The upper limit of $50 M_\odot$ was chosen to be consistent with
  current LIGO compact binary coalescence searches, and it excludes
  ``by construction'' intermediate-mass BH searches, discussed e.g.
  in
  \cite{2017arXiv170404628T}. Moreover, pair instability and pulsation
  pair instability in massive helium cores
  \cite{2002ApJ...567..532H,2017ApJ...836..244W} may inhibit the
  formation of 1g BHs with masses larger than $\sim 50 M_\odot$
  \cite{2016A&A...594A..97B}. If multiple mergers occur through mass
  segregation in stellar clusters, the more massive objects will tend
  to form binaries, thus increasing the component masses of 1g+2g and
  2g+2g populations. Our $50 M_\odot$ upper mass limit is therefore
  conservative, because physical mechanisms such as pair instabilities
  and mass segregation would further separate the mass distributions
  of populations involving multiple mergers and make them more easily
  distinguishable.}

Given the great uncertainties on the spin magnitude and orientation of
binary BHs
\cite{2008ApJ...682..474B,2015PhR...548....1M,2016Natur.534..512B,2016ApJ...832L...2R},
in all three cases we assume the dimensionless spin magnitudes
$\chi_{1,2}$ to be uniformly distributed in $[0,1]$, and their
directions to be isotropically distributed.\footnote{Rodriguez \emph{et
  al.} \cite{2016ApJ...832L...2R} argued that massive field binaries
  should typically have aligned spins because ``heavy'' BHs receive
  small supernova kicks that are unable to tilt the orbit
  \cite{2000ApJ...541..319K,2013PhRvD..87j4028G}, {while} the spins
  of massive binaries produced in dense stellar environments should be
  isotropically distributed. A more detailed investigation of the
  correlation between spin alignment and binary BH formation requires
  astrophysical modeling that is beyond the scope of this paper (see
  e.g. \cite{2008ApJ...682..474B,2013PhRvD..87j4028G}).}  We are only
interested in the global statistical properties of the
population. Since isotropic spin distributions stay isotropic under
precession and gravitational radiation reaction
\cite{2007ApJ...661L.147B,2015PhRvD..92f4016G}, the assumption of
isotropy will hold also at the small separations relevant for GW
observations. For this reason there is no need to carry out
post-Newtonian evolutions of the spin distributions for individual
binaries of the kind discussed
in~\cite{2015PhRvL.114h1103K,2015PhRvD..92f4016G,2016PhRvD..93l4066G}.

\subsection{The 2g+2g population}

In order to construct the 2g+2g population we use the following
procedure. We randomly extract two binaries from a given 1g+1g
population. For these binaries, we estimate the final mass $M_f$ and
spin $\chi_{\rm f}$ of the merger remnant using the numerical relativity
fitting formulas of
Refs.~\cite{2012ApJ...758...63B,2009ApJ...704L..40B}\footnote{There
  are several alternative fitting formulas for the final masses and
  spins
  \cite{2014PhRvD..89j4052L,2014PhRvD..90j4004H,2015PhRvD..92b4022Z,2016PhRvD..93d4006H,
    2016ApJ...825L..19H,2016arXiv161009713H,2017PhRvD..95f4024J}.  The
  difference between different prescriptions is smaller than
  measurement errors in GW observations, and therefore the choice of a
  particular fitting formula is of no consequence for our present
  purpose.}  as implemented in \cite{2016PhRvD..93l4066G}.  These
masses and spins are used as input for the second round of binary
mergers.

To perform meaningful comparison with the 1g+1g model described above,
we again restrict our population to binaries with component masses in
the range $[5,50] M_\odot$, because this is the mass range targeted by
LIGO compact binary coalescence searches.

\subsection{The 1g+2g population}

The 1g+2g distribution is the obvious mixture of the two: we draw one
binary from the 1g+1g distribution, merge it to obtain a 2g BH, and
then consider the merger of this 2g BH with a 1g BH.

\begin{figure*}[t]
\includegraphics[width=\textwidth,page=1]{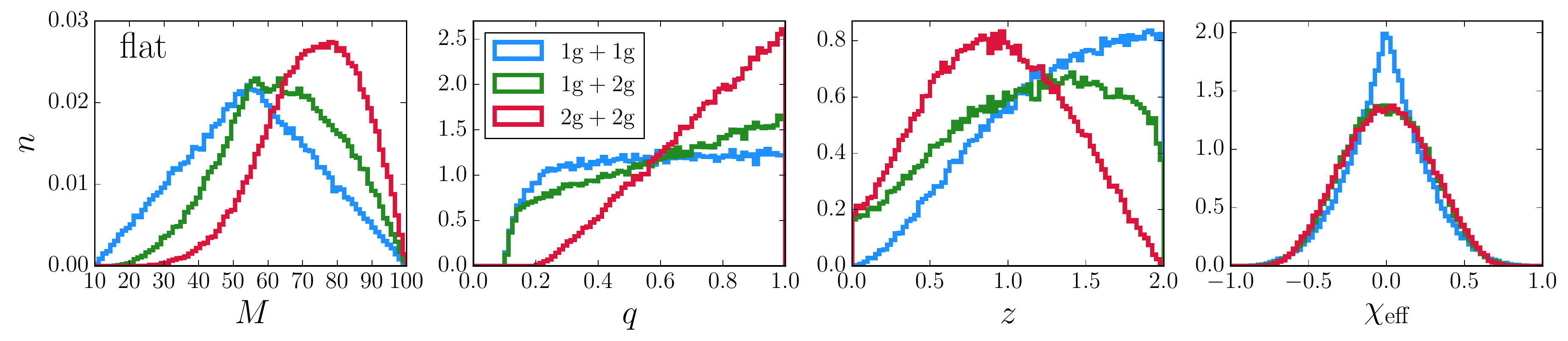}
\includegraphics[width=\textwidth,page=2]{theodist_vec.pdf}
\includegraphics[width=\textwidth,page=3]{theodist_vec.pdf}
\caption{Theoretical distribution of the observable parameters
  ${\bf u}=\{M,\, q,\, z,\, \chi_{\rm eff} \}$ for 1g+1g (blue), 1g+2g
  (green) and 2g+2g (red) populations, assuming the ``flat'' (top),
  ``log'' (middle), and ``power law'' (bottom) mass
  distributions. 
    }
\label{fig:mergindis}
\end{figure*}

\begin{figure*}[t]
\includegraphics[width=\textwidth,page=1]{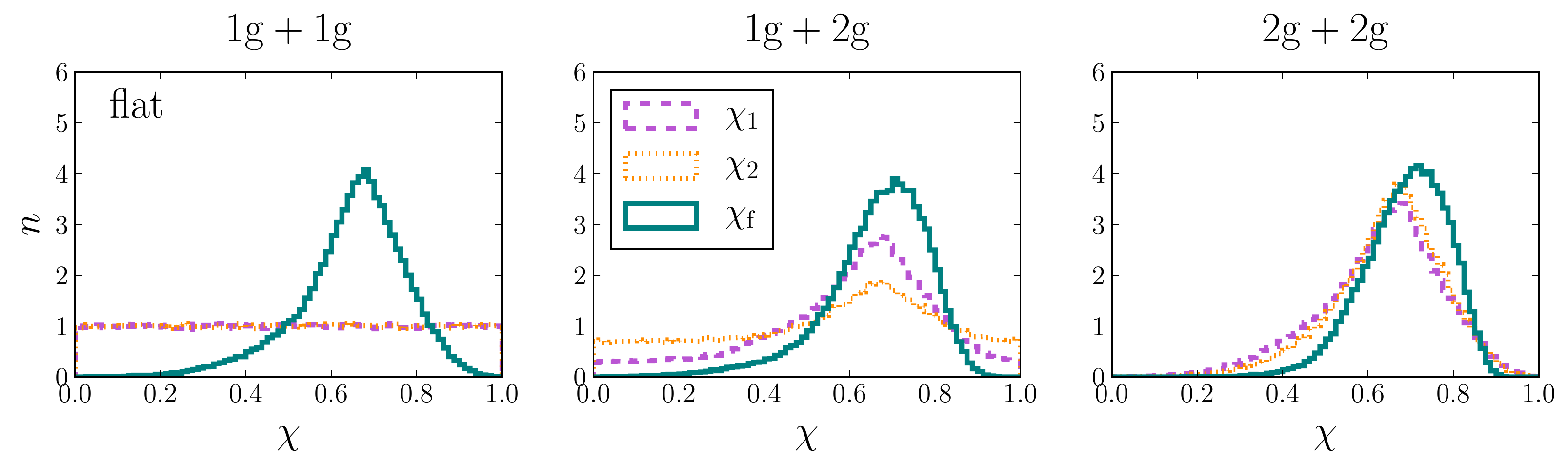}
\includegraphics[width=\textwidth,page=2]{theodist_spinonly.pdf}
\includegraphics[width=\textwidth,page=3]{theodist_spinonly.pdf}
\caption{Spin magnitude distributions for primary ($\chi_1$),
  secondary ($\chi_2$), and postmerger ($\chi_{\rm f}$) BH spins in
  each of the various models used in this paper. On average, mergers
  tend to produce BH spins clustered around $\sim 0.7$, quite
  independently of the progenitor parameters (cf. Fig. 3 and the
  left panels in Figs. 4 and 5 of Ref.~\cite{2008ApJ...684..822B}).}
\label{fig:spinonly}
\end{figure*}

\subsection{Redshift distribution}
\label{redshiftdistribution}
 
The redshift distribution of BH mergers in the three different
populations should be different, because on average 2g mergers are
expected to happen later than 1g mergers.  We can estimate the delay
times between the formation and merger of a BH binary using the
quadrupole formula
\begin{equation}
\frac{da}{dt} = -\frac{64}{5} \frac{q}{(1+q)^2} \frac{M^3}{a^3}
                \frac{G^3}{c^5}\,,
\end{equation}
with the result
\begin{equation}
t= \int_{a}^0 \frac{dt}{da'} da' = \frac{5}{256}\frac{(1+q)^2}{q} \frac{a^4}{M^3} \frac{c^5}{G^3}\,.
\end{equation}
If the binary initial separations $a$ are drawn from a log-flat
distribution (i.e., $dn/da \propto 1/a$), the distribution of the
merger times is also log-flat (cf.~\cite{2016arXiv161201524B}):
\begin{align}
\frac{dn}{dt} =\frac{dn}{da} \frac{da}{dt} \propto \frac{1}{a^4}\propto \frac{1}{t}\,.
\end{align}

The ``lookback time'' $t_L$ is given by {\cite{1999astro.ph..5116H}}
\begin{align}
t_L=\frac{1}{H_0}\int_0^z \frac{dz}{(1+z)\sqrt{\Omega_M (1+z)^3 + \Omega_\Lambda}}\,,
\end{align}
where we assume $\Omega_k=0$, $\Omega_M=0.307$, $\Omega_\Lambda=0.693$ and $H_0 = 67.7\, {\rm km s}^{-1} {\rm Mpc}^{-1}$ \cite{2016A&A...594A..13P}.
From the lookback time we can compute the time $t_L(z_1) - t_L(z_2)$
necessary for the Universe to evolve from redshift $z_1$ to redshift
$z_2$.

We distribute the 1g+1g sources uniformly in comoving volume with
redshifts $z<2$. For the 1g+2g population, we assume that 2g BHs
formed at some redshift $\tilde z$ drawn from the same distribution
used for 1g+1g binaries. We then extract a delay time $t_D$ from a
flat distribution in $\log(t_D)$ in the range
$t_D\in[10^{-4}$~Gyrs$,\,t_L(\tilde z)]$. The lower limit is very
conservative, and it roughly corresponds to the merger time for a
$10 M_\odot$ BH binary evolving from an initial orbital separation
$a=10 R_\odot$. The redshift $z$ of a 1g+2g merger is then given by
the numerical solution of the equation
\begin{align}
t_L(\tilde z) -t_L(z) = t_D\,. 
\label{solvez}
\end{align}

Finally, for the 2g+2g population we extract two values $\tilde z_1$,
$\tilde z_2$ from the 1g+1g distribution. The redshift $z$ of a 2g+2g
merger follows again from a numerical solution of Eq.~(\ref{solvez}),
with the difference that now we set
$\tilde z = \min(\tilde z_1,\tilde z_2)$.

{In Sec.~\ref{caveats} we will discuss how time delay
  prescriptions affect our results.}

\subsection{Measurable parameters}
\label{measurableparams}

For concreteness and simplicity, we will characterize each binary by
its total mass $M=m_1+m_2$, mass ratio $q=m_2/m_1\leq 1$, redshift $z$
and ``effective spin'' \cite{2016PhRvL.116x1102A}
\begin{align}
\chi_{\rm eff} = \frac{1}{M} \left( \frac{\mathbf{S_1}}{m_1} + \frac{\mathbf{S_2}}{m_2}\right)\cdot \mathbf{\hat L}\,.
\label{chieffdefinition}
\end{align}
The effective spin (a mass-weighted sum of the projection of the spins
{$\mathbf{S_i}=m_i^2 \chi_i \mathbf{\hat S_i}$} along the orbital angular momentum $\mathbf{L}$) is a
constant of the motion in post-Newtonian evolutions, at least at 2PN
order \cite{2008PhRvD..78d4021R, 2015PhRvD..92f4016G}. It is also the
easiest spin parameter to measure \cite{2016PhRvD..93h4042P,2016PhRvL.116x1102A}.

Let us introduce a vector ${\bf u}$ whose components are the
observable variables to use in our statistical analysis,
i.e. 
\begin{align} 
{\bf u}=\{M,\, q,\, z,\, \chi_{\rm eff} \}\,.  
\end{align}
The components of this vector will be labeled by an index $j=1,..., J$
such that $u_1=M$, $u_2=q$, etcetera; a capital Latin index $J$ will
denote the dimensionality of the vector $\bf u$, i.e. the number of
observables considered in the analysis.  Each binary in our catalog is
characterized by a specific set of observable properties
$\bar u^{(i)}$, where the superscript index $(i=1,...,I)$ labels
entries in our synthetic catalog.

The theoretical distributions of measurable source parameters
${\bf u}=\{M,\, q,\, z,\, \chi_{\rm eff} \}$ for 1g+1g, 1g+2g and
2g+2g events are compared in Fig.~\ref{fig:mergindis}. Each row
corresponds to one of the three mass distributions described in
Sec.~\ref{1g1gpop}.

The mass distributions have some noteworthy features. First of all,
and quite obviously, 2g BHs have higher component masses.  Therefore
the total mass is higher when 2g BHs are present (for any given
assumption on the mass distribution), and this effect is most notable
for the 2g+2g distributions. Mergers also tend to increase the number
of comparable-mass binaries, in part because of the fixed mass range
for the component masses ($m_i\in [5,50]M_\odot$).
For the ``power law'' mass function, the mass ratio of the 1g+2g
population peaks at $q=0.5$. This is because the mass distribution of
the primary BH is strongly peaked at the low end of the range (i.e.,
at $\sim 5M_\odot$), so many 2g binaries are nearly equal mass, with
component masses close to $5M_\odot$.

Redshift distributions also follow the expected trend: most 1g+1g
events occur at large redshift, whereas mergers involving one or two
2g BHs occur (on average) at smaller redshift, {because there is a
  time delay between the formation of 1g BHs via core collapse and
  their subsequent merger.}

The most striking differences are found in the distributions of
individual spins. To better illustrate this point, in
Fig.~\ref{fig:spinonly} we show the distribution of the individual
BH spins $(\chi_1,\,\chi_2)$, as well as the distribution of the spin
of the remnant $\chi_{\rm f}$.  As discussed in
\cite{2008ApJ...684..822B}, from a statistical point of view the
effect of mergers is to ``cluster'' BH spins around
$\chi_{\rm f}\sim 0.7$, quite independently of the progenitor
parameters. While the 1g+1g spin magnitudes are uniform in the range
$[0,\,1]$ by construction, spin distributions become peaked at
$\sim 0.7$ when 2g BHs are involved.  This clustering is evident in
the distribution of primary spins $\chi_1$ for the 1g+2g and 2g+2g
cases, and in the distribution of secondary spins $\chi_2$ for the
2g+2g case.  For the 1g+2g population, the peak at $\chi_2\sim 0.7$ is
less pronounced.  This is because the lower-mass BH is most likely 1g,
and the spin distribution of 1g BHs is by construction uniform in
$[0,\,1]$.

Unfortunately low-SNR GW observations of merger events are not very
sensitive to $\chi_1$ and $\chi_2$,
but rather to the effective spin $\chi_{\rm eff}$ defined in
Eq.~(\ref{chieffdefinition}). The right column of
Fig.~\ref{fig:mergindis} shows that the effect of mergers is
considerably smeared out in $\chi_{\rm eff}$, but more binaries with
$\chi_{\rm eff}\sim 0$ are expected if all sources are 1g BHs.
Measurements of $\chi_{\rm eff}$ may still be sufficient to
distinguish between different populations, especially when comparing
1g+1g against either 1g+2g or 2g+2g.  Discriminating between BH
progenitors should be considerably easier with future detectors, when
high-SNR events will allow for more precise measurements of $\chi_1$,
$\chi_2$ and $\chi_{\rm f}$
\cite{2016PhRvL.117j1102B,2016PhRvD..94l1501V,2017PhRvD..95f4052V} .

\subsection{Single detections}

{In the rest of this paper we will study how statistical inference from
  several detections can be used to constrain the underlying BH
  population. However, it is possible that \emph{single}
  detections with specific parameters can already provide smoking gun
  evidence for the occurrence of multiple mergers.}

{One possibility, as mentioned in Sec.~\ref{1g1gpop}, is that
  pair instabilities may prevent the formation of 1g BHs with masses
  sensibly above
  $\sim 50
  M_\odot$~\cite{2002ApJ...567..532H,2017ApJ...836..244W,2016A&A...594A..97B}. If
  this is indeed the case, a single detection of a merging BH binary
  where one of the components has mass larger than $50 M_\odot$ would
  indicate the occurrence of multiple mergers. This argument, however,
  relies on two crucial assumptions: (i) that 1g BHs always form from
  stellar collapse, while more exotic formation channels
  (e.g. involving primordial BHs) may produce massive BHs without
  invoking multiple mergers; (ii) that pair instabilities in core
  collapse do indeed prevent the formation of massive BHs. Pair
  instabilities, pair instability pulsations and the exact value of
  the maximum BH mass that can be produced via core collapse are all
  topics of current research~\cite{2016A&A...594A..97B}.}

{
  Another possibility involves accurate measurements of the component
  spins through the detection of a single nearby, non face-on binary
  merger with comparable, low masses and many precession cycles in the
  LIGO band. Unfortunately, parameter estimation studies suggest that
  current-generation detectors could allow dimensionless spin
  measurement errors $\sim 0.3$ in best-case scenarios
  \cite{2017PhRvD..95f4053V}. Errors of this magnitude are comparable
  to the width of the peaks in the spins distributions shown in
  Fig.~\ref{fig:spinonly} and there is significant uncertainty in the
  spin magnitude distribution of astrophysical BHs, so it seems
  unlikely that \emph{single} spin measurements may allow us to tell
  apart 1g BHs from 2g BHs, at least in the near future.}

\section{Observable distributions}
\label{sec:obsdistr}
From the {theoretical distributions} described in Sec.~\ref{sec:distributions}, we construct \emph{observable distributions} by (i) selecting detectable binaries according to a detection statistic, such as a threshold in the signal-to-noise ratio (SNR), and (ii) folding in measurement errors.

\subsection{Detection probability}

We first assign a detection probability ${\kappa}^{(i)}<1$ to each binary in our catalogs. This number takes into account the detector sensitivity and antenna pattern, as well as the (random) sky position  of the source. 
We compute ${\kappa}^{(i)}$ following the procedure outlined in Ref.~\cite{2015ApJ...806..263D}, where an astrophysical catalog of binaries produced using the {\sc Startrack} population synthesis code was filtered to produce similar catalogs of observable binaries for a specific set of GW detectors. This procedure is briefly reviewed below.

Each binary produces a GW strain $h(t)$ and an expectation value for the SNR 
\begin{equation}
\rho^2 = 4 \int_0^\infty\frac{|\tilde h(f)|}{S_n(f)} df\,,
\end{equation}
where $S_n(f)$ is the noise power spectral density of the detector and
$\tilde h(f)$ is the Fourier transform of the strain $h(t)$. The
strain is computed using the IMRPhenomC waveform
model~\cite{2010PhRvD..82f4016S}. In this paper we consider noise
power spectral densities for the first AdLIGO observing run (O1), the
Advanced LIGO design sensitivity~\cite{2016LRR....19....1A}, A+
(Advanced LIGO with squeezing)
and Voyager (the most advanced instrument that can be hosted in
facilities similar to LIGO) \cite{dcc_instrument}.

For any binary in our catalog we can compute $\rho_{\rm opt}$, i.e. the single-detector SNR for a binary that is optimally located and oriented in the sky. We then select those binaries in the catalog that are above a detection threshold $\rho_{\rm opt}\geq \rho_{\rm thr} = 8$. This criterion has often been used as a simple, reasonable proxy for a more realistic calculation of GW detection rates in multi-detector networks \cite{2010CQGra..27q3001A,2015ApJ...806..263D}. Then we compute the detection probability as
\begin{align}
\kappa^{(i)}= P\left(w^{(i)}\right)\,,
\end{align}
where the function $P(w^{(i)})$ is the cumulative distribution function for the projection parameter $w^{(i)}\equiv \rho_{\rm thr}/\rho^{(i)}_{\rm opt}$.  This cumulative distribution function takes into account the geometrical ``peanut factor''  that characterizes the sensitivity of the detector to the source sky location, inclination and polarization (see \cite{2015ApJ...806..263D} and references therein). Roughly speaking, $w^{(i)}=1$ means that the source is in a ``blind spot'' of the detector, while $w^{(i)}=0$ in the high-SNR limit.  
A tabulated version of $P(w^{(i)})$ is publicly available.\footnote{\href{http://www.phy.olemiss.edu/~berti/research}{www.phy.olemiss.edu/$\sim$berti/research}} We use standard spline interpolation to compute this function for generic values of $w^{(i)}$.

\begin{figure*}
\centering
\includegraphics[width=\textwidth]{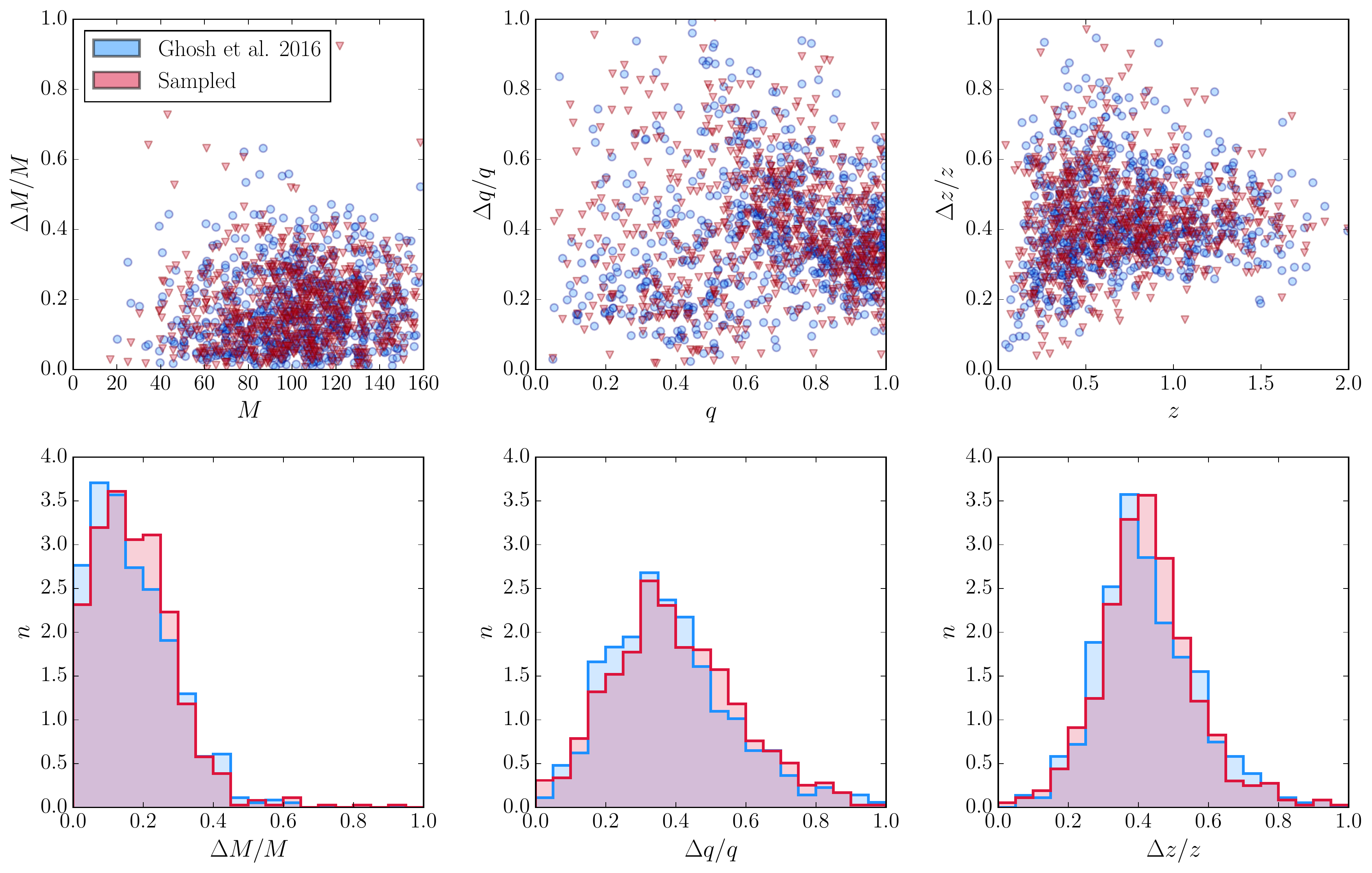}
\caption{Blue: relative errors on the total mass $M$ (left), mass ratio
  $q$ (middle) and redshift $z$ (right) as computed in
  Ref.~\cite{2016PhRvD..94j4070G}. Red: resampling of these data,
  obtained as described in Sec.~\ref{sec:errors}. The top panels show scatter
  plots of the relative error on each parameter as a function of the
  value of that parameter for the source. The bottom panels show the
  same information as a histogram. }
\label{fig:bettererrors}
\end{figure*}

\subsection{Measurement errors}
\label{sec:errors}

Ideally we should compute measurement errors for each binary in the catalog using Markov-Chain Monte Carlo methods, and use the obtained posteriors to perform model selection. This is computationally expensive, and unnecessary from the point of view of our proof-of-principle analysis. For our present purpose we adopt a much simpler prescription, described below.

We build on a study by Ghosh \emph{et al.} \cite{2016PhRvD..94j4070G}, who computed BH binary measurement errors using the \textsc{lalinference} code \cite{2015PhRvD..91d2003V} (see also \cite{2009PhRvD..79h4032A,2013PhRvD..87j4003L,2013PhRvD..87b4004C,2015PhRvD..92b2002G} for more work on the subject). In particular, we use their results for aligned-spin BH binaries detected by a network of 3 advanced detectors. Their data set provides 1$\sigma$ errors on several quantities, including the total mass $M$, mass ratio $q$ and redshift $z$. These are shown in blue in Fig.~\ref{fig:bettererrors}. 

The data set is too sparse to perform an efficient binning and interpolation in three dimensions ($M,\,q,\,z$). In order to partially account for the expected degeneracies (e.g., close binaries will generally have smaller errors on the masses), we adopt the following procedure. Consider a binary in our catalog with parameters $({\bar M,\,\bar q,\,\bar z})$. To estimate measurement errors on the parameters of this binary, we consider the 5 ``closest'' binaries in the data set of Ref.~\cite{2016PhRvD..94j4070G}, and compute the average and standard deviation of their measurement errors. Here ``closest'' is defined in the following sense: given the maximum and minimum value of each of the three parameters ($M,\,q,\,z$), we rescale their actual values so that these parameters are distributed in a cube of size one; then we compute the Euclidean distance between binaries in this cube. The average and standard deviation from the 5 closest binaries are then used to extract the measurement errors ${\sigma_{\bar M}, \sigma_{\bar q}, \sigma_{\bar z}}$ from a normal distribution. The red dots and histograms in Fig.~\ref{fig:bettererrors} show the measurement errors obtained from this resampling. The obtained distributions look remarkably close to the original data. Errors on the redshift are slightly overestimated, so (if anything) our resampling procedure seems to yield conservative predictions. Estimates for the errors on $\chi_{\rm eff}$ were not computed in  Ref.~\cite{2016PhRvD..94j4070G}, so we assume $\sigma_{\chi_{\rm eff}}=0.1$ for all binaries measured by LIGO at design sensitivity. {This rough estimate is quite conservative, and it is consistent with measurement errors in the first GW detections \cite{2016PhRvX...6d1015A}.}

Ref.~\cite{2016PhRvD..94j4070G} computed parameter estimation errors for the LIGO-Virgo network at design sensitivity. Fisher matrix arguments \cite{1995PhRvD..52..848P}  suggest that the capabilities of other detectors can be estimated rescaling the errors on the total mass, mass ratio, luminosity distance and $\chi_{\rm eff}$ by the ratio of SNRs, i.e. 
\begin{align}
\sigma_{\rm Detector} =\sigma_{\rm LIGO} \frac{\rho_{\rm LIGO}} {\rho_{\rm Detector}}\,.
\end{align}
Luminosity distance and redshift are related by
\begin{align}
D_L = \frac{1+z}{H_0} \int_0^z \frac{dz}{\sqrt{\Omega_M (1+z)^3 + \Omega_\Lambda}}\,,
\end{align}
(where we use units such that $c=1$), so that
\begin{align}
\frac{d D_L}{dz} = \frac{D_L}{1+z}+ \frac{1+z}{H_0 \sqrt{\Omega_M (1+z)^3 + \Omega_\Lambda}}\,.
\end{align}
The error on the redshift $\sigma_z$ is related to the error on $D_L$ by  
\begin{align}
 \left( \frac{\sigma_z}{D_L} \frac{d D_L}{dz}\right)^{2}=   \left( \frac{\sigma_{D_L}}{D_L}\right)^2 +\left( \frac{\sigma_{H_0}}{H_0}\right)^2
\end{align} 
where we assumed
$\sigma_{\Omega_\Lambda}\simeq 0$ (see
e.g. \cite{2002MNRAS.331..805H,2005PhRvD..71h4025B}). Given recent
discrepancies in the determination of $H_0$, we assume
${\sigma_{H_0}}/{H_0}=0.1$~\cite{2016PhLB..761..242D,2016ApJ...826...56R,2016JCAP...10..019B}.

\begin{figure*}
\includegraphics[width=\textwidth,page=1]{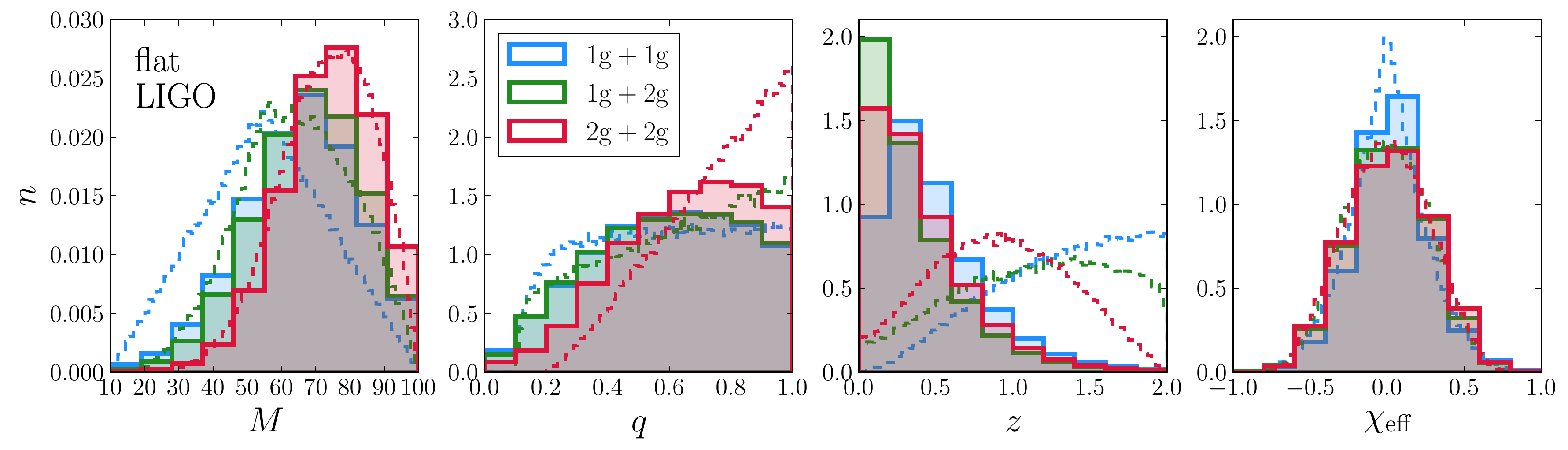}
\includegraphics[width=\textwidth,page=2]{filterdist_flat}
\caption{Observable distributions for  Advanced LIGO at design sensitivity (top) and Voyager (bottom). All plots refer to the ``flat'' mass distribution. In each panel, dashed lines show the theoretical distribution for the 1g+1g (blue), 1g+2g (green) and 2g+2g (red) populations; {these are the same curves shown in Fig.~\ref{fig:mergindis}. Following the same color scheme, solid shaded histograms} show the ``observed'' population, consisting of events that pass the SNR threshold and that include measurement errors.
  \label{fig:filterdist_flat}}
\end{figure*}

\subsection{Binning}

Recall that each binary is characterized by a vector of observable parameters ${\bf u}=\{u_1,\dots,u_J \}$.  If (for simplicity) we momentarily neglect measurement errors, the observable distribution is just a sum of Dirac deltas centered at $\bar {\bf u}^{(i)}$, and each delta is weighted by the detection probability ${\kappa}^{(i)}$:
\begin{align}
\tilde r({\bf u},\lambda) = \frac{{\sum_{i=1}^I {\kappa}^{(i)}}  \prod_{j=1}^J \delta\left(u_j - \bar u^{(i)}_j\right) }{\sum_{i=1}^I {\kappa}^{(i)}}\,,
\label{noerrors}
\end{align}
where $\lambda$ labels the model (cf. Sec.~\ref{sec:modsel}) and the denominator ensures normalization. 
Using the procedure described in Sec.~\ref{sec:errors} we can obtain estimates of the  1$\sigma$ errors on the measurement of each parameter. The $i$th binary in the catalog now has estimated parameters $\bar u^{(i)}$ with errors $\sigma^{(i)} = \sigma({\bar u^{(i)}})$. 
Assuming that errors are normally distributed and neglecting degeneracies,
we can substitute the Dirac deltas of Eq.~(\ref{noerrors}) with Gaussian distributions:
\begin{align}
\tilde r({\bf u},\lambda) = \frac{{\sum_{i=1}^I {\kappa}^{(i)}}  \prod_{j=1}^J 
\mathcal{N}\left(u_j; \bar u_j^{(i)},\sigma^{(i)}\right)
}{\sum_{i=1}^I {\kappa}^{(i)}}\;,
\label{witherrors}
\end{align}
where
\begin{align}
\mathcal{N}\left(u_j; \bar u_j^{(i)},\sigma^{(i)}\right) = \frac{1}{\sigma^{(i)} \sqrt{2\pi}}\exp{\left(-\frac{u_j - \bar u_j^{(i)}}{2\sigma^{2\,(i)}}\right)}\;.
\end{align}

Next, we need to bin the distributions $\tilde r({\bf u},\lambda)$. In each direction $j$, we construct bins $k_j$ with extrema $b_{k_j}$ and $B_{k_j}$, i.e. $u_j \in (b_{k_j}, B_{k_j})$. The function  $\tilde r({\bf u},\lambda)$ in each multi-dimensional bin $\{k_1,...,k_J\}$ is given by the integral 
\begin{align}
\tilde r_{k_1,...,k_J} (\lambda) = \int_{b_{k_1}}^{B_{k_1}} du_1 ... \int_{b_{k_J}}^{B_{k_J}} du_J \; \tilde r({\bf u},\,\lambda)
\nonumber
\\
=\frac{{\sum_{i=1}^I {\kappa}^{(i)}}  \prod_{j=1}^J   \int_{b_{k_j}}^{B_{k_j}}  \mathcal{N}\left(u_j; \bar u_j^{(i)},\sigma^{(i)}\right) du_j  }{\sum_{i=1}^I {\kappa}^{(i)}  }\;. \label{binrates}
\end{align}
In practice, we spread each source over multiple bins because of measurement errors (see~\cite{2010PhRvD..81j4014G,2011PhRvD..83d4036S} for a similar approach in the LISA context).
Eq.~(\ref{binrates}) is correctly normalized to $1$ only if the bins $k_j$ span the entire support of $\tilde r(\mathbf{u},\lambda)$. When substituting Dirac deltas with Gaussian distributions we are adding support in the whole range $[-\infty,+\infty]$ for each of the $u_j$'s, and inevitably we end up using a finite range. 
For simplicity, we just renormalize  $\tilde r_{k_1,...,k_J}(\lambda)$ such that 
\begin{align}
\sum_{k_1} ... \sum_{k_J} \tilde r_{k_1,...,k_J} = 1\,.
\end{align}

From now on we will identify the bins by a multi-index variable $k=\{k_1,...,k_J\}$, so (for example) we can write  $\sum_k f_k \equiv\sum_{k_1} ... \sum_{k_J} f_{k_1,...,k_J} $ for any binned quantity $f$.

\subsection{Putting the pieces together}

Examples of observable distributions are given in Fig.~\ref{fig:filterdist_flat} for Advanced LIGO at design sensitivity (top) and Voyager (bottom) assuming the ``flat'' mass function. In each panel, dashed lines show the theoretical distribution for the 1g+1g, 1g+2g and 2g+2g populations, as already presented in Fig.~\ref{fig:mergindis}. The histograms show the \emph{observable} population, i.e. the distribution of detectable binaries, where the measured parameters take into account also measurement errors. Some trends are visible. 

Let us first focus on the top row, which refers to observations with Advanced LIGO at design sensitivity.
It is clear that binaries with larger total mass and lower redshift produce stronger signals, and therefore they are more likely to be detected. In particular, Advanced LIGO can hardly detect any binaries at redshift $z\gtrsim 1$. The distribution of $\chi_{\rm eff}$ also shows a mild excess of observable events with $\chi_{\rm eff}\simeq 0$ for the 1g+1g population with respect to the 1g+2g and 2g+2g populations, suggesting that measurements of $\chi_{\rm eff}$ can indeed help to discriminate between populations. 

The bottom row of Fig.~\ref{fig:filterdist_flat} shows that the increased sensitivity of a Voyager-like detector has two main effects: it makes observable distributions in each of the parameters much closer to the corresponding theoretical distributions, and (quite importantly) it extends the reach of the detector to high-$z$ binaries. We obviously expect that more sensitive detectors will allow better discrimination between the different populations.%

The 2g+2g population presents a peak at $M\sim 80 M_\odot$ and $q\sim 1$. Equal-mass binaries  of $\sim 40 M_\odot+40 M_\odot$ can only be detected by Advanced LIGO at design sensitivity if they are located at very small redshift (cf. e.g. \cite{2016ApJ...818L..22A}). This explains the significant drop in the number of observed events as $q\to 1$. The effect is strongly mitigated in Voyager, because the instrument is more sensitive at low frequency.

\section{Statistical tools}
\label{sec:modsel}

In this section we briefly introduce statistical tools to perform Bayesian model selection.
We  label models by a parameter $\lambda$ that can be either discrete (if we want to distinguish two competing models $A$ and $B$) or continuous (if want to measure the ``mixing fraction'' between competing models that best describes the data). 

\begin{table*}
\renewcommand{\arraystretch}{1.3}

\begin{tabular}{ll|c|c|c}
& & $\quad1{\rm g}\!+\!1{\rm g}$ vs. $2{\rm g}\!+\!2{\rm g}\quad$ & $\quad1{\rm g}\!+\!1{\rm g}$ vs. $1{\rm g}\!+\!2{\rm g}\quad$ &  $\quad1{\rm g}\!+\!2{\rm g}$ vs. $2{\rm g}\!+\!2{\rm g}\quad$ \\
\hline
O1 LIGO  & $\;\;$flat$\;\;$ & 12.7 (15.8) & 2.0 (2.0) &6.4 (7.6)\\
& $\;\;$log$\;\;$ & 3.3 (3.5) &0.9 (0.9) & 3.5 (3.8) \\
& $\;\;$power law$\;\;$ & 0.7 (1.0) & 1.3 (1.6) &0.6 (0.6) \\
Ad. LIGO (design)& $\;\;$flat$\;\;$ & 30.2 (37.8) & 1.4 (3.7) & 21.9 (10.11) \\
&$\;\;$log$\;\;$ & 4.3 (7.0) & 0.6 (1.4) & 6.9 (5.1) \\
& $\;\;$power law$\;\;$ & 0.6 (1.7) & 1.0 (3.8) & 0.6 (0.5) \\
\end{tabular}
\caption{Odds ratios from the three O1 observations (GW150914, GW151226 and LVT151012) and from hypothetical observations of the same events at Advanced LIGO design sensitivity. {Odds ratios in parentheses were computed omitting all redshift information, i.e. considering the 3-dimensional vector of observables ${\bf u}=\{M,\, q,\, \chi_{\rm eff} \}$.}}
\label{O1oddstable}
\end{table*}

\subsection{Number of observations}

Our goal is to infer which model $\lambda$ best describes a set of data. As explained above, our binned distributions $\tilde r_k(\lambda)$ are normalized. To compare our models with the data we need an extra parameter $N(\lambda)$, the total number of observations predicted by model $\lambda$. We write  
\begin{align}
r_k(\lambda)= N(\lambda) \;\tilde r_k(\lambda)\,.
\label{rNtilde}
\end{align}

As for the individual binary parameters, we introduce an array ${\bf d}$ whose elements are the single observations $d^{(i)}$, which in turn are $J-$dimensional arrays. We bin the array ${\bf d}$ on the same grid used for the catalogs to obtain binned values $d_k$.

The likelihood of obtaining a data set $d_k$  from model $\lambda$ is given by
\begin{align}
p({\bf d} | \lambda) = \prod_k  \frac{({r_k(\lambda))}^{d_k} e^{-r_k(\lambda)}}{d_k!}\,.
\label{likelihood_notmar}
\end{align}

In our analysis the total number of observation does not contain information about the given model (this may not be the case for more realistic scenarios, where different models predict different merging rates: see e.g. \cite{2015ApJ...810...58S}). We therefore marginalize the likelihood over $N(\lambda)$. Plugging Eq.~(\ref{rNtilde}) into Eq.~(\ref{likelihood_notmar}) one obtains   \cite{2011PhRvD..83d4036S}
\begin{align}
p({\bf d}| \lambda) = \left( \prod_k   \frac{({\tilde r_k(\lambda))}^{d_k} e^{- \tilde r_k(\lambda)}}{d_k!} \right) \left( N(\lambda)^{\sum_k d_k} e^{-N(\lambda)} \right)\,,
\end{align}
and consequently the marginalized likelihood is 
\begin{align}
\tilde p ({\bf d}| \lambda) = \left( \prod_k   \frac{({\tilde r_k(\lambda))}^{d_k} e^{- \tilde r_k(\lambda)}}{d_k!} \right)\sum_N \left( N^{\sum_k d_k} e^{-N} \right)\,.
\label{marglikelihood}
\end{align}
Note that the term $\sum_N \left( N^{\sum_k n_k} e^{-N} \right)$ is a multiplicative coefficient that only depends on the data ${\bf d}$, and not on the model $\lambda$. This term can be ignored because, as we will see below, we are only interested in likelihood ratios, not in the likelihoods themselves.

From now on, to simplify notation, we will drop the tilde on $p$ and assume that likelihoods are always marginalized over the total number of events.

\subsection{Model selection }
\label{modselsec}

Let us first look at model comparison between \emph{pure} models, so that $\lambda$ is a discrete variable. Given models $\lambda=A$ and $\lambda=B$, their \emph{odds ratio} is defined as
\begin{align}
O_{AB} = \frac{p({\bf d}|A) \pi(A)} {p({\bf d}|B) \pi(B)}\,,
\end{align}
where $\pi$ is the prior probability assigned to each of the two models. The simplest assumption on the priors is $\pi(A)= \pi(B)=1/2$, such that the odds ratio reduces to the likelihood ratio. If $O_{AB}\gg 1$  ($O_{AB}\ll 1$) the data favors model $A$ ($B$). The probability of model $A$  is 
\begin{align}
p_A= \frac{O_{AB}}{1+O_{AB}} = \frac{p({\bf d}|A)}{p({\bf d}|A)+p({\bf d}|B)}\,,
\end{align}
and the probability of model $B$ is $p_B=1-p_A$. Sometimes $\sigma$-levels are used  to quantify the significance of a discrete model comparison, in analogy with Gaussian measurements. The expression relating the odds ratio $\mathcal{O}$ and $\sigma$ is
\begin{align}
\mathcal{O}= \frac{1}{1- 2 \erf(\sigma)}\,.
\end{align}

We can also assume that the data are represented by a mixture of two or more models, and assess whether the data themselves are informative about the underlying model mixing fractions. Each pure model $m$ enters the mixed model with a weight $f_m$, such that $\sum f_m=1$. Model comparison is equivalent to Bayesian inference on the parameters $\lambda=\{{f_1, f_2,\dots}\}$, as described by the posterior distribution
\begin{align}
p(\lambda | {\bf d})   = \frac{p( {\bf d} | \lambda)  \pi(\lambda)}{\int p( {\bf d} | \lambda)  \pi(\lambda) d\lambda}\,.
\label{bayesmixed}
\end{align}
As before, $\pi(\lambda)$ is the prior assigned to each  mixed model. We choose  $\pi(\lambda)$ to be uniformly distributed on the surface $\sum f_m=1$.\footnote{For instance, for a mixture of three models $\lambda=\{f_1,f_2,f_3\}$ the equation $\sum f_m=1$ describes a 2-dimensional surface $S$ of area $\sqrt{3}/2$. The uniform prior on $S$ is given by 
$\pi(f_1,f_2,f_3) = 2/ \sqrt{3}$, so that $\iint_S \pi \,dS=1$.}
From a computational point of view, we first draw values of $\lambda$ from the uniform prior, and then we produce a statistical sample distributed according to $p({\bf d}|\lambda)$ using a standard Monte Carlo hit-or-miss algorithm.

\section{Results}
\label{sec:results}

So far we have outlined a procedure to build a set of ``synthetic'' GW observations of merging BH binaries (along with their associated errors) from simple astrophysical considerations. We now wish to understand whether these observations can be used to distinguish between different populations using Bayesian model selection (see e.g. \cite{2010PhRvD..81j4014G,2011CQGra..28i4018G,2011PhRvD..83d4036S,2011PhRvD..84j1501B,2015ApJ...810...58S} for previous studies of this problem in different contexts).

\subsection{LIGO O1 data}
\label{O1data}

We first apply our model comparison tool to the three LIGO O1 observations. The data set $\mathbf{d}$ consists of the maximum likelihood values  provided in Ref.~\cite{2016PhRvX...6d1015A}:
\begin{itemize}
\item  GW150914:

$M=65.3 M_\odot$, $q=0.81$, $z=0.090$, $\chi_{\rm eff}=-0.06$.

\item  GW151226: 

$M=21.8 M_\odot$, $q=0.52$, $z=0.094$, {$\chi_{\rm eff}=0.21$}.

\item  LVT151012: 

$M=37 M_\odot$, $q=0.57$, $z=0.201$, $\chi_{\rm eff}=0.03$.

\end{itemize}
As stressed above, measurements errors are included in this analysis at the level of the catalogs, by spreading each source over multiple bins. A more in-depth study should make use of the posterior distribution of the observed parameters obtained through dedicated parameter-estimation pipelines.

Performing model selection as described in the previous sections and using the O1 sensitivity curve, we obtain the odds ratio reported in Table~\ref{O1oddstable}.
We also repeat the same exercise assuming the anticipated noise power spectral density of Advanced LIGO at design sensitivity. This basically answers the question: ``what if the O1 observations had been carried out with a better detector?''

As shown in Table~\ref{O1oddstable}, most of the odds ratios are in the range $0.3\lesssim \mathcal{O} \lesssim 3$, corresponding to $1\sigma$. This simply indicates that three observations are not enough to perform a meaningful statistical analysis. However some of the comparisons return odds ratios $\mathcal{O}\sim 10$, approaching $2\sigma$ evidence. When this happens (i) the 1g+1g population seems to be preferred, and (ii) the odds become higher for a more sensitive detector like Voyager. In these cases the algorithm seems to capture real statistical differences between the catalogs, that become more pronounced when more binaries are detected and measurement errors get smaller.

As a note of caution, we stress here that such discrete model comparison analyses can only tell us which of two competing models better describes a given data set, \emph{not} which model is correct. For instance, our results in Table~\ref{O1oddstable} show some dependence on the underlying mass distribution. This could be due either to the low dimensionality of the statistical sample (cf. Sec.~\ref{simulatedpure} below), or to the fact that none of the three mass distributions faithfully describes the observations. {To bracket uncertainties in the time delay prescription (cf. Sec.~\ref{redshiftdistribution}), Table~\ref{O1oddstable} also lists odds ratios computed omitting all redshift information. This calculation shows that assumptions on the time delays do not significantly affect our conclusions, given the limited statistics currently available.} It will be straightforward to update our analysis with higher statistics and better motivated BH binary formation models when more data become available.

\subsection{Simulated data: Pure models}
\label{simulatedpure}

\begin{figure*}
\includegraphics[page=1,width=\textwidth]{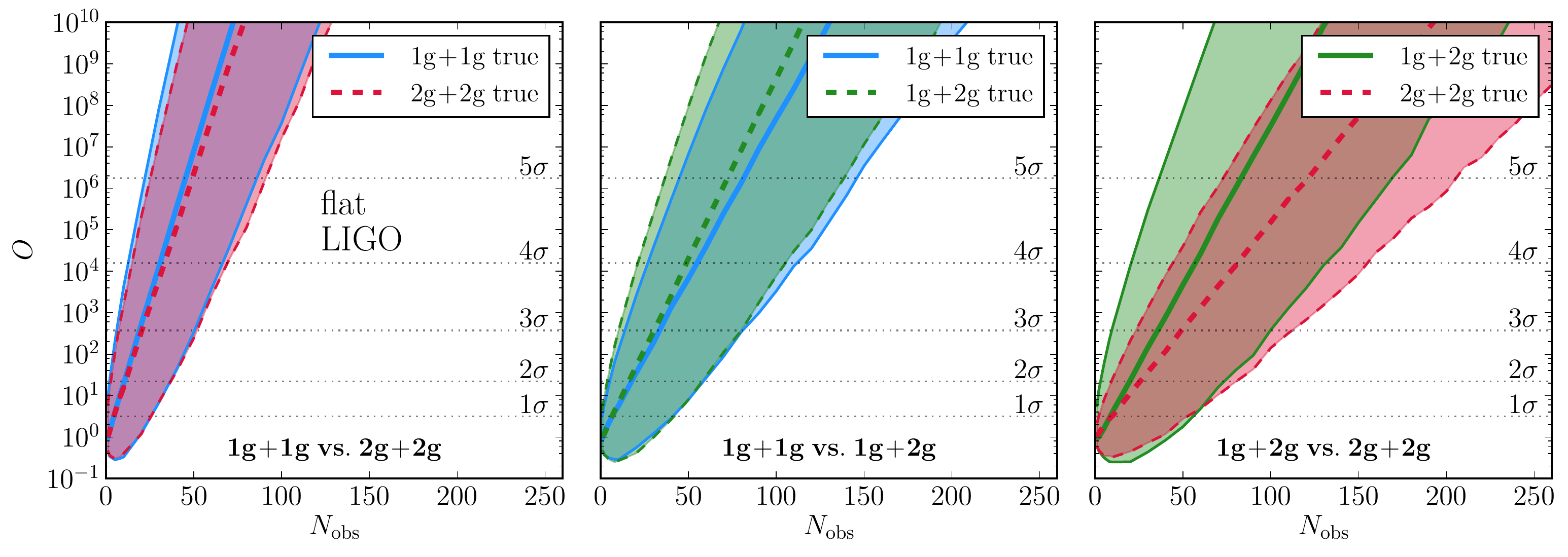}\\
\includegraphics[page=2,width=\textwidth]{oddsnobs_flat.pdf}
\caption{Number of events that are necessary to distinguish
  populations for Advanced LIGO at design sensitivity (top) and
  Voyager (bottom). The median odds ratio (thick lines) and 90\%
  confidence intervals to identify each model as true are plotted as
  functions of the number of observations $N_{\rm obs}$.}
  \label{oddsnobs}
\end{figure*}

\begin{table*}
\renewcommand{\arraystretch}{1.3}
\begin{tabular}{ll|ccc|ccc|ccc|ccc|ccc|ccc}
 \multicolumn{2}{c|}{$N_{\rm obs}$ at $5\sigma$}& \multicolumn{3}{c|}{$\;\;$  {\bf 1g1gT}/2g2g$\;\;$ } & \multicolumn{3}{c|}{$\;\;${\bf 2g2gT}/1g1g$\;\;$} & \multicolumn{3}{c|}{$\;\;${\bf 1g1gT}/1g2g$\;\;$} & \multicolumn{3}{c|}{$\;\;${\bf 1g2gT}/1g1g$\;\;$} & \multicolumn{3}{c|}{$\;\;${\bf 1g2gT}/2g2g$\;\;$} & \multicolumn{3}{c}{$\;\;${\bf 2g2gT}/1g2g$\;\;$}\\
&& $\;5\%\,$ & $\,50\%$ & $\,95\%\;$  & $\;5\%\,$ & $\,50\%$ & $\,95\%\;$   &  $\;5\%\,$ & $\,50\%$ & $\,95\%\;$&  $\;5\%\,$ & $\,50\%$ & $\,95\%\;$&  $\;5\%\,$ & $\,50\%$ & $\,95\%\;$ &  $\;5\%\,$ & $\,50\%$ & $\,95\%\;$  \\
\hline
LIGO O1 & $\;\;$flat$\;$ & 27& 53& 100& 31& 57& 103& 40& 76& 143& 44& 80& 146& 50& 105& 204& 77& 133& 233\\
 & $\;\;$log$\;$ & 27& 52& 94& 25& 50& 94& 30& 58& 106& 29& 56& 106& 42& 86& 165& 59& 104& 182\\
 & $\;\;$power law$\;$ & 14& 29& 57& 19& 35& 64& 7& 17& 34& 13& 23& 41& 31& 61& 114& 35& 64& 117\\
Ad.~LIGO & $\;\;$flat$\;$ & 23& 46& 86& 26& 50& 91& 45& 82& 146& 37& 73& 139& 37& 83& 170& 73& 122& 206\\
 & $\;\;$log$\;$ & 20& 41& 79& 24& 45& 83& 41& 73& 132& 33& 66& 122& 26& 56& 112& 48& 81& 138\\
 & $\;\;$power law$\;$ & 20& 39& 72& 18& 37& 70& 10& 21& 40& 11& 22& 41& 15& 31& 61& 20& 37& 67\\
 A+ & $\;\;$flat$\;$ & 18& 39& 75& 22& 43& 79& 46& 83& 149& 34& 69& 136& 34& 80& 165& 75& 123& 211\\
 & $\;\;$log$\;$ & 16& 34& 65& 19& 38& 69& 41& 73& 131& 30& 62& 120& 22& 51& 107& 50& 81& 136\\
 & $\;\;$power law$\;$ & 17& 35& 67& 17& 34& 65& 10& 22& 41& 10& 21& 40& 12& 27& 52& 20& 35& 61\\
 Voyager & $\;\;$flat$\;$ & 6& 15& 33& 10& 21& 39& 34& 69& 128& 27& 62& 122& 13& 36& 80& 36& 61& 102\\
 & $\;\;$log$\;$ & 4& 11& 25& 8& 17& 32& 25& 53& 102& 20& 51& 101& 8& 23& 51& 26& 44& 73\\
 & $\;\;$power law$\;$ & 5& 13& 26& 7& 16& 31& 9& 19& 37& 7& 18& 36& 4& 11& 24& 12& 21& 35\\
\end{tabular}

\caption{Number of observations needed to distinguish populations at $5\sigma$ with $5\%$, $50\%$ and $95\%$ probability. The ``true'' model is marked by a {\bf T} in the column header. For instance, in column {\bf 1g1gT}/2g2g we compare models 1g+1g and 2g+2g when observations are drawn from the 1g+1g catalog.}
\label{conflevel}
\end{table*}

The results of Sec.~\ref{O1data} show, not surprisingly, that more than three observations are needed to discriminate between different models. In order to estimate the capabilities of larger data sets and more sensitive detectors, here we perform model selection on simulated observations.  Our main goal is to estimate  how many observations are needed to distinguish a pair of models with a given confidence level.

Given a model $\lambda_{\rm true}$, we extract the number of events per bin $d_k$ assuming a Poisson distribution 
\begin{align}
p(d_k) = \frac{{r_k(\lambda_{\rm true})}^{d_k} e^{-r_k(\lambda_{\rm true})}}{d_k!}\;. \label{nipoisson}
\end{align}
Here the total number of observation $N_{\rm obs}=N(\lambda_{\rm true})$ is a free parameter that we need to specify. We expect model comparison to be easier/harder if more/less observations are available. 
This statement is made more quantitative in Fig.~\ref{oddsnobs} and Table~\ref{conflevel}. 

Figure~\ref{oddsnobs} shows the odds ratio distribution obtained from several realization of $N_{\rm obs}$ observations.  For each pair of models we plot $\mathcal{O}_{AB}$ (when $A$ is the true model) and $\mathcal{O}_{BA}$ (when $B$ is the true model), thus addressing how easy (or hard) it is to identify any of the models \emph{if it is correct}. Thick lines mark the median odds, while the shaded areas encompass  90\% of the realizations (i.e., they cover the range between the 5th and the 95th percentiles).	

The odds ratio $\mathcal{O}$ increases roughly exponentially with the number of observations $N_{\rm obs}$, so our ability to distinguish between different models should rapidly improve in the coming years. Table~\ref{conflevel} shows that in $5\%$ of the realizations, as few as $\sim 20$ detections are enough to discriminate the 1g+1g population from the 2g+2g population at $5\sigma$ with Advanced LIGO at design sensitivity, while $N_{\rm obs}\sim 80$ observations are necessary to achieve $5\sigma$ confidence in 95\% of the realizations.

Model selection involving the 1g+2g population typically requires a larger number of observations. This is clear when comparing the left panels of Fig.~\ref{oddsnobs} to the middle and right panels. In both the (1g+1g vs. 1g+2g) and (1g+2g vs. 2g+2g) comparisons the odds ratio grows (roughly) exponentially, but with smaller slope compared to the (1g+1g vs. 2g+2g) case. However the slope (and the odds ratio $\mathcal{O}$) is larger when 1g+2g is the true model: it is slightly easier to mistake a 1g+1g (or 2g+2g) population for a 1g+2g population than vice versa.

Model comparison is easier with more sensitive detectors. For example, distinguishing 1g+1g from 2g+2g at $5\sigma$ in 90\% of the realizations requires only $\sim 30$ Voyager observations (instead of $\sim 80$ for Advanced LIGO at design sensitivity).

In Sec.~\ref{O1data}, where only three observations were considered, the results were greatly dependent on the assumed mass distribution. Table~\ref{conflevel} shows that this dependence becomes much weaker when more observations are available and/or the instrumental sensitivity improves. This is largely due to the discriminating power of the redshift distribution of the events, which becomes more relevant when high-$z$ binaries become detectable (cf. Fig.~\ref{fig:filterdist_flat}).

\subsection{Simulated data: Mixed models}

\begin{figure*}[p]
\centering
\includegraphics[width=0.9\textwidth,page=1]{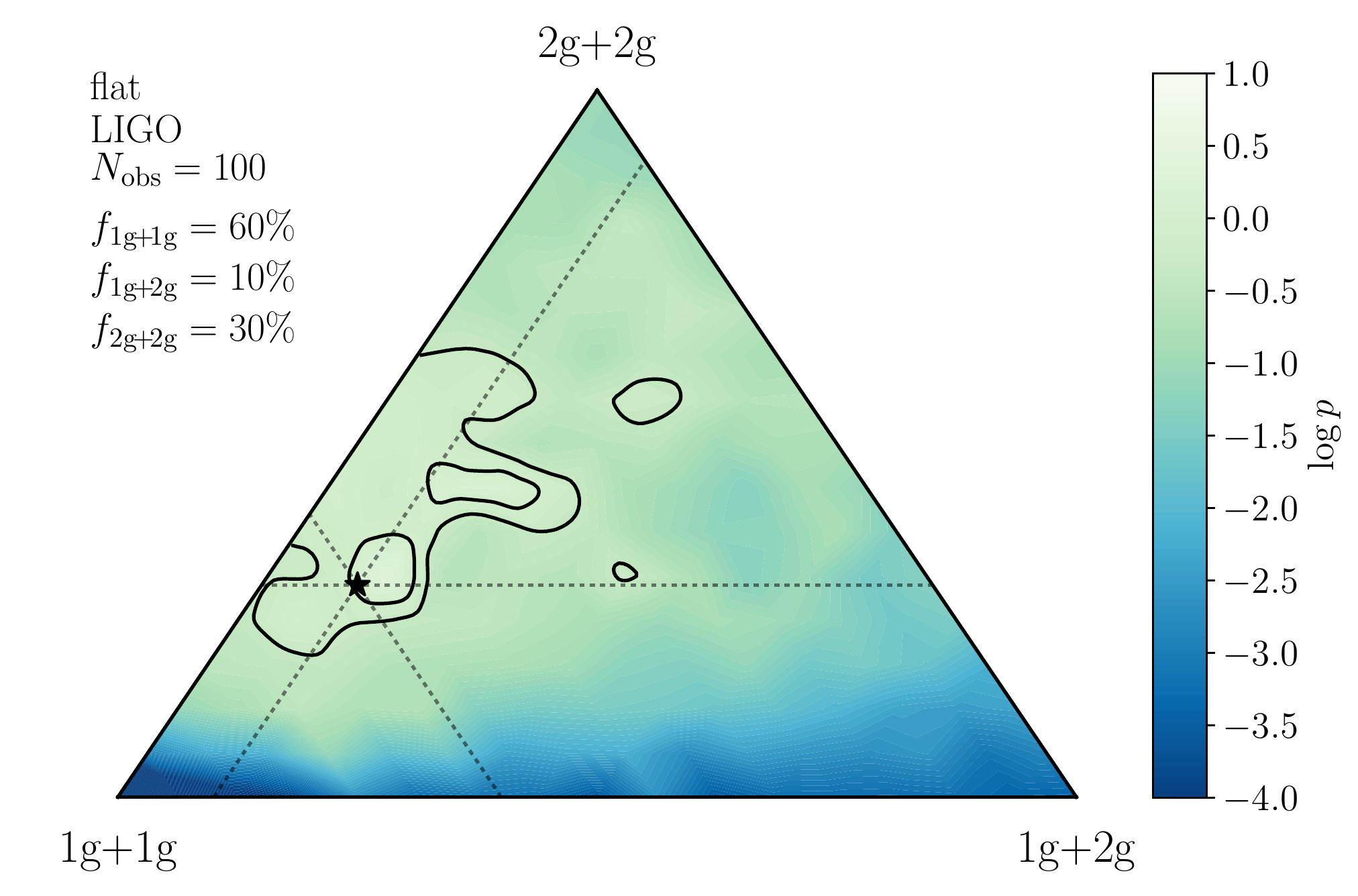}\\
\vspace{0.5cm}
\includegraphics[width=0.24\textwidth,page=1]{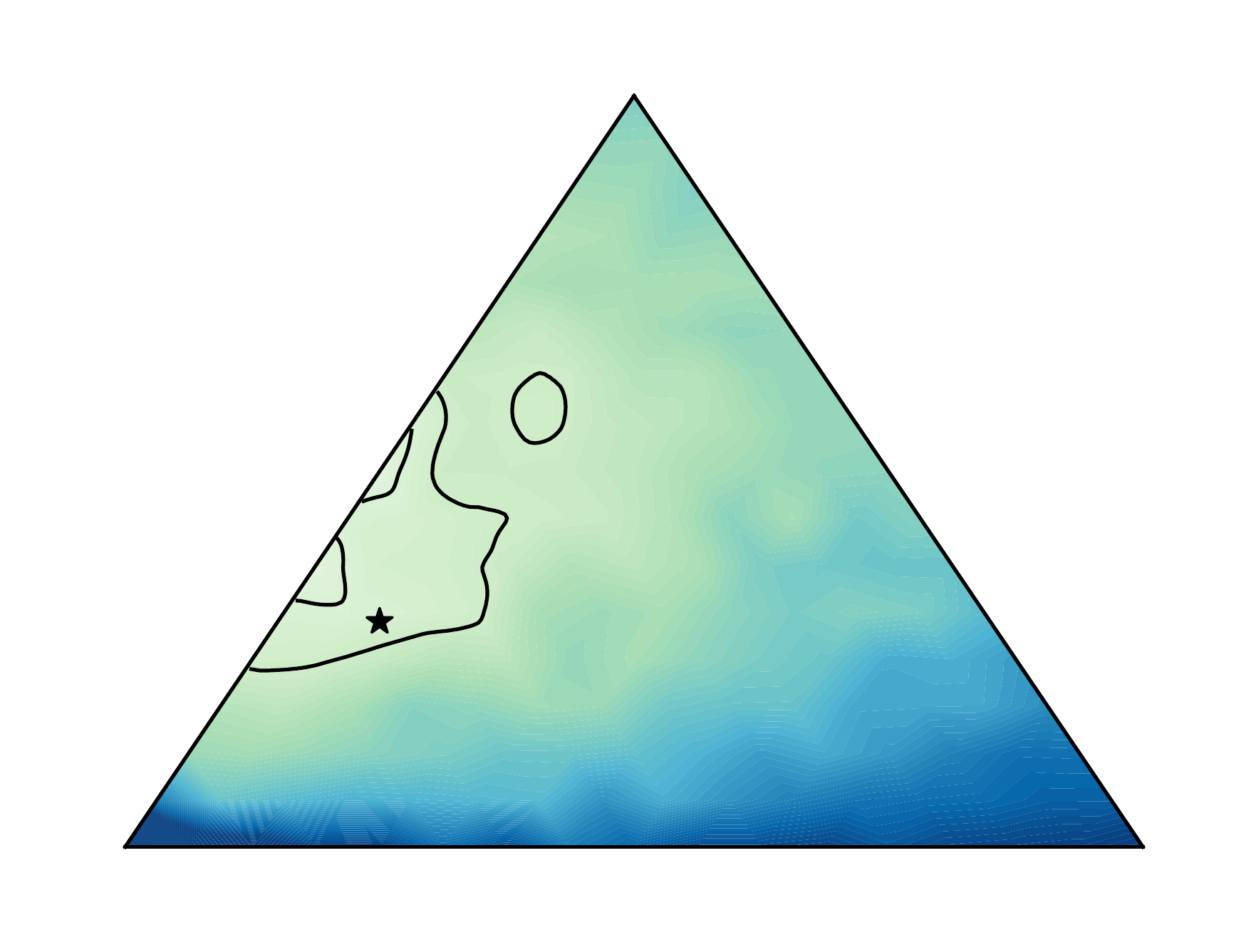}
\includegraphics[width=0.24\textwidth,page=2]{mixfractions_simplex_off.pdf}
\includegraphics[width=0.24\textwidth,page=3]{mixfractions_simplex_off.pdf}
\includegraphics[width=0.24\textwidth,page=4]{mixfractions_simplex_off.pdf}
\includegraphics[width=0.24\textwidth,page=5]{mixfractions_simplex_off.pdf}
\includegraphics[width=0.24\textwidth,page=6]{mixfractions_simplex_off.pdf}
\includegraphics[width=0.24\textwidth,page=7]{mixfractions_simplex_off.pdf}
\includegraphics[width=0.24\textwidth,page=8]{mixfractions_simplex_off.pdf}
\vspace{1cm}
\caption{Posterior distribution of the mixing fraction between the 1g+1g, 2g+2g and 1g+2g pure models. Each triangle shows the model space defined by $\sum f =1$ for a given realization of $N_{\rm obs}=100$ observations. The corners correspond to the three pure models. The black star marks the ``true'' injected value of the mixing fractions. Each of the injected mixing fractions is constant along one of the dashed lines.  The log-likelihood is shown in the color map: lighter regions are more likely than darker regions. Solid black contours mark the 50\% and 90\% confidence regions.}
\label{mixedfigure}
\end{figure*}
 
Let us now turn to a more ambitious task. As anticipated in Sec.~\ref{modselsec}, we now consider a population of binaries consisting of a \emph{mixture} of the three pure models 1g+1g, 1g+2g and 2g+2g. The task is to measure their mixing fraction, i.e. to determine how many binaries belong to each of the three pure populations. This is computationally expensive, as it requires many evaluations of the likelihood defined in Eq.~(\ref{bayesmixed}) through Monte Carlo methods.
 
 As a proof of principle, in Fig.~\ref{mixedfigure} we show results  for a specific choice of the mixing parameters. Simulated observations are drawn from a model\footnote{{Our injected fraction of 2g BHs was chosen only for illustrative purposes. It is higher than current estimates of merger rates in nuclear clusters, which are favorable environments for multiple merger events~\cite{2016ApJ...831..187A}.}} where $60\%$ of the binaries are 1g+1g, $10\%$ are 2g+2g, and $30\%$ are 1g+2g: 
\begin{align} 
\lambda_{\rm true}\equiv 
\big\{f_{\rm 1g+1g}, f_{\rm 1g+2g}, f_{\rm 2g+2g}\big\}=
\big\{0.6, 0.1, 0.3\big\}\,. 
\end{align}
For concreteness we assume the ``flat'' mass prescription and consider several realizations of $N_{\rm obs}=100$ observations performed with the Advanced LIGO network at design sensitivity. Each of the triangles in Fig.~\ref{mixedfigure} shows the surface $f_{\rm 1g+1g}+f_{\rm 1g+2g}+f_{\rm 2g+2g}=1$. The color coding corresponds to the values of the posterior $p(\lambda | {\bf d})$. Pure models lie on the corners of this ``Dalitz plot,'' while the star marks the injected fraction.

As expected, measuring mixing fractions is sensibly harder than performing discrete model comparison, and it is going to require many more observations (a similar result was obtained in Ref.~\cite{2011PhRvD..83d4036S}). The injected fractions are recovered only in some of the realizations, suggesting that these data points are probably not enough to confidently perform the measurement.

In any case, we can note some trends. Most (but not all) of the realizations assign a rather low probability to the region where $f_{2g+2g}\sim 0$. Whenever a few 2g+2g events are present, their properties are sensibly different from those involving 1g BHs, and therefore the 2g+2g population can be identified relatively easily. Although we may be unlucky and estimate mixing fractions which are sensibly different from their true values, our model comparison algorithm does return a statistically consistent result. Out of 1000 realizations, we find that the correct mixing fraction is identified within the 50\% (90\%) confidence interval in $57\%$ (25\%) of the cases. The relatively small number of observations is likely to be one of the main reasons for this relatively pessimistic result: if we assume $N_{\rm obs}=1000$, the correct mixing fraction is identified within the 50\% (90\%) confidence interval in $90\%$ (77\%) of the cases.

In conclusion, this preliminary study shows that measuring mixing fractions will be challenging in the near future. Estimating mixing fractions with high confidence may require several hundreds (if not thousands) of observations.

\subsection{Caveats on mass functions and time delays}
\label{caveats}

We have shown that, given a sufficient number of detected events, it
is possible to distinguish a given 1g+1g BH population from a variant of
the \emph{same} population where repeated mergers occur. Here we
discuss how uncertainties in the assumed 1g+1g mass distribution and in
time delay prescriptions may bias our conclusions.

\begin{table*}
\centering
\renewcommand{\arraystretch}{1.3}

\begin{tabular}{@{\hskip 0.2in}c@{\hskip 0.2in}|@{\hskip 0.15in}rcl@{\hskip 0.2in}|@{\hskip 0.15in}c@{\hskip 0.2in}|l}
Injection & &Test& & Preferred \\
\hline
\hline
flat 1g+1g & flat 1g+1g&~vs~~&flat 2g+2g & flat 1g+1g & \cmark T\\
 & log 1g+1g&~vs~~&log 2g+2g & not significant \\
  & power law 1g+1g&~vs~~&power law 2g+2g & power law 1g+1g &\cmark\\
flat 2g+2g & flat 1g+1g&~vs~~&flat 2g+2g & flat 2g+2g & \cmark T\\
  & log 1g+1g&~vs~~&log 2g+2g & log 2g+2g &\cmark \\
  & power law 1g+1g&~vs~~&power law 2g+2g & power law 1g+1g & \xmark\\
\hline
log 1g+1g & flat 1g+1g&~vs~~&flat 2g+2g & flat 1g+1g&\cmark \\
 & log 1g+1g&~vs~~&log 2g+2g & log 1g+1g &\cmark T\\
 & power law 1g+1g&~vs~~&power law 2g+2g & power law 1g+1g &\cmark\\
log 2g+2g & flat 1g+1g&~vs~~&flat 2g+2g & flat 1g+1g &\xmark \\
  & log 1g+1g&~vs~~&log 2g+2g & log 2g+2g &\cmark\\
  & power law 1g+1g&~vs~~&power law 2g+2g & power law 1g+1g &\xmark\\
\hline
power law 1g+1g & flat 1g+1g&~vs~~&flat 2g+2g & flat 1g+1g &\cmark\\
 & log 1g+1g&~vs~~&log 2g+2g & log 1g+1g &\cmark\\
  & power law 1g+1g&~vs~~&power law 2g+2g & power law 1g+1g &\cmark T\\
power law 2g+2g & flat 1g+1g&~vs~~&flat 2g+2g & flat 1g+1g &\xmark\\
  & log 1g+1g&~vs~~&log 2g+2g & log 1g+1g &\xmark\\
  & power law 1g+1g&~vs~~&power law 2g+2g & power law 2g+2g &\cmark T\\
\end{tabular}
\renewcommand{\arraystretch}{1}
\caption{{Model comparison tests between populations characterized by different merger generations \emph{and} mass distributions using the Advanced LIGO sensitivity curve. For each injected distribution and model comparison we report the preferred population in the limit where $N_{\rm obs}\to\infty$ (in practice we use $N_{\rm obs} =10^3$). The true population (T) is correctly identified whenever it is among those tested. While most of the comparisons correctly identify the merger generation (rows denoted by a  check mark \cmark), in some cases making the wrong assumption on the underlying mass distribution prevents a correct identification (rows denoted by a cross \xmark). In one case (second row) we obtained odds ratios consistent with one even when  $N_{\rm obs}\to\infty$, so that no conclusions can be drawn and the comparison is marked as ``not significant.''} {In all other cases the behavior of $\mathcal{O}(N_{\rm obs})$ is qualitatively similar to Fig.~\ref{oddsnobs}, i.e. the odds ratio grows exponentially until populations can be distinguished at $5\sigma$.}}
\label{differentmasses}
\end{table*}

In Table~\ref{differentmasses} we perform pure model comparisons
between BH binary populations that differ in \emph{both} merger
generation (1g+1g, 1g+2g, 2g+2g) scenario and in the assumed mass
distribution. As shown in Sec.~\ref{simulatedpure}, the true
distribution is correctly identified whenever it is among those
tested. When the injected population is not among those being
compared, differences in the assumed mass distribution can sometimes
dominate over differences induced by the occurrence of subsequent
mergers. For instance, if injected 2g+2g observations assuming the
``flat'' mass distributions are examined assuming a ``power law'' mass
model, one would erroneously conclude that the observed population is
1g+1g, rather than 2g+2g. 

{However this should not be a problem in practice, because the mass
distribution should soon be well constrained by
observations. Realistic astrophysical scenarios typically predict a
small fraction of multiple mergers, i.e., a small fraction of 2g
events. Even remaining theory-agnostic, this anticipation is already
(although inconclusively) supported by the data. Our Table
\ref{O1oddstable} suggests that 1g+1g mergers may already be favored
over 2g scenarios. 
So, in practice, 
there are
theoretical and (hopefully soon) experimental reason to assume that
the majority of detected events have a 1g+1g origin. In this very
plausible scenario, the model selection procedure can be
``bootstrapped'' as follows:}

\begin{itemize}
\item[{(i)}] {The mass distribution is inferred from a large enough
  number of detections, assuming that most events are 1g+1g;}

{\item[(ii)] This observationally inferred mass distribution can be
  used to replace our ``toy'' mass distributions (flat, log or power
  law) for 1g+1g BHs, and the 2g distributions can be constructed
  through hierarchical mergers as described earlier;}

{\item[(iii)] We can now look at all measurable properties of the
  population to determine whether some (presumably small) fraction of
  events has a 2g origin.}
\end{itemize}

{Table~\ref{differentmasses} shows that step (i) above does not present
problems. Indeed, according to Table~\ref{differentmasses}, while it
is indeed possible to wrongly rule in favor of 1g+1g BHs given 2g+2g
injections, the converse is unlikely: if model selection favors 2g+2g
BHs, the injected data never belong to a 1g+1g population with a
different mass spectrum.}

To quantify the importance of time delays, we repeated all the
comparisons shown in Table~\ref{conflevel} excluding redshift
information, i.e. taking ${\bf u}=\{M,\, q,\,\chi_{\rm eff} \}$ as our
vector of observable quantities. We find that the correct population
is always identified. The odds still grow exponentially with the
number of observations, although with somewhat shallower slopes. This
is expected, because the statistical analysis is performed using less
information. Omitting redshift information does not significantly
affect the 1g+1g vs. 2g+2g and 1g+2g vs. 2g+2g comparisons, but it
plays a more important role in the 1g+1g vs 1g+2g comparisons. This is
because (as illustrated in Fig.~\ref{fig:filterdist_flat}) the mass
distributions are very similar for these populations, which are
therefore harder to distinguish if redshifts are ignored. For
instance, while Fig.~\ref{oddsnobs} shows that $\sim 40$ observations
are enough to distinguish the 1g+1g and 1g+2g ``flat'' populations at
$3\sigma$ with LIGO in 50\% of the realizations, up to $\sim 200$
events will be necessary to reach the same conclusion in the absence
of redshift information.

\section{Conclusions}	
\label{sec:conclusions}

The main result of this paper is that GW observations of merging stellar-mass BH binaries can be used to gather information about their progenitors. Starting from simple, physically motivated populations of  ``first generation'' (1g) BHs born from stellar collapse, we construct populations where merging binaries include ``second generation'' (2g) BHs, whose masses and spins are computed using numerical relativity fitting formulas.
Then we use Bayesian model selection to determine whether current or future ground-based GW interferometers can distinguish different populations. If 2g BHs occur in nature, it should be possible to recover evidence for their existence from GW data; otherwise, the data can be used to constrain astrophysical models that produce 2g BHs. 

As a first application of our Bayesian model selection framework, we perform model selection using the two confirmed detections (GW150914 and GW151226) and the LVT151012 trigger from Advanced LIGO's first observing run. It is quite remarkable that, even with only three data points, some of the comparisons show odds ratios as high as $\sim 10$ in favor of 1g BHs.
As expected, model selection performance improves with more observations and more sensitive detectors. Indeed, as shown in Fig.~\ref {oddsnobs}, the Bayesian odds ratio for comparisons between two pure models scales (roughly) exponentially with the number of observations. Depending on the actual realization, $\sim 20$--$200$ Advanced LIGO observations at design sensitivity should allow us to discriminate which of the three populations is favored by the data at $5\sigma$ confidence level in one-to-one comparisons. Instrumental upgrades will bring this number down to $15$--$200$ observations for A+, and $5$--$100$ for Voyager. 

More realistically, astrophysical populations of merging binaries will be a mixture of all three populations (1g+1g, 1g+2g, 2g+2g), and the real experimental task is to determine the relative mixing fractions. Using simulated data, we construct synthetic catalogs assuming a mixture of models for the different BH generations, and attempt to measure the mixing fractions using Bayesian inference. Our preliminary results suggest that this is a much more challenging task: recovering the mixing fractions may require several hundreds (if not thousands) of observations.

This work should be regarded as a proof-of-principle study that can (and should) be extended in several directions. Our simple models are not supposed to be astrophysically realistic: they were developed solely to show that, at least in principle, GW observations could provide information on the occurrence of multiple stellar-mass BH mergers. The inclusion of detailed spin alignment models and more realistic mass distributions (see e.g. \cite{2016arXiv161101157K}), preferably with input from population synthesis codes, is an important topic for future investigation.

As illustrated in Sec.~\ref{measurableparams} (see, in particular, Fig.~\ref{fig:spinonly}), the spin magnitudes of the merging BHs are very sensitive to their merger history. This is also true for the massive BH binaries observable by LISA: see e.g. \cite{2008ApJ...684..822B,2014ApJ...794..104S}. Unlike BHs born from stellar collapse, the spin distribution of postmerger BHs should be strongly peaked at $\chi_{\rm f}\sim 0.7$. In this paper we only considered measurements of the ``effective spin'' $\chi_{\rm eff}$, because this is the spin parameter that enters at lowest PN order in the gravitational waveform. This is a very conservative approach. As shown in Fig.~\ref {fig:mergindis}, the ``memory effect'' encoded in the spin magnitudes is largely washed out in this variable. Measurements of the individual spin magnitudes should be possible by considering better waveform models or higher SNR signals: in this sense, our predictions should be regarded as conservative. Moreover, high-SNR ringdown observations will allow measurements of the final (postmerger) spin $\chi_{\rm f}$ within a few percent \cite{2016PhRvL.117j1102B}. These measurements could also be used to identify the progenitors of merging BHs.

The model selection framework developed in this paper is complementary to other studies, which usually focus on discriminating specific astrophysical formation channels (e.g., field binaries vs. dynamical formation scenarios  \cite{2015ApJ...810...58S,2017CQGra..34cLT01V,2016ApJ...832L...2R,2016PhRvD..94f4020N,2016ApJ...830L..18B,2017MNRAS.465.4375N}, but see also \cite{2016MNRAS.457.4499H} for work on intermediate-mass BHs). {We focused on using statistical distributions consisting of several observations, but it is possible that single events may be smoking guns for (or against) multiple merger scenarios, at the price of making stronger assumptions on the formation mechanism of 1g BHs. For example, binaries with component masses above the pair-instability gap~\cite{2016A&A...594A..97B} can point to the occurrence of multiple mergers is we assume that 1g BHs are formed via core collapse, and if we are confident about the upper mass limit on 1g BHs set by pair instabilities.} We hope that our approach will spark more studies of the astrophysical information encoded in present and future GW data sets.

\vspace{0.2cm} 

While completing our study we learned that Maya Fishbach, Daniel Holz and Ben Farr have been pursuing a similar investigation \cite{2017arXiv170306869F}. Their work nicely complements our own, as they focus on the spin distributions and address the detectability of more than two generations of mergers.

\acknowledgements

We thank Archisman Ghosh, Walter Del Pozzo and Parameswaran Ajith for
sharing parameter estimation data from
Ref.~\cite{2016PhRvD..94j4070G}. We also thank Daniel Holz, Maya
Fishbach, Ben Farr, Leo Stein and Chris Moore for discussions. D.G. is
supported by NASA through Einstein Postdoctoral Fellowship Grant
No. PF6-170152 awarded by the Chandra X-ray Center, which is operated
by the Smithsonian Astrophysical Observatory for NASA under Contract
NAS8-03060. E.B. was supported by NSF Grant No. PHY-1607130 and by FCT
contract IF/00797/2014/CP1214/CT0012 under the IF2014 Program.
This work was supported by the H2020-MSCA-RISE-2015 Grant No. StronGrHEP-690904.
Computations were performed on the Caltech computer cluster ``Wheeler,'' supported by the Sherman Fairchild Foundation and Caltech. Partial support is acknowledged by NSF CAREER Award PHY-1151197. 
This research made use of python packages \textsc{astropy} \cite{2013A&A...558A..33A} and  \textsc{matplotlib}
\cite{2007CSE.....9...90H}.
\bibliography{paper2gen}

\begin{thebibliography}{103}%
\makeatletter
\providecommand \@ifxundefined [1]{%
 \@ifx{#1\undefined}
}%
\providecommand \@ifnum [1]{%
 \ifnum #1\expandafter \@firstoftwo
 \else \expandafter \@secondoftwo
 \fi
}%
\providecommand \@ifx [1]{%
 \ifx #1\expandafter \@firstoftwo
 \else \expandafter \@secondoftwo
 \fi
}%
\providecommand \natexlab [1]{#1}%
\providecommand \enquote  [1]{``#1''}%
\providecommand \bibnamefont  [1]{#1}%
\providecommand \bibfnamefont [1]{#1}%
\providecommand \citenamefont [1]{#1}%
\providecommand \href@noop [0]{\@secondoftwo}%
\providecommand \href [0]{\begingroup \@sanitize@url \@href}%
\providecommand \@href[1]{\@@startlink{#1}\@@href}%
\providecommand \@@href[1]{\endgroup#1\@@endlink}%
\providecommand \@sanitize@url [0]{\catcode `\\12\catcode `\$12\catcode
  `\&12\catcode `\#12\catcode `\^12\catcode `\_12\catcode `\%12\relax}%
\providecommand \@@startlink[1]{}%
\providecommand \@@endlink[0]{}%
\providecommand \url  [0]{\begingroup\@sanitize@url \@url }%
\providecommand \@url [1]{\endgroup\@href {#1}{\urlprefix }}%
\providecommand \urlprefix  [0]{URL }%
\providecommand \Eprint [0]{\href }%
\providecommand \doibase [0]{http://dx.doi.org/}%
\providecommand \selectlanguage [0]{\@gobble}%
\providecommand \bibinfo  [0]{\@secondoftwo}%
\providecommand \bibfield  [0]{\@secondoftwo}%
\providecommand \translation [1]{[#1]}%
\providecommand \BibitemOpen [0]{}%
\providecommand \bibitemStop [0]{}%
\providecommand \bibitemNoStop [0]{.\EOS\space}%
\providecommand \EOS [0]{\spacefactor3000\relax}%
\providecommand \BibitemShut  [1]{\csname bibitem#1\endcsname}%
\let\auto@bib@innerbib\@empty
\bibitem [{\citenamefont {{Abbott {\it et al.} (LIGO Scientific Collaboration
  and Virgo Collaboration)}}(2016{\natexlab{a}})}]{2016PhRvL.116f1102A}%
  \BibitemOpen
  \bibfield  {author} {\bibinfo {author} {\bibfnamefont {B.~P.}\ \bibnamefont
  {{Abbott {\it et al.} (LIGO Scientific Collaboration and Virgo
  Collaboration)}}},\ }\href {\doibase 10.1103/PhysRevLett.116.061102}
  {\bibfield  {journal} {\bibinfo  {journal} {\prl}\ }\textbf {\bibinfo
  {volume} {116}},\ \bibinfo {eid} {061102} (\bibinfo {year}
  {2016}{\natexlab{a}})},\ \Eprint {http://arxiv.org/abs/1602.03837}
  {arXiv:1602.03837 [gr-qc]} \BibitemShut {NoStop}%
\bibitem [{\citenamefont {{Abbott {\it et al.} (LIGO Scientific Collaboration
  and Virgo Collaboration)}}(2016{\natexlab{b}})}]{2016arXiv160604855T}%
  \BibitemOpen
  \bibfield  {author} {\bibinfo {author} {\bibfnamefont {B.~P.}\ \bibnamefont
  {{Abbott {\it et al.} (LIGO Scientific Collaboration and Virgo
  Collaboration)}}},\ }\href {\doibase 10.1103/PhysRevLett.116.241103}
  {\bibfield  {journal} {\bibinfo  {journal} {\prl}\ }\textbf {\bibinfo
  {volume} {116}},\ \bibinfo {eid} {241103} (\bibinfo {year}
  {2016}{\natexlab{b}})},\ \Eprint {http://arxiv.org/abs/1606.04855}
  {arXiv:1606.04855 [gr-qc]} \BibitemShut {NoStop}%
\bibitem [{\citenamefont {{Abbott {\it et al.} (LIGO Scientific Collaboration
  and Virgo Collaboration)}}(2016{\natexlab{c}})}]{2016PhRvX...6d1015A}%
  \BibitemOpen
  \bibfield  {author} {\bibinfo {author} {\bibfnamefont {B.~P.}\ \bibnamefont
  {{Abbott {\it et al.} (LIGO Scientific Collaboration and Virgo
  Collaboration)}}},\ }\href {\doibase 10.1103/PhysRevX.6.041015} {\bibfield
  {journal} {\bibinfo  {journal} {Physical Review X}\ }\textbf {\bibinfo
  {volume} {6}},\ \bibinfo {eid} {041015} (\bibinfo {year}
  {2016}{\natexlab{c}})},\ \Eprint {http://arxiv.org/abs/1606.04856}
  {arXiv:1606.04856 [gr-qc]} \BibitemShut {NoStop}%
\bibitem [{\citenamefont {{Abbott {\it et al.} (LIGO Scientific Collaboration
  and Virgo Collaboration)}}(2016{\natexlab{d}})}]{2016ApJ...818L..22A}%
  \BibitemOpen
  \bibfield  {author} {\bibinfo {author} {\bibfnamefont {B.~P.}\ \bibnamefont
  {{Abbott {\it et al.} (LIGO Scientific Collaboration and Virgo
  Collaboration)}}},\ }\href {\doibase 10.3847/2041-8205/818/2/L22} {\bibfield
  {journal} {\bibinfo  {journal} {\apjl}\ }\textbf {\bibinfo {volume} {818}},\
  \bibinfo {eid} {L22} (\bibinfo {year} {2016}{\natexlab{d}})},\ \Eprint
  {http://arxiv.org/abs/1602.03846} {arXiv:1602.03846 [astro-ph.HE]}
  \BibitemShut {NoStop}%
\bibitem [{\citenamefont {{Postnov}}\ and\ \citenamefont
  {{Yungelson}}(2014)}]{2014LRR....17....3P}%
  \BibitemOpen
  \bibfield  {author} {\bibinfo {author} {\bibfnamefont {K.~A.}\ \bibnamefont
  {{Postnov}}}\ and\ \bibinfo {author} {\bibfnamefont {L.~R.}\ \bibnamefont
  {{Yungelson}}},\ }\href {\doibase 10.12942/lrr-2014-3} {\bibfield  {journal}
  {\bibinfo  {journal} {Living Reviews in Relativity}\ }\textbf {\bibinfo
  {volume} {17}},\ \bibinfo {eid} {3} (\bibinfo {year} {2014})},\ \Eprint
  {http://arxiv.org/abs/1403.4754} {arXiv:1403.4754 [astro-ph.HE]} \BibitemShut
  {NoStop}%
\bibitem [{\citenamefont {{Benacquista}}\ and\ \citenamefont
  {{Downing}}(2013)}]{2013LRR....16....4B}%
  \BibitemOpen
  \bibfield  {author} {\bibinfo {author} {\bibfnamefont {M.~J.}\ \bibnamefont
  {{Benacquista}}}\ and\ \bibinfo {author} {\bibfnamefont {J.~M.~B.}\
  \bibnamefont {{Downing}}},\ }\href {\doibase 10.12942/lrr-2013-4} {\bibfield
  {journal} {\bibinfo  {journal} {Living Reviews in Relativity}\ }\textbf
  {\bibinfo {volume} {16}},\ \bibinfo {eid} {4} (\bibinfo {year} {2013})},\
  \Eprint {http://arxiv.org/abs/1110.4423} {arXiv:1110.4423 [astro-ph.SR]}
  \BibitemShut {NoStop}%
\bibitem [{\citenamefont {{Mapelli}}\ \emph {et~al.}(2009)\citenamefont
  {{Mapelli}}, \citenamefont {{Colpi}},\ and\ \citenamefont
  {{Zampieri}}}]{2009MNRAS.395L..71M}%
  \BibitemOpen
  \bibfield  {author} {\bibinfo {author} {\bibfnamefont {M.}~\bibnamefont
  {{Mapelli}}}, \bibinfo {author} {\bibfnamefont {M.}~\bibnamefont {{Colpi}}},
  \ and\ \bibinfo {author} {\bibfnamefont {L.}~\bibnamefont {{Zampieri}}},\
  }\href {\doibase 10.1111/j.1745-3933.2009.00645.x} {\bibfield  {journal}
  {\bibinfo  {journal} {\mnras}\ }\textbf {\bibinfo {volume} {395}},\ \bibinfo
  {pages} {L71} (\bibinfo {year} {2009})},\ \Eprint
  {http://arxiv.org/abs/0902.3540} {arXiv:0902.3540 [astro-ph.HE]} \BibitemShut
  {NoStop}%
\bibitem [{\citenamefont {{Belczynski}}\ \emph {et~al.}(2010)\citenamefont
  {{Belczynski}}, \citenamefont {{Dominik}}, \citenamefont {{Bulik}},
  \citenamefont {{O'Shaughnessy}}, \citenamefont {{Fryer}},\ and\ \citenamefont
  {{Holz}}}]{2010ApJ...715L.138B}%
  \BibitemOpen
  \bibfield  {author} {\bibinfo {author} {\bibfnamefont {K.}~\bibnamefont
  {{Belczynski}}}, \bibinfo {author} {\bibfnamefont {M.}~\bibnamefont
  {{Dominik}}}, \bibinfo {author} {\bibfnamefont {T.}~\bibnamefont {{Bulik}}},
  \bibinfo {author} {\bibfnamefont {R.}~\bibnamefont {{O'Shaughnessy}}},
  \bibinfo {author} {\bibfnamefont {C.}~\bibnamefont {{Fryer}}}, \ and\
  \bibinfo {author} {\bibfnamefont {D.~E.}\ \bibnamefont {{Holz}}},\ }\href
  {\doibase 10.1088/2041-8205/715/2/L138} {\bibfield  {journal} {\bibinfo
  {journal} {\apjl}\ }\textbf {\bibinfo {volume} {715}},\ \bibinfo {pages}
  {L138} (\bibinfo {year} {2010})},\ \Eprint {http://arxiv.org/abs/1004.0386}
  {arXiv:1004.0386 [astro-ph.HE]} \BibitemShut {NoStop}%
\bibitem [{\citenamefont {{Dominik}}\ \emph {et~al.}(2012)\citenamefont
  {{Dominik}}, \citenamefont {{Belczynski}}, \citenamefont {{Fryer}},
  \citenamefont {{Holz}}, \citenamefont {{Berti}}, \citenamefont {{Bulik}},
  \citenamefont {{Mandel}},\ and\ \citenamefont
  {{O'Shaughnessy}}}]{2012ApJ...759...52D}%
  \BibitemOpen
  \bibfield  {author} {\bibinfo {author} {\bibfnamefont {M.}~\bibnamefont
  {{Dominik}}}, \bibinfo {author} {\bibfnamefont {K.}~\bibnamefont
  {{Belczynski}}}, \bibinfo {author} {\bibfnamefont {C.}~\bibnamefont
  {{Fryer}}}, \bibinfo {author} {\bibfnamefont {D.~E.}\ \bibnamefont {{Holz}}},
  \bibinfo {author} {\bibfnamefont {E.}~\bibnamefont {{Berti}}}, \bibinfo
  {author} {\bibfnamefont {T.}~\bibnamefont {{Bulik}}}, \bibinfo {author}
  {\bibfnamefont {I.}~\bibnamefont {{Mandel}}}, \ and\ \bibinfo {author}
  {\bibfnamefont {R.}~\bibnamefont {{O'Shaughnessy}}},\ }\href {\doibase
  10.1088/0004-637X/759/1/52} {\bibfield  {journal} {\bibinfo  {journal}
  {\apj}\ }\textbf {\bibinfo {volume} {759}},\ \bibinfo {eid} {52} (\bibinfo
  {year} {2012})},\ \Eprint {http://arxiv.org/abs/1202.4901} {arXiv:1202.4901
  [astro-ph.HE]} \BibitemShut {NoStop}%
\bibitem [{\citenamefont {{Dominik}}\ \emph {et~al.}(2013)\citenamefont
  {{Dominik}}, \citenamefont {{Belczynski}}, \citenamefont {{Fryer}},
  \citenamefont {{Holz}}, \citenamefont {{Berti}}, \citenamefont {{Bulik}},
  \citenamefont {{Mandel}},\ and\ \citenamefont
  {{O'Shaughnessy}}}]{2013ApJ...779...72D}%
  \BibitemOpen
  \bibfield  {author} {\bibinfo {author} {\bibfnamefont {M.}~\bibnamefont
  {{Dominik}}}, \bibinfo {author} {\bibfnamefont {K.}~\bibnamefont
  {{Belczynski}}}, \bibinfo {author} {\bibfnamefont {C.}~\bibnamefont
  {{Fryer}}}, \bibinfo {author} {\bibfnamefont {D.~E.}\ \bibnamefont {{Holz}}},
  \bibinfo {author} {\bibfnamefont {E.}~\bibnamefont {{Berti}}}, \bibinfo
  {author} {\bibfnamefont {T.}~\bibnamefont {{Bulik}}}, \bibinfo {author}
  {\bibfnamefont {I.}~\bibnamefont {{Mandel}}}, \ and\ \bibinfo {author}
  {\bibfnamefont {R.}~\bibnamefont {{O'Shaughnessy}}},\ }\href {\doibase
  10.1088/0004-637X/779/1/72} {\bibfield  {journal} {\bibinfo  {journal}
  {\apj}\ }\textbf {\bibinfo {volume} {779}},\ \bibinfo {eid} {72} (\bibinfo
  {year} {2013})},\ \Eprint {http://arxiv.org/abs/1308.1546} {arXiv:1308.1546
  [astro-ph.HE]} \BibitemShut {NoStop}%
\bibitem [{\citenamefont {{Spera}}\ \emph {et~al.}(2015)\citenamefont
  {{Spera}}, \citenamefont {{Mapelli}},\ and\ \citenamefont
  {{Bressan}}}]{2015MNRAS.451.4086S}%
  \BibitemOpen
  \bibfield  {author} {\bibinfo {author} {\bibfnamefont {M.}~\bibnamefont
  {{Spera}}}, \bibinfo {author} {\bibfnamefont {M.}~\bibnamefont {{Mapelli}}},
  \ and\ \bibinfo {author} {\bibfnamefont {A.}~\bibnamefont {{Bressan}}},\
  }\href {\doibase 10.1093/mnras/stv1161} {\bibfield  {journal} {\bibinfo
  {journal} {\mnras}\ }\textbf {\bibinfo {volume} {451}},\ \bibinfo {pages}
  {4086} (\bibinfo {year} {2015})},\ \Eprint {http://arxiv.org/abs/1505.05201}
  {arXiv:1505.05201 [astro-ph.SR]} \BibitemShut {NoStop}%
\bibitem [{\citenamefont {{Dominik}}\ \emph {et~al.}(2015)\citenamefont
  {{Dominik}}, \citenamefont {{Berti}}, \citenamefont {{O'Shaughnessy}},
  \citenamefont {{Mandel}}, \citenamefont {{Belczynski}}, \citenamefont
  {{Fryer}}, \citenamefont {{Holz}}, \citenamefont {{Bulik}},\ and\
  \citenamefont {{Pannarale}}}]{2015ApJ...806..263D}%
  \BibitemOpen
  \bibfield  {author} {\bibinfo {author} {\bibfnamefont {M.}~\bibnamefont
  {{Dominik}}}, \bibinfo {author} {\bibfnamefont {E.}~\bibnamefont {{Berti}}},
  \bibinfo {author} {\bibfnamefont {R.}~\bibnamefont {{O'Shaughnessy}}},
  \bibinfo {author} {\bibfnamefont {I.}~\bibnamefont {{Mandel}}}, \bibinfo
  {author} {\bibfnamefont {K.}~\bibnamefont {{Belczynski}}}, \bibinfo {author}
  {\bibfnamefont {C.}~\bibnamefont {{Fryer}}}, \bibinfo {author} {\bibfnamefont
  {D.~E.}\ \bibnamefont {{Holz}}}, \bibinfo {author} {\bibfnamefont
  {T.}~\bibnamefont {{Bulik}}}, \ and\ \bibinfo {author} {\bibfnamefont
  {F.}~\bibnamefont {{Pannarale}}},\ }\href {\doibase
  10.1088/0004-637X/806/2/263} {\bibfield  {journal} {\bibinfo  {journal}
  {\apj}\ }\textbf {\bibinfo {volume} {806}},\ \bibinfo {eid} {263} (\bibinfo
  {year} {2015})},\ \Eprint {http://arxiv.org/abs/1405.7016} {arXiv:1405.7016
  [astro-ph.HE]} \BibitemShut {NoStop}%
\bibitem [{\citenamefont {{Belczynski}}\ \emph
  {et~al.}(2016{\natexlab{a}})\citenamefont {{Belczynski}}, \citenamefont
  {{Holz}}, \citenamefont {{Bulik}},\ and\ \citenamefont
  {{O'Shaughnessy}}}]{2016arXiv160204531B}%
  \BibitemOpen
  \bibfield  {author} {\bibinfo {author} {\bibfnamefont {K.}~\bibnamefont
  {{Belczynski}}}, \bibinfo {author} {\bibfnamefont {D.~E.}\ \bibnamefont
  {{Holz}}}, \bibinfo {author} {\bibfnamefont {T.}~\bibnamefont {{Bulik}}}, \
  and\ \bibinfo {author} {\bibfnamefont {R.}~\bibnamefont {{O'Shaughnessy}}},\
  }\href {\doibase 10.1038/nature18322} {\bibfield  {journal} {\bibinfo
  {journal} {\nat}\ }\textbf {\bibinfo {volume} {534}},\ \bibinfo {pages} {512}
  (\bibinfo {year} {2016}{\natexlab{a}})},\ \Eprint
  {http://arxiv.org/abs/1602.04531} {arXiv:1602.04531 [astro-ph.HE]}
  \BibitemShut {NoStop}%
\bibitem [{\citenamefont {{Belczynski}}\ \emph
  {et~al.}(2016{\natexlab{b}})\citenamefont {{Belczynski}}, \citenamefont
  {{Repetto}}, \citenamefont {{Holz}}, \citenamefont {{O'Shaughnessy}},
  \citenamefont {{Bulik}}, \citenamefont {{Berti}}, \citenamefont {{Fryer}},\
  and\ \citenamefont {{Dominik}}}]{2016ApJ...819..108B}%
  \BibitemOpen
  \bibfield  {author} {\bibinfo {author} {\bibfnamefont {K.}~\bibnamefont
  {{Belczynski}}}, \bibinfo {author} {\bibfnamefont {S.}~\bibnamefont
  {{Repetto}}}, \bibinfo {author} {\bibfnamefont {D.~E.}\ \bibnamefont
  {{Holz}}}, \bibinfo {author} {\bibfnamefont {R.}~\bibnamefont
  {{O'Shaughnessy}}}, \bibinfo {author} {\bibfnamefont {T.}~\bibnamefont
  {{Bulik}}}, \bibinfo {author} {\bibfnamefont {E.}~\bibnamefont {{Berti}}},
  \bibinfo {author} {\bibfnamefont {C.}~\bibnamefont {{Fryer}}}, \ and\
  \bibinfo {author} {\bibfnamefont {M.}~\bibnamefont {{Dominik}}},\ }\href
  {\doibase 10.3847/0004-637X/819/2/108} {\bibfield  {journal} {\bibinfo
  {journal} {\apj}\ }\textbf {\bibinfo {volume} {819}},\ \bibinfo {eid} {108}
  (\bibinfo {year} {2016}{\natexlab{b}})},\ \Eprint
  {http://arxiv.org/abs/1510.04615} {arXiv:1510.04615 [astro-ph.HE]}
  \BibitemShut {NoStop}%
\bibitem [{\citenamefont {{Rodriguez}}\ \emph {et~al.}(2015)\citenamefont
  {{Rodriguez}}, \citenamefont {{Morscher}}, \citenamefont {{Pattabiraman}},
  \citenamefont {{Chatterjee}}, \citenamefont {{Haster}},\ and\ \citenamefont
  {{Rasio}}}]{2015PhRvL.115e1101R}%
  \BibitemOpen
  \bibfield  {author} {\bibinfo {author} {\bibfnamefont {C.~L.}\ \bibnamefont
  {{Rodriguez}}}, \bibinfo {author} {\bibfnamefont {M.}~\bibnamefont
  {{Morscher}}}, \bibinfo {author} {\bibfnamefont {B.}~\bibnamefont
  {{Pattabiraman}}}, \bibinfo {author} {\bibfnamefont {S.}~\bibnamefont
  {{Chatterjee}}}, \bibinfo {author} {\bibfnamefont {C.-J.}\ \bibnamefont
  {{Haster}}}, \ and\ \bibinfo {author} {\bibfnamefont {F.~A.}\ \bibnamefont
  {{Rasio}}},\ }\href {\doibase 10.1103/PhysRevLett.115.051101} {\bibfield
  {journal} {\bibinfo  {journal} {\prl}\ }\textbf {\bibinfo {volume} {115}},\
  \bibinfo {eid} {051101} (\bibinfo {year} {2015})},\ \Eprint
  {http://arxiv.org/abs/1505.00792} {arXiv:1505.00792 [astro-ph.HE]}
  \BibitemShut {NoStop}%
\bibitem [{\citenamefont {{Rodriguez}}\ \emph
  {et~al.}(2016{\natexlab{a}})\citenamefont {{Rodriguez}}, \citenamefont
  {{Chatterjee}},\ and\ \citenamefont {{Rasio}}}]{2016PhRvD..93h4029R}%
  \BibitemOpen
  \bibfield  {author} {\bibinfo {author} {\bibfnamefont {C.~L.}\ \bibnamefont
  {{Rodriguez}}}, \bibinfo {author} {\bibfnamefont {S.}~\bibnamefont
  {{Chatterjee}}}, \ and\ \bibinfo {author} {\bibfnamefont {F.~A.}\
  \bibnamefont {{Rasio}}},\ }\href {\doibase 10.1103/PhysRevD.93.084029}
  {\bibfield  {journal} {\bibinfo  {journal} {\prd}\ }\textbf {\bibinfo
  {volume} {93}},\ \bibinfo {eid} {084029} (\bibinfo {year}
  {2016}{\natexlab{a}})},\ \Eprint {http://arxiv.org/abs/1602.02444}
  {arXiv:1602.02444 [astro-ph.HE]} \BibitemShut {NoStop}%
\bibitem [{\citenamefont {{Chatterjee}}\ \emph {et~al.}(2017)\citenamefont
  {{Chatterjee}}, \citenamefont {{Rodriguez}},\ and\ \citenamefont
  {{Rasio}}}]{2016arXiv160300884C}%
  \BibitemOpen
  \bibfield  {author} {\bibinfo {author} {\bibfnamefont {S.}~\bibnamefont
  {{Chatterjee}}}, \bibinfo {author} {\bibfnamefont {C.~L.}\ \bibnamefont
  {{Rodriguez}}}, \ and\ \bibinfo {author} {\bibfnamefont {F.~A.}\ \bibnamefont
  {{Rasio}}},\ }\href {\doibase 10.3847/1538-4357/834/1/68} {\bibfield
  {journal} {\bibinfo  {journal} {\apj}\ }\textbf {\bibinfo {volume} {834}},\
  \bibinfo {eid} {68} (\bibinfo {year} {2017})},\ \Eprint
  {http://arxiv.org/abs/1603.00884} {arXiv:1603.00884} \BibitemShut {NoStop}%
\bibitem [{\citenamefont {{Rodriguez}}\ \emph
  {et~al.}(2016{\natexlab{b}})\citenamefont {{Rodriguez}}, \citenamefont
  {{Haster}}, \citenamefont {{Chatterjee}}, \citenamefont {{Kalogera}},\ and\
  \citenamefont {{Rasio}}}]{2016ApJ...824L...8R}%
  \BibitemOpen
  \bibfield  {author} {\bibinfo {author} {\bibfnamefont {C.~L.}\ \bibnamefont
  {{Rodriguez}}}, \bibinfo {author} {\bibfnamefont {C.-J.}\ \bibnamefont
  {{Haster}}}, \bibinfo {author} {\bibfnamefont {S.}~\bibnamefont
  {{Chatterjee}}}, \bibinfo {author} {\bibfnamefont {V.}~\bibnamefont
  {{Kalogera}}}, \ and\ \bibinfo {author} {\bibfnamefont {F.~A.}\ \bibnamefont
  {{Rasio}}},\ }\href {\doibase 10.3847/2041-8205/824/1/L8} {\bibfield
  {journal} {\bibinfo  {journal} {\apjl}\ }\textbf {\bibinfo {volume} {824}},\
  \bibinfo {eid} {L8} (\bibinfo {year} {2016}{\natexlab{b}})},\ \Eprint
  {http://arxiv.org/abs/1604.04254} {arXiv:1604.04254 [astro-ph.HE]}
  \BibitemShut {NoStop}%
\bibitem [{\citenamefont {{Remillard}}\ and\ \citenamefont
  {{McClintock}}(2006)}]{2006ARA&A..44...49R}%
  \BibitemOpen
  \bibfield  {author} {\bibinfo {author} {\bibfnamefont {R.~A.}\ \bibnamefont
  {{Remillard}}}\ and\ \bibinfo {author} {\bibfnamefont {J.~E.}\ \bibnamefont
  {{McClintock}}},\ }\href {\doibase 10.1146/annurev.astro.44.051905.092532}
  {\bibfield  {journal} {\bibinfo  {journal} {\araa}\ }\textbf {\bibinfo
  {volume} {44}},\ \bibinfo {pages} {49} (\bibinfo {year} {2006})},\ \Eprint
  {http://arxiv.org/abs/astro-ph/0606352} {astro-ph/0606352} \BibitemShut
  {NoStop}%
\bibitem [{\citenamefont {{Narayan}}\ and\ \citenamefont
  {{McClintock}}(2013)}]{2013arXiv1312.6698N}%
  \BibitemOpen
  \bibfield  {author} {\bibinfo {author} {\bibfnamefont {R.}~\bibnamefont
  {{Narayan}}}\ and\ \bibinfo {author} {\bibfnamefont {J.~E.}\ \bibnamefont
  {{McClintock}}},\ }\href@noop {} {\bibfield  {journal} {\bibinfo  {journal}
  {ArXiv e-prints}\ } (\bibinfo {year} {2013})},\ \Eprint
  {http://arxiv.org/abs/1312.6698} {arXiv:1312.6698 [astro-ph.HE]} \BibitemShut
  {NoStop}%
\bibitem [{\citenamefont {{Wen}}(2003)}]{2003ApJ...598..419W}%
  \BibitemOpen
  \bibfield  {author} {\bibinfo {author} {\bibfnamefont {L.}~\bibnamefont
  {{Wen}}},\ }\href {\doibase 10.1086/378794} {\bibfield  {journal} {\bibinfo
  {journal} {\apj}\ }\textbf {\bibinfo {volume} {598}},\ \bibinfo {pages} {419}
  (\bibinfo {year} {2003})},\ \Eprint {http://arxiv.org/abs/astro-ph/0211492}
  {astro-ph/0211492} \BibitemShut {NoStop}%
\bibitem [{\citenamefont {{Antonini}}\ \emph {et~al.}(2014)\citenamefont
  {{Antonini}}, \citenamefont {{Murray}},\ and\ \citenamefont
  {{Mikkola}}}]{2014ApJ...781...45A}%
  \BibitemOpen
  \bibfield  {author} {\bibinfo {author} {\bibfnamefont {F.}~\bibnamefont
  {{Antonini}}}, \bibinfo {author} {\bibfnamefont {N.}~\bibnamefont
  {{Murray}}}, \ and\ \bibinfo {author} {\bibfnamefont {S.}~\bibnamefont
  {{Mikkola}}},\ }\href {\doibase 10.1088/0004-637X/781/1/45} {\bibfield
  {journal} {\bibinfo  {journal} {\apj}\ }\textbf {\bibinfo {volume} {781}},\
  \bibinfo {eid} {45} (\bibinfo {year} {2014})},\ \Eprint
  {http://arxiv.org/abs/1308.3674} {arXiv:1308.3674 [astro-ph.HE]} \BibitemShut
  {NoStop}%
\bibitem [{\citenamefont {{Antonini}}\ \emph {et~al.}(2016)\citenamefont
  {{Antonini}}, \citenamefont {{Chatterjee}}, \citenamefont {{Rodriguez}},
  \citenamefont {{Morscher}}, \citenamefont {{Pattabiraman}}, \citenamefont
  {{Kalogera}},\ and\ \citenamefont {{Rasio}}}]{2016ApJ...816...65A}%
  \BibitemOpen
  \bibfield  {author} {\bibinfo {author} {\bibfnamefont {F.}~\bibnamefont
  {{Antonini}}}, \bibinfo {author} {\bibfnamefont {S.}~\bibnamefont
  {{Chatterjee}}}, \bibinfo {author} {\bibfnamefont {C.~L.}\ \bibnamefont
  {{Rodriguez}}}, \bibinfo {author} {\bibfnamefont {M.}~\bibnamefont
  {{Morscher}}}, \bibinfo {author} {\bibfnamefont {B.}~\bibnamefont
  {{Pattabiraman}}}, \bibinfo {author} {\bibfnamefont {V.}~\bibnamefont
  {{Kalogera}}}, \ and\ \bibinfo {author} {\bibfnamefont {F.~A.}\ \bibnamefont
  {{Rasio}}},\ }\href {\doibase 10.3847/0004-637X/816/2/65} {\bibfield
  {journal} {\bibinfo  {journal} {\apj}\ }\textbf {\bibinfo {volume} {816}},\
  \bibinfo {eid} {65} (\bibinfo {year} {2016})},\ \Eprint
  {http://arxiv.org/abs/1509.05080} {arXiv:1509.05080} \BibitemShut {NoStop}%
\bibitem [{\citenamefont {{Kinugawa}}\ \emph {et~al.}(2014)\citenamefont
  {{Kinugawa}}, \citenamefont {{Inayoshi}}, \citenamefont {{Hotokezaka}},
  \citenamefont {{Nakauchi}},\ and\ \citenamefont
  {{Nakamura}}}]{2014MNRAS.442.2963K}%
  \BibitemOpen
  \bibfield  {author} {\bibinfo {author} {\bibfnamefont {T.}~\bibnamefont
  {{Kinugawa}}}, \bibinfo {author} {\bibfnamefont {K.}~\bibnamefont
  {{Inayoshi}}}, \bibinfo {author} {\bibfnamefont {K.}~\bibnamefont
  {{Hotokezaka}}}, \bibinfo {author} {\bibfnamefont {D.}~\bibnamefont
  {{Nakauchi}}}, \ and\ \bibinfo {author} {\bibfnamefont {T.}~\bibnamefont
  {{Nakamura}}},\ }\href {\doibase 10.1093/mnras/stu1022} {\bibfield  {journal}
  {\bibinfo  {journal} {\mnras}\ }\textbf {\bibinfo {volume} {442}},\ \bibinfo
  {pages} {2963} (\bibinfo {year} {2014})},\ \Eprint
  {http://arxiv.org/abs/1402.6672} {arXiv:1402.6672 [astro-ph.HE]} \BibitemShut
  {NoStop}%
\bibitem [{\citenamefont {{Hartwig}}\ \emph {et~al.}(2016)\citenamefont
  {{Hartwig}}, \citenamefont {{Volonteri}}, \citenamefont {{Bromm}},
  \citenamefont {{Klessen}}, \citenamefont {{Barausse}}, \citenamefont
  {{Magg}},\ and\ \citenamefont {{Stacy}}}]{2016MNRAS.460L..74H}%
  \BibitemOpen
  \bibfield  {author} {\bibinfo {author} {\bibfnamefont {T.}~\bibnamefont
  {{Hartwig}}}, \bibinfo {author} {\bibfnamefont {M.}~\bibnamefont
  {{Volonteri}}}, \bibinfo {author} {\bibfnamefont {V.}~\bibnamefont
  {{Bromm}}}, \bibinfo {author} {\bibfnamefont {R.~S.}\ \bibnamefont
  {{Klessen}}}, \bibinfo {author} {\bibfnamefont {E.}~\bibnamefont
  {{Barausse}}}, \bibinfo {author} {\bibfnamefont {M.}~\bibnamefont {{Magg}}},
  \ and\ \bibinfo {author} {\bibfnamefont {A.}~\bibnamefont {{Stacy}}},\ }\href
  {\doibase 10.1093/mnrasl/slw074} {\bibfield  {journal} {\bibinfo  {journal}
  {\mnras}\ }\textbf {\bibinfo {volume} {460}},\ \bibinfo {pages} {L74}
  (\bibinfo {year} {2016})},\ \Eprint {http://arxiv.org/abs/1603.05655}
  {arXiv:1603.05655} \BibitemShut {NoStop}%
\bibitem [{\citenamefont {{Belczynski}}\ \emph
  {et~al.}(2016{\natexlab{c}})\citenamefont {{Belczynski}}, \citenamefont
  {{Ryu}}, \citenamefont {{Perna}}, \citenamefont {{Berti}}, \citenamefont
  {{Tanaka}},\ and\ \citenamefont {{Bulik}}}]{2016arXiv161201524B}%
  \BibitemOpen
  \bibfield  {author} {\bibinfo {author} {\bibfnamefont {K.}~\bibnamefont
  {{Belczynski}}}, \bibinfo {author} {\bibfnamefont {T.}~\bibnamefont {{Ryu}}},
  \bibinfo {author} {\bibfnamefont {R.}~\bibnamefont {{Perna}}}, \bibinfo
  {author} {\bibfnamefont {E.}~\bibnamefont {{Berti}}}, \bibinfo {author}
  {\bibfnamefont {T.~L.}\ \bibnamefont {{Tanaka}}}, \ and\ \bibinfo {author}
  {\bibfnamefont {T.}~\bibnamefont {{Bulik}}},\ }\href@noop {} {\bibfield
  {journal} {\bibinfo  {journal} {ArXiv e-prints}\ } (\bibinfo {year}
  {2016}{\natexlab{c}})},\ \Eprint {http://arxiv.org/abs/1612.01524}
  {arXiv:1612.01524 [astro-ph.HE]} \BibitemShut {NoStop}%
\bibitem [{\citenamefont {{Mandel}}\ and\ \citenamefont {{de
  Mink}}(2016)}]{2016MNRAS.458.2634M}%
  \BibitemOpen
  \bibfield  {author} {\bibinfo {author} {\bibfnamefont {I.}~\bibnamefont
  {{Mandel}}}\ and\ \bibinfo {author} {\bibfnamefont {S.~E.}\ \bibnamefont {{de
  Mink}}},\ }\href {\doibase 10.1093/mnras/stw379} {\bibfield  {journal}
  {\bibinfo  {journal} {\mnras}\ }\textbf {\bibinfo {volume} {458}},\ \bibinfo
  {pages} {2634} (\bibinfo {year} {2016})},\ \Eprint
  {http://arxiv.org/abs/1601.00007} {arXiv:1601.00007 [astro-ph.HE]}
  \BibitemShut {NoStop}%
\bibitem [{\citenamefont {{de Mink}}\ and\ \citenamefont
  {{Mandel}}(2016)}]{2016MNRAS.460.3545D}%
  \BibitemOpen
  \bibfield  {author} {\bibinfo {author} {\bibfnamefont {S.~E.}\ \bibnamefont
  {{de Mink}}}\ and\ \bibinfo {author} {\bibfnamefont {I.}~\bibnamefont
  {{Mandel}}},\ }\href {\doibase 10.1093/mnras/stw1219} {\bibfield  {journal}
  {\bibinfo  {journal} {\mnras}\ }\textbf {\bibinfo {volume} {460}},\ \bibinfo
  {pages} {3545} (\bibinfo {year} {2016})},\ \Eprint
  {http://arxiv.org/abs/1603.02291} {arXiv:1603.02291 [astro-ph.HE]}
  \BibitemShut {NoStop}%
\bibitem [{\citenamefont {{Marchant}}\ \emph {et~al.}(2016)\citenamefont
  {{Marchant}}, \citenamefont {{Langer}}, \citenamefont {{Podsiadlowski}},
  \citenamefont {{Tauris}},\ and\ \citenamefont
  {{Moriya}}}]{2016A&A...588A..50M}%
  \BibitemOpen
  \bibfield  {author} {\bibinfo {author} {\bibfnamefont {P.}~\bibnamefont
  {{Marchant}}}, \bibinfo {author} {\bibfnamefont {N.}~\bibnamefont
  {{Langer}}}, \bibinfo {author} {\bibfnamefont {P.}~\bibnamefont
  {{Podsiadlowski}}}, \bibinfo {author} {\bibfnamefont {T.~M.}\ \bibnamefont
  {{Tauris}}}, \ and\ \bibinfo {author} {\bibfnamefont {T.~J.}\ \bibnamefont
  {{Moriya}}},\ }\href {\doibase 10.1051/0004-6361/201628133} {\bibfield
  {journal} {\bibinfo  {journal} {\aap}\ }\textbf {\bibinfo {volume} {588}},\
  \bibinfo {eid} {A50} (\bibinfo {year} {2016})},\ \Eprint
  {http://arxiv.org/abs/1601.03718} {arXiv:1601.03718 [astro-ph.SR]}
  \BibitemShut {NoStop}%
\bibitem [{\citenamefont {{Bird}}\ \emph {et~al.}(2016)\citenamefont {{Bird}},
  \citenamefont {{Cholis}}, \citenamefont {{Mu{\~n}oz}}, \citenamefont
  {{Ali-Ha{\"\i}moud}}, \citenamefont {{Kamionkowski}}, \citenamefont
  {{Kovetz}}, \citenamefont {{Raccanelli}},\ and\ \citenamefont
  {{Riess}}}]{2016PhRvL.116t1301B}%
  \BibitemOpen
  \bibfield  {author} {\bibinfo {author} {\bibfnamefont {S.}~\bibnamefont
  {{Bird}}}, \bibinfo {author} {\bibfnamefont {I.}~\bibnamefont {{Cholis}}},
  \bibinfo {author} {\bibfnamefont {J.~B.}\ \bibnamefont {{Mu{\~n}oz}}},
  \bibinfo {author} {\bibfnamefont {Y.}~\bibnamefont {{Ali-Ha{\"\i}moud}}},
  \bibinfo {author} {\bibfnamefont {M.}~\bibnamefont {{Kamionkowski}}},
  \bibinfo {author} {\bibfnamefont {E.~D.}\ \bibnamefont {{Kovetz}}}, \bibinfo
  {author} {\bibfnamefont {A.}~\bibnamefont {{Raccanelli}}}, \ and\ \bibinfo
  {author} {\bibfnamefont {A.~G.}\ \bibnamefont {{Riess}}},\ }\href {\doibase
  10.1103/PhysRevLett.116.201301} {\bibfield  {journal} {\bibinfo  {journal}
  {\prl}\ }\textbf {\bibinfo {volume} {116}},\ \bibinfo {eid} {201301}
  (\bibinfo {year} {2016})},\ \Eprint {http://arxiv.org/abs/1603.00464}
  {arXiv:1603.00464} \BibitemShut {NoStop}%
\bibitem [{\citenamefont {{Cholis}}\ \emph {et~al.}(2016)\citenamefont
  {{Cholis}}, \citenamefont {{Kovetz}}, \citenamefont {{Ali-Ha{\"\i}moud}},
  \citenamefont {{Bird}}, \citenamefont {{Kamionkowski}}, \citenamefont
  {{Mu{\~n}oz}},\ and\ \citenamefont {{Raccanelli}}}]{2016PhRvD..94h4013C}%
  \BibitemOpen
  \bibfield  {author} {\bibinfo {author} {\bibfnamefont {I.}~\bibnamefont
  {{Cholis}}}, \bibinfo {author} {\bibfnamefont {E.~D.}\ \bibnamefont
  {{Kovetz}}}, \bibinfo {author} {\bibfnamefont {Y.}~\bibnamefont
  {{Ali-Ha{\"\i}moud}}}, \bibinfo {author} {\bibfnamefont {S.}~\bibnamefont
  {{Bird}}}, \bibinfo {author} {\bibfnamefont {M.}~\bibnamefont
  {{Kamionkowski}}}, \bibinfo {author} {\bibfnamefont {J.~B.}\ \bibnamefont
  {{Mu{\~n}oz}}}, \ and\ \bibinfo {author} {\bibfnamefont {A.}~\bibnamefont
  {{Raccanelli}}},\ }\href {\doibase 10.1103/PhysRevD.94.084013} {\bibfield
  {journal} {\bibinfo  {journal} {\prd}\ }\textbf {\bibinfo {volume} {94}},\
  \bibinfo {eid} {084013} (\bibinfo {year} {2016})},\ \Eprint
  {http://arxiv.org/abs/1606.07437} {arXiv:1606.07437 [astro-ph.HE]}
  \BibitemShut {NoStop}%
\bibitem [{\citenamefont {{Antonini}}\ and\ \citenamefont
  {{Rasio}}(2016)}]{2016ApJ...831..187A}%
  \BibitemOpen
  \bibfield  {author} {\bibinfo {author} {\bibfnamefont {F.}~\bibnamefont
  {{Antonini}}}\ and\ \bibinfo {author} {\bibfnamefont {F.~A.}\ \bibnamefont
  {{Rasio}}},\ }\href {\doibase 10.3847/0004-637X/831/2/187} {\bibfield
  {journal} {\bibinfo  {journal} {\apj}\ }\textbf {\bibinfo {volume} {831}},\
  \bibinfo {eid} {187} (\bibinfo {year} {2016})},\ \Eprint
  {http://arxiv.org/abs/1606.04889} {arXiv:1606.04889 [astro-ph.HE]}
  \BibitemShut {NoStop}%
\bibitem [{\citenamefont {{Mapelli}}(2016)}]{2016MNRAS.459.3432M}%
  \BibitemOpen
  \bibfield  {author} {\bibinfo {author} {\bibfnamefont {M.}~\bibnamefont
  {{Mapelli}}},\ }\href {\doibase 10.1093/mnras/stw869} {\bibfield  {journal}
  {\bibinfo  {journal} {\mnras}\ }\textbf {\bibinfo {volume} {459}},\ \bibinfo
  {pages} {3432} (\bibinfo {year} {2016})},\ \Eprint
  {http://arxiv.org/abs/1604.03559} {arXiv:1604.03559} \BibitemShut {NoStop}%
\bibitem [{\citenamefont {{O'Leary}}\ \emph {et~al.}(2016)\citenamefont
  {{O'Leary}}, \citenamefont {{Meiron}},\ and\ \citenamefont
  {{Kocsis}}}]{2016ApJ...824L..12O}%
  \BibitemOpen
  \bibfield  {author} {\bibinfo {author} {\bibfnamefont {R.~M.}\ \bibnamefont
  {{O'Leary}}}, \bibinfo {author} {\bibfnamefont {Y.}~\bibnamefont {{Meiron}}},
  \ and\ \bibinfo {author} {\bibfnamefont {B.}~\bibnamefont {{Kocsis}}},\
  }\href {\doibase 10.3847/2041-8205/824/1/L12} {\bibfield  {journal} {\bibinfo
   {journal} {\apjl}\ }\textbf {\bibinfo {volume} {824}},\ \bibinfo {eid} {L12}
  (\bibinfo {year} {2016})},\ \Eprint {http://arxiv.org/abs/1602.02809}
  {arXiv:1602.02809 [astro-ph.HE]} \BibitemShut {NoStop}%
\bibitem [{\citenamefont {{Merritt}}\ \emph {et~al.}(2004)\citenamefont
  {{Merritt}}, \citenamefont {{Milosavljevi{\'c}}}, \citenamefont {{Favata}},
  \citenamefont {{Hughes}},\ and\ \citenamefont
  {{Holz}}}]{2004ApJ...607L...9M}%
  \BibitemOpen
  \bibfield  {author} {\bibinfo {author} {\bibfnamefont {D.}~\bibnamefont
  {{Merritt}}}, \bibinfo {author} {\bibfnamefont {M.}~\bibnamefont
  {{Milosavljevi{\'c}}}}, \bibinfo {author} {\bibfnamefont {M.}~\bibnamefont
  {{Favata}}}, \bibinfo {author} {\bibfnamefont {S.~A.}\ \bibnamefont
  {{Hughes}}}, \ and\ \bibinfo {author} {\bibfnamefont {D.~E.}\ \bibnamefont
  {{Holz}}},\ }\href {\doibase 10.1086/421551} {\bibfield  {journal} {\bibinfo
  {journal} {\apjl}\ }\textbf {\bibinfo {volume} {607}},\ \bibinfo {pages} {L9}
  (\bibinfo {year} {2004})},\ \Eprint {http://arxiv.org/abs/astro-ph/0402057}
  {astro-ph/0402057} \BibitemShut {NoStop}%
\bibitem [{\citenamefont {{McKernan}}\ \emph {et~al.}(2017)\citenamefont
  {{McKernan}}, \citenamefont {{Ford}}, \citenamefont {{Bellovary}},
  \citenamefont {{Leigh}}, \citenamefont {{Haiman}}, \citenamefont {{Kocsis}},
  \citenamefont {{Lyra}}, \citenamefont {{MacLow}}, \citenamefont {{Metzger}},
  \citenamefont {{O'Dowd}}, \citenamefont {{Endlich}},\ and\ \citenamefont
  {{Rosen}}}]{2017arXiv170207818M}%
  \BibitemOpen
  \bibfield  {author} {\bibinfo {author} {\bibfnamefont {B.}~\bibnamefont
  {{McKernan}}}, \bibinfo {author} {\bibfnamefont {K.~E.~S.}\ \bibnamefont
  {{Ford}}}, \bibinfo {author} {\bibfnamefont {J.}~\bibnamefont {{Bellovary}}},
  \bibinfo {author} {\bibfnamefont {N.~W.~C.}\ \bibnamefont {{Leigh}}},
  \bibinfo {author} {\bibfnamefont {Z.}~\bibnamefont {{Haiman}}}, \bibinfo
  {author} {\bibfnamefont {B.}~\bibnamefont {{Kocsis}}}, \bibinfo {author}
  {\bibfnamefont {W.}~\bibnamefont {{Lyra}}}, \bibinfo {author} {\bibfnamefont
  {M.-M.}\ \bibnamefont {{MacLow}}}, \bibinfo {author} {\bibfnamefont
  {B.}~\bibnamefont {{Metzger}}}, \bibinfo {author} {\bibfnamefont
  {M.}~\bibnamefont {{O'Dowd}}}, \bibinfo {author} {\bibfnamefont
  {S.}~\bibnamefont {{Endlich}}}, \ and\ \bibinfo {author} {\bibfnamefont
  {D.~J.}\ \bibnamefont {{Rosen}}},\ }\href@noop {} {\bibfield  {journal}
  {\bibinfo  {journal} {ArXiv e-prints}\ } (\bibinfo {year} {2017})},\ \Eprint
  {http://arxiv.org/abs/1702.07818} {arXiv:1702.07818 [astro-ph.HE]}
  \BibitemShut {NoStop}%
\bibitem [{\citenamefont {{Bellovary}}\ \emph {et~al.}(2016)\citenamefont
  {{Bellovary}}, \citenamefont {{Mac Low}}, \citenamefont {{McKernan}},\ and\
  \citenamefont {{Ford}}}]{2016ApJ...819L..17B}%
  \BibitemOpen
  \bibfield  {author} {\bibinfo {author} {\bibfnamefont {J.~M.}\ \bibnamefont
  {{Bellovary}}}, \bibinfo {author} {\bibfnamefont {M.-M.}\ \bibnamefont {{Mac
  Low}}}, \bibinfo {author} {\bibfnamefont {B.}~\bibnamefont {{McKernan}}}, \
  and\ \bibinfo {author} {\bibfnamefont {K.~E.~S.}\ \bibnamefont {{Ford}}},\
  }\href {\doibase 10.3847/2041-8205/819/2/L17} {\bibfield  {journal} {\bibinfo
   {journal} {\apjl}\ }\textbf {\bibinfo {volume} {819}},\ \bibinfo {eid} {L17}
  (\bibinfo {year} {2016})},\ \Eprint {http://arxiv.org/abs/1511.00005}
  {arXiv:1511.00005} \BibitemShut {NoStop}%
\bibitem [{\citenamefont {{Clesse}}\ and\ \citenamefont
  {{Garc{\'{\i}}a-Bellido}}(2017)}]{2017PDU....15..142C}%
  \BibitemOpen
  \bibfield  {author} {\bibinfo {author} {\bibfnamefont {S.}~\bibnamefont
  {{Clesse}}}\ and\ \bibinfo {author} {\bibfnamefont {J.}~\bibnamefont
  {{Garc{\'{\i}}a-Bellido}}},\ }\href {\doibase 10.1016/j.dark.2016.10.002}
  {\bibfield  {journal} {\bibinfo  {journal} {Physics of the Dark Universe}\
  }\textbf {\bibinfo {volume} {15}},\ \bibinfo {pages} {142} (\bibinfo {year}
  {2017})},\ \Eprint {http://arxiv.org/abs/1603.05234} {arXiv:1603.05234}
  \BibitemShut {NoStop}%
\bibitem [{\citenamefont {{Kushnir}}\ \emph {et~al.}(2016)\citenamefont
  {{Kushnir}}, \citenamefont {{Zaldarriaga}}, \citenamefont {{Kollmeier}},\
  and\ \citenamefont {{Waldman}}}]{2016MNRAS.462..844K}%
  \BibitemOpen
  \bibfield  {author} {\bibinfo {author} {\bibfnamefont {D.}~\bibnamefont
  {{Kushnir}}}, \bibinfo {author} {\bibfnamefont {M.}~\bibnamefont
  {{Zaldarriaga}}}, \bibinfo {author} {\bibfnamefont {J.~A.}\ \bibnamefont
  {{Kollmeier}}}, \ and\ \bibinfo {author} {\bibfnamefont {R.}~\bibnamefont
  {{Waldman}}},\ }\href {\doibase 10.1093/mnras/stw1684} {\bibfield  {journal}
  {\bibinfo  {journal} {\mnras}\ }\textbf {\bibinfo {volume} {462}},\ \bibinfo
  {pages} {844} (\bibinfo {year} {2016})},\ \Eprint
  {http://arxiv.org/abs/1605.03839} {arXiv:1605.03839 [astro-ph.HE]}
  \BibitemShut {NoStop}%
\bibitem [{\citenamefont {{Berti}}\ and\ \citenamefont
  {{Volonteri}}(2008)}]{2008ApJ...684..822B}%
  \BibitemOpen
  \bibfield  {author} {\bibinfo {author} {\bibfnamefont {E.}~\bibnamefont
  {{Berti}}}\ and\ \bibinfo {author} {\bibfnamefont {M.}~\bibnamefont
  {{Volonteri}}},\ }\href {\doibase 10.1086/590379} {\bibfield  {journal}
  {\bibinfo  {journal} {\apj}\ }\textbf {\bibinfo {volume} {684}},\ \bibinfo
  {eid} {822-828} (\bibinfo {year} {2008})},\ \Eprint
  {http://arxiv.org/abs/0802.0025} {arXiv:0802.0025} \BibitemShut {NoStop}%
\bibitem [{\citenamefont {{Abbott {\it et al.} (LIGO Scientific Collaboration
  and Virgo Collaboration)}}(2017)}]{2017arXiv170404628T}%
  \BibitemOpen
  \bibfield  {author} {\bibinfo {author} {\bibfnamefont {B.~P.}\ \bibnamefont
  {{Abbott {\it et al.} (LIGO Scientific Collaboration and Virgo
  Collaboration)}}},\ }\href@noop {} {\bibfield  {journal} {\bibinfo  {journal}
  {ArXiv e-prints}\ } (\bibinfo {year} {2017})},\ \Eprint
  {http://arxiv.org/abs/1704.04628} {arXiv:1704.04628 [gr-qc]} \BibitemShut
  {NoStop}%
\bibitem [{\citenamefont {{Heger}}\ and\ \citenamefont
  {{Woosley}}(2002)}]{2002ApJ...567..532H}%
  \BibitemOpen
  \bibfield  {author} {\bibinfo {author} {\bibfnamefont {A.}~\bibnamefont
  {{Heger}}}\ and\ \bibinfo {author} {\bibfnamefont {S.~E.}\ \bibnamefont
  {{Woosley}}},\ }\href {\doibase 10.1086/338487} {\bibfield  {journal}
  {\bibinfo  {journal} {\apj}\ }\textbf {\bibinfo {volume} {567}},\ \bibinfo
  {pages} {532} (\bibinfo {year} {2002})},\ \Eprint
  {http://arxiv.org/abs/astro-ph/0107037} {astro-ph/0107037} \BibitemShut
  {NoStop}%
\bibitem [{\citenamefont {{Woosley}}(2017)}]{2017ApJ...836..244W}%
  \BibitemOpen
  \bibfield  {author} {\bibinfo {author} {\bibfnamefont {S.~E.}\ \bibnamefont
  {{Woosley}}},\ }\href {\doibase 10.3847/1538-4357/836/2/244} {\bibfield
  {journal} {\bibinfo  {journal} {\apj}\ }\textbf {\bibinfo {volume} {836}},\
  \bibinfo {eid} {244} (\bibinfo {year} {2017})},\ \Eprint
  {http://arxiv.org/abs/1608.08939} {arXiv:1608.08939 [astro-ph.HE]}
  \BibitemShut {NoStop}%
\bibitem [{\citenamefont {{Belczynski}}\ \emph
  {et~al.}(2016{\natexlab{d}})\citenamefont {{Belczynski}}, \citenamefont
  {{Heger}}, \citenamefont {{Gladysz}}, \citenamefont {{Ruiter}}, \citenamefont
  {{Woosley}}, \citenamefont {{Wiktorowicz}}, \citenamefont {{Chen}},
  \citenamefont {{Bulik}}, \citenamefont {{O'Shaughnessy}}, \citenamefont
  {{Holz}}, \citenamefont {{Fryer}},\ and\ \citenamefont
  {{Berti}}}]{2016A&A...594A..97B}%
  \BibitemOpen
  \bibfield  {author} {\bibinfo {author} {\bibfnamefont {K.}~\bibnamefont
  {{Belczynski}}}, \bibinfo {author} {\bibfnamefont {A.}~\bibnamefont
  {{Heger}}}, \bibinfo {author} {\bibfnamefont {W.}~\bibnamefont {{Gladysz}}},
  \bibinfo {author} {\bibfnamefont {A.~J.}\ \bibnamefont {{Ruiter}}}, \bibinfo
  {author} {\bibfnamefont {S.}~\bibnamefont {{Woosley}}}, \bibinfo {author}
  {\bibfnamefont {G.}~\bibnamefont {{Wiktorowicz}}}, \bibinfo {author}
  {\bibfnamefont {H.-Y.}\ \bibnamefont {{Chen}}}, \bibinfo {author}
  {\bibfnamefont {T.}~\bibnamefont {{Bulik}}}, \bibinfo {author} {\bibfnamefont
  {R.}~\bibnamefont {{O'Shaughnessy}}}, \bibinfo {author} {\bibfnamefont
  {D.~E.}\ \bibnamefont {{Holz}}}, \bibinfo {author} {\bibfnamefont {C.~L.}\
  \bibnamefont {{Fryer}}}, \ and\ \bibinfo {author} {\bibfnamefont
  {E.}~\bibnamefont {{Berti}}},\ }\href {\doibase 10.1051/0004-6361/201628980}
  {\bibfield  {journal} {\bibinfo  {journal} {\aap}\ }\textbf {\bibinfo
  {volume} {594}},\ \bibinfo {eid} {A97} (\bibinfo {year}
  {2016}{\natexlab{d}})},\ \Eprint {http://arxiv.org/abs/1607.03116}
  {arXiv:1607.03116 [astro-ph.HE]} \BibitemShut {NoStop}%
\bibitem [{\citenamefont {{Belczynski}}\ \emph {et~al.}(2008)\citenamefont
  {{Belczynski}}, \citenamefont {{Taam}}, \citenamefont {{Rantsiou}},\ and\
  \citenamefont {{van der Sluys}}}]{2008ApJ...682..474B}%
  \BibitemOpen
  \bibfield  {author} {\bibinfo {author} {\bibfnamefont {K.}~\bibnamefont
  {{Belczynski}}}, \bibinfo {author} {\bibfnamefont {R.~E.}\ \bibnamefont
  {{Taam}}}, \bibinfo {author} {\bibfnamefont {E.}~\bibnamefont {{Rantsiou}}},
  \ and\ \bibinfo {author} {\bibfnamefont {M.}~\bibnamefont {{van der
  Sluys}}},\ }\href {\doibase 10.1086/589609} {\bibfield  {journal} {\bibinfo
  {journal} {\apj}\ }\textbf {\bibinfo {volume} {682}},\ \bibinfo {eid}
  {474-486} (\bibinfo {year} {2008})},\ \Eprint
  {http://arxiv.org/abs/astro-ph/0703131} {astro-ph/0703131} \BibitemShut
  {NoStop}%
\bibitem [{\citenamefont {{Miller}}\ and\ \citenamefont
  {{Miller}}(2015)}]{2015PhR...548....1M}%
  \BibitemOpen
  \bibfield  {author} {\bibinfo {author} {\bibfnamefont {M.~C.}\ \bibnamefont
  {{Miller}}}\ and\ \bibinfo {author} {\bibfnamefont {J.~M.}\ \bibnamefont
  {{Miller}}},\ }\href {\doibase 10.1016/j.physrep.2014.09.003} {\bibfield
  {journal} {\bibinfo  {journal} {\physrep}\ }\textbf {\bibinfo {volume}
  {548}},\ \bibinfo {pages} {1} (\bibinfo {year} {2015})},\ \Eprint
  {http://arxiv.org/abs/1408.4145} {arXiv:1408.4145 [astro-ph.HE]} \BibitemShut
  {NoStop}%
\bibitem [{\citenamefont {{Belczynski}}\ \emph
  {et~al.}(2016{\natexlab{e}})\citenamefont {{Belczynski}}, \citenamefont
  {{Holz}}, \citenamefont {{Bulik}},\ and\ \citenamefont
  {{O'Shaughnessy}}}]{2016Natur.534..512B}%
  \BibitemOpen
  \bibfield  {author} {\bibinfo {author} {\bibfnamefont {K.}~\bibnamefont
  {{Belczynski}}}, \bibinfo {author} {\bibfnamefont {D.~E.}\ \bibnamefont
  {{Holz}}}, \bibinfo {author} {\bibfnamefont {T.}~\bibnamefont {{Bulik}}}, \
  and\ \bibinfo {author} {\bibfnamefont {R.}~\bibnamefont {{O'Shaughnessy}}},\
  }\href {\doibase 10.1038/nature18322} {\bibfield  {journal} {\bibinfo
  {journal} {\nat}\ }\textbf {\bibinfo {volume} {534}},\ \bibinfo {pages} {512}
  (\bibinfo {year} {2016}{\natexlab{e}})},\ \Eprint
  {http://arxiv.org/abs/1602.04531} {arXiv:1602.04531 [astro-ph.HE]}
  \BibitemShut {NoStop}%
\bibitem [{\citenamefont {{Rodriguez}}\ \emph
  {et~al.}(2016{\natexlab{c}})\citenamefont {{Rodriguez}}, \citenamefont
  {{Zevin}}, \citenamefont {{Pankow}}, \citenamefont {{Kalogera}},\ and\
  \citenamefont {{Rasio}}}]{2016ApJ...832L...2R}%
  \BibitemOpen
  \bibfield  {author} {\bibinfo {author} {\bibfnamefont {C.~L.}\ \bibnamefont
  {{Rodriguez}}}, \bibinfo {author} {\bibfnamefont {M.}~\bibnamefont
  {{Zevin}}}, \bibinfo {author} {\bibfnamefont {C.}~\bibnamefont {{Pankow}}},
  \bibinfo {author} {\bibfnamefont {V.}~\bibnamefont {{Kalogera}}}, \ and\
  \bibinfo {author} {\bibfnamefont {F.~A.}\ \bibnamefont {{Rasio}}},\ }\href
  {\doibase 10.3847/2041-8205/832/1/L2} {\bibfield  {journal} {\bibinfo
  {journal} {\apjl}\ }\textbf {\bibinfo {volume} {832}},\ \bibinfo {eid} {L2}
  (\bibinfo {year} {2016}{\natexlab{c}})},\ \Eprint
  {http://arxiv.org/abs/1609.05916} {arXiv:1609.05916 [astro-ph.HE]}
  \BibitemShut {NoStop}%
\bibitem [{\citenamefont {{Kalogera}}(2000)}]{2000ApJ...541..319K}%
  \BibitemOpen
  \bibfield  {author} {\bibinfo {author} {\bibfnamefont {V.}~\bibnamefont
  {{Kalogera}}},\ }\href {\doibase 10.1086/309400} {\bibfield  {journal}
  {\bibinfo  {journal} {\apj}\ }\textbf {\bibinfo {volume} {541}},\ \bibinfo
  {pages} {319} (\bibinfo {year} {2000})},\ \Eprint
  {http://arxiv.org/abs/astro-ph/9911417} {astro-ph/9911417} \BibitemShut
  {NoStop}%
\bibitem [{\citenamefont {{Gerosa}}\ \emph {et~al.}(2013)\citenamefont
  {{Gerosa}}, \citenamefont {{Kesden}}, \citenamefont {{Berti}}, \citenamefont
  {{O'Shaughnessy}},\ and\ \citenamefont {{Sperhake}}}]{2013PhRvD..87j4028G}%
  \BibitemOpen
  \bibfield  {author} {\bibinfo {author} {\bibfnamefont {D.}~\bibnamefont
  {{Gerosa}}}, \bibinfo {author} {\bibfnamefont {M.}~\bibnamefont {{Kesden}}},
  \bibinfo {author} {\bibfnamefont {E.}~\bibnamefont {{Berti}}}, \bibinfo
  {author} {\bibfnamefont {R.}~\bibnamefont {{O'Shaughnessy}}}, \ and\ \bibinfo
  {author} {\bibfnamefont {U.}~\bibnamefont {{Sperhake}}},\ }\href {\doibase
  10.1103/PhysRevD.87.104028} {\bibfield  {journal} {\bibinfo  {journal}
  {\prd}\ }\textbf {\bibinfo {volume} {87}},\ \bibinfo {eid} {104028} (\bibinfo
  {year} {2013})},\ \Eprint {http://arxiv.org/abs/1302.4442} {arXiv:1302.4442
  [gr-qc]} \BibitemShut {NoStop}%
\bibitem [{\citenamefont {{Bogdanovi{\'c}}}\ \emph {et~al.}(2007)\citenamefont
  {{Bogdanovi{\'c}}}, \citenamefont {{Reynolds}},\ and\ \citenamefont
  {{Miller}}}]{2007ApJ...661L.147B}%
  \BibitemOpen
  \bibfield  {author} {\bibinfo {author} {\bibfnamefont {T.}~\bibnamefont
  {{Bogdanovi{\'c}}}}, \bibinfo {author} {\bibfnamefont {C.~S.}\ \bibnamefont
  {{Reynolds}}}, \ and\ \bibinfo {author} {\bibfnamefont {M.~C.}\ \bibnamefont
  {{Miller}}},\ }\href {\doibase 10.1086/518769} {\bibfield  {journal}
  {\bibinfo  {journal} {\apjl}\ }\textbf {\bibinfo {volume} {661}},\ \bibinfo
  {pages} {L147} (\bibinfo {year} {2007})},\ \Eprint
  {http://arxiv.org/abs/astro-ph/0703054} {astro-ph/0703054} \BibitemShut
  {NoStop}%
\bibitem [{\citenamefont {{Gerosa}}\ \emph {et~al.}(2015)\citenamefont
  {{Gerosa}}, \citenamefont {{Kesden}}, \citenamefont {{Sperhake}},
  \citenamefont {{Berti}},\ and\ \citenamefont
  {{O'Shaughnessy}}}]{2015PhRvD..92f4016G}%
  \BibitemOpen
  \bibfield  {author} {\bibinfo {author} {\bibfnamefont {D.}~\bibnamefont
  {{Gerosa}}}, \bibinfo {author} {\bibfnamefont {M.}~\bibnamefont {{Kesden}}},
  \bibinfo {author} {\bibfnamefont {U.}~\bibnamefont {{Sperhake}}}, \bibinfo
  {author} {\bibfnamefont {E.}~\bibnamefont {{Berti}}}, \ and\ \bibinfo
  {author} {\bibfnamefont {R.}~\bibnamefont {{O'Shaughnessy}}},\ }\href
  {\doibase 10.1103/PhysRevD.92.064016} {\bibfield  {journal} {\bibinfo
  {journal} {\prd}\ }\textbf {\bibinfo {volume} {92}},\ \bibinfo {eid} {064016}
  (\bibinfo {year} {2015})},\ \Eprint {http://arxiv.org/abs/1506.03492}
  {arXiv:1506.03492 [gr-qc]} \BibitemShut {NoStop}%
\bibitem [{\citenamefont {{Kesden}}\ \emph {et~al.}(2015)\citenamefont
  {{Kesden}}, \citenamefont {{Gerosa}}, \citenamefont {{O'Shaughnessy}},
  \citenamefont {{Berti}},\ and\ \citenamefont
  {{Sperhake}}}]{2015PhRvL.114h1103K}%
  \BibitemOpen
  \bibfield  {author} {\bibinfo {author} {\bibfnamefont {M.}~\bibnamefont
  {{Kesden}}}, \bibinfo {author} {\bibfnamefont {D.}~\bibnamefont {{Gerosa}}},
  \bibinfo {author} {\bibfnamefont {R.}~\bibnamefont {{O'Shaughnessy}}},
  \bibinfo {author} {\bibfnamefont {E.}~\bibnamefont {{Berti}}}, \ and\
  \bibinfo {author} {\bibfnamefont {U.}~\bibnamefont {{Sperhake}}},\ }\href
  {\doibase 10.1103/PhysRevLett.114.081103} {\bibfield  {journal} {\bibinfo
  {journal} {\prl}\ }\textbf {\bibinfo {volume} {114}},\ \bibinfo {eid}
  {081103} (\bibinfo {year} {2015})},\ \Eprint {http://arxiv.org/abs/1411.0674}
  {arXiv:1411.0674 [gr-qc]} \BibitemShut {NoStop}%
\bibitem [{\citenamefont {{Gerosa}}\ and\ \citenamefont
  {{Kesden}}(2016)}]{2016PhRvD..93l4066G}%
  \BibitemOpen
  \bibfield  {author} {\bibinfo {author} {\bibfnamefont {D.}~\bibnamefont
  {{Gerosa}}}\ and\ \bibinfo {author} {\bibfnamefont {M.}~\bibnamefont
  {{Kesden}}},\ }\href {\doibase 10.1103/PhysRevD.93.124066} {\bibfield
  {journal} {\bibinfo  {journal} {\prd}\ }\textbf {\bibinfo {volume} {93}},\
  \bibinfo {eid} {124066} (\bibinfo {year} {2016})},\ \Eprint
  {http://arxiv.org/abs/1605.01067} {arXiv:1605.01067 [astro-ph.HE]}
  \BibitemShut {NoStop}%
\bibitem [{\citenamefont {{Barausse}}\ \emph {et~al.}(2012)\citenamefont
  {{Barausse}}, \citenamefont {{Morozova}},\ and\ \citenamefont
  {{Rezzolla}}}]{2012ApJ...758...63B}%
  \BibitemOpen
  \bibfield  {author} {\bibinfo {author} {\bibfnamefont {E.}~\bibnamefont
  {{Barausse}}}, \bibinfo {author} {\bibfnamefont {V.}~\bibnamefont
  {{Morozova}}}, \ and\ \bibinfo {author} {\bibfnamefont {L.}~\bibnamefont
  {{Rezzolla}}},\ }\href {\doibase 10.1088/0004-637X/758/1/63} {\bibfield
  {journal} {\bibinfo  {journal} {\apj}\ }\textbf {\bibinfo {volume} {758}},\
  \bibinfo {eid} {63} (\bibinfo {year} {2012})},\ \Eprint
  {http://arxiv.org/abs/1206.3803} {arXiv:1206.3803 [gr-qc]} \BibitemShut
  {NoStop}%
\bibitem [{\citenamefont {{Barausse}}\ and\ \citenamefont
  {{Rezzolla}}(2009)}]{2009ApJ...704L..40B}%
  \BibitemOpen
  \bibfield  {author} {\bibinfo {author} {\bibfnamefont {E.}~\bibnamefont
  {{Barausse}}}\ and\ \bibinfo {author} {\bibfnamefont {L.}~\bibnamefont
  {{Rezzolla}}},\ }\href {\doibase 10.1088/0004-637X/704/1/L40} {\bibfield
  {journal} {\bibinfo  {journal} {\apjl}\ }\textbf {\bibinfo {volume} {704}},\
  \bibinfo {pages} {L40} (\bibinfo {year} {2009})},\ \Eprint
  {http://arxiv.org/abs/0904.2577} {arXiv:0904.2577 [gr-qc]} \BibitemShut
  {NoStop}%
\bibitem [{\citenamefont {{Lousto}}\ and\ \citenamefont
  {{Zlochower}}(2014)}]{2014PhRvD..89j4052L}%
  \BibitemOpen
  \bibfield  {author} {\bibinfo {author} {\bibfnamefont {C.~O.}\ \bibnamefont
  {{Lousto}}}\ and\ \bibinfo {author} {\bibfnamefont {Y.}~\bibnamefont
  {{Zlochower}}},\ }\href {\doibase 10.1103/PhysRevD.89.104052} {\bibfield
  {journal} {\bibinfo  {journal} {\prd}\ }\textbf {\bibinfo {volume} {89}},\
  \bibinfo {eid} {104052} (\bibinfo {year} {2014})},\ \Eprint
  {http://arxiv.org/abs/1312.5775} {arXiv:1312.5775 [gr-qc]} \BibitemShut
  {NoStop}%
\bibitem [{\citenamefont {{Healy}}\ \emph {et~al.}(2014)\citenamefont
  {{Healy}}, \citenamefont {{Lousto}},\ and\ \citenamefont
  {{Zlochower}}}]{2014PhRvD..90j4004H}%
  \BibitemOpen
  \bibfield  {author} {\bibinfo {author} {\bibfnamefont {J.}~\bibnamefont
  {{Healy}}}, \bibinfo {author} {\bibfnamefont {C.~O.}\ \bibnamefont
  {{Lousto}}}, \ and\ \bibinfo {author} {\bibfnamefont {Y.}~\bibnamefont
  {{Zlochower}}},\ }\href {\doibase 10.1103/PhysRevD.90.104004} {\bibfield
  {journal} {\bibinfo  {journal} {\prd}\ }\textbf {\bibinfo {volume} {90}},\
  \bibinfo {eid} {104004} (\bibinfo {year} {2014})},\ \Eprint
  {http://arxiv.org/abs/1406.7295} {arXiv:1406.7295 [gr-qc]} \BibitemShut
  {NoStop}%
\bibitem [{\citenamefont {{Zlochower}}\ and\ \citenamefont
  {{Lousto}}(2015)}]{2015PhRvD..92b4022Z}%
  \BibitemOpen
  \bibfield  {author} {\bibinfo {author} {\bibfnamefont {Y.}~\bibnamefont
  {{Zlochower}}}\ and\ \bibinfo {author} {\bibfnamefont {C.~O.}\ \bibnamefont
  {{Lousto}}},\ }\href {\doibase 10.1103/PhysRevD.92.024022} {\bibfield
  {journal} {\bibinfo  {journal} {\prd}\ }\textbf {\bibinfo {volume} {92}},\
  \bibinfo {eid} {024022} (\bibinfo {year} {2015})},\ \Eprint
  {http://arxiv.org/abs/1503.07536} {arXiv:1503.07536 [gr-qc]} \BibitemShut
  {NoStop}%
\bibitem [{\citenamefont {{Husa}}\ \emph {et~al.}(2016)\citenamefont {{Husa}},
  \citenamefont {{Khan}}, \citenamefont {{Hannam}}, \citenamefont
  {{P{\"u}rrer}}, \citenamefont {{Ohme}}, \citenamefont {{Forteza}},\ and\
  \citenamefont {{Boh{\'e}}}}]{2016PhRvD..93d4006H}%
  \BibitemOpen
  \bibfield  {author} {\bibinfo {author} {\bibfnamefont {S.}~\bibnamefont
  {{Husa}}}, \bibinfo {author} {\bibfnamefont {S.}~\bibnamefont {{Khan}}},
  \bibinfo {author} {\bibfnamefont {M.}~\bibnamefont {{Hannam}}}, \bibinfo
  {author} {\bibfnamefont {M.}~\bibnamefont {{P{\"u}rrer}}}, \bibinfo {author}
  {\bibfnamefont {F.}~\bibnamefont {{Ohme}}}, \bibinfo {author} {\bibfnamefont
  {X.~J.}\ \bibnamefont {{Forteza}}}, \ and\ \bibinfo {author} {\bibfnamefont
  {A.}~\bibnamefont {{Boh{\'e}}}},\ }\href {\doibase
  10.1103/PhysRevD.93.044006} {\bibfield  {journal} {\bibinfo  {journal}
  {\prd}\ }\textbf {\bibinfo {volume} {93}},\ \bibinfo {eid} {044006} (\bibinfo
  {year} {2016})},\ \Eprint {http://arxiv.org/abs/1508.07250} {arXiv:1508.07250
  [gr-qc]} \BibitemShut {NoStop}%
\bibitem [{\citenamefont {{Hofmann}}\ \emph {et~al.}(2016)\citenamefont
  {{Hofmann}}, \citenamefont {{Barausse}},\ and\ \citenamefont
  {{Rezzolla}}}]{2016ApJ...825L..19H}%
  \BibitemOpen
  \bibfield  {author} {\bibinfo {author} {\bibfnamefont {F.}~\bibnamefont
  {{Hofmann}}}, \bibinfo {author} {\bibfnamefont {E.}~\bibnamefont
  {{Barausse}}}, \ and\ \bibinfo {author} {\bibfnamefont {L.}~\bibnamefont
  {{Rezzolla}}},\ }\href {\doibase 10.3847/2041-8205/825/2/L19} {\bibfield
  {journal} {\bibinfo  {journal} {\apjl}\ }\textbf {\bibinfo {volume} {825}},\
  \bibinfo {eid} {L19} (\bibinfo {year} {2016})},\ \Eprint
  {http://arxiv.org/abs/1605.01938} {arXiv:1605.01938 [gr-qc]} \BibitemShut
  {NoStop}%
\bibitem [{\citenamefont {{Healy}}\ and\ \citenamefont
  {{Lousto}}(2017)}]{2016arXiv161009713H}%
  \BibitemOpen
  \bibfield  {author} {\bibinfo {author} {\bibfnamefont {J.}~\bibnamefont
  {{Healy}}}\ and\ \bibinfo {author} {\bibfnamefont {C.~O.}\ \bibnamefont
  {{Lousto}}},\ }\href {\doibase 10.1103/PhysRevD.95.024037} {\bibfield
  {journal} {\bibinfo  {journal} {\prd}\ }\textbf {\bibinfo {volume} {95}},\
  \bibinfo {eid} {024037} (\bibinfo {year} {2017})},\ \Eprint
  {http://arxiv.org/abs/1610.09713} {arXiv:1610.09713 [gr-qc]} \BibitemShut
  {NoStop}%
\bibitem [{\citenamefont {{Jim{\'e}nez-Forteza}}\ \emph
  {et~al.}(2017)\citenamefont {{Jim{\'e}nez-Forteza}}, \citenamefont
  {{Keitel}}, \citenamefont {{Husa}}, \citenamefont {{Hannam}}, \citenamefont
  {{Khan}},\ and\ \citenamefont {{P{\"u}rrer}}}]{2017PhRvD..95f4024J}%
  \BibitemOpen
  \bibfield  {author} {\bibinfo {author} {\bibfnamefont {X.}~\bibnamefont
  {{Jim{\'e}nez-Forteza}}}, \bibinfo {author} {\bibfnamefont {D.}~\bibnamefont
  {{Keitel}}}, \bibinfo {author} {\bibfnamefont {S.}~\bibnamefont {{Husa}}},
  \bibinfo {author} {\bibfnamefont {M.}~\bibnamefont {{Hannam}}}, \bibinfo
  {author} {\bibfnamefont {S.}~\bibnamefont {{Khan}}}, \ and\ \bibinfo {author}
  {\bibfnamefont {M.}~\bibnamefont {{P{\"u}rrer}}},\ }\href {\doibase
  10.1103/PhysRevD.95.064024} {\bibfield  {journal} {\bibinfo  {journal}
  {\prd}\ }\textbf {\bibinfo {volume} {95}},\ \bibinfo {eid} {064024} (\bibinfo
  {year} {2017})},\ \Eprint {http://arxiv.org/abs/1611.00332} {arXiv:1611.00332
  [gr-qc]} \BibitemShut {NoStop}%
\bibitem [{\citenamefont {{Hogg}}(1999)}]{1999astro.ph..5116H}%
  \BibitemOpen
  \bibfield  {author} {\bibinfo {author} {\bibfnamefont {D.~W.}\ \bibnamefont
  {{Hogg}}},\ }\href@noop {} {\bibfield  {journal} {\bibinfo  {journal} {ArXiv
  e-prints}\ } (\bibinfo {year} {1999})},\ \Eprint
  {http://arxiv.org/abs/astro-ph/9905116} {astro-ph/9905116} \BibitemShut
  {NoStop}%
\bibitem [{\citenamefont {{Ade {\it et al.} (Planck
  Collaboration)}}(2016)}]{2016A&A...594A..13P}%
  \BibitemOpen
  \bibfield  {author} {\bibinfo {author} {\bibfnamefont {P.~A.~R.}\
  \bibnamefont {{Ade {\it et al.} (Planck Collaboration)}}},\ }\href {\doibase
  10.1051/0004-6361/201525830} {\bibfield  {journal} {\bibinfo  {journal}
  {\aap}\ }\textbf {\bibinfo {volume} {594}},\ \bibinfo {eid} {A13} (\bibinfo
  {year} {2016})},\ \Eprint {http://arxiv.org/abs/1502.01589}
  {arXiv:1502.01589} \BibitemShut {NoStop}%
\bibitem [{\citenamefont {{Abbott {\it et al.} (LIGO Scientific Collaboration
  and Virgo Collaboration)}}(2016{\natexlab{e}})}]{2016PhRvL.116x1102A}%
  \BibitemOpen
  \bibfield  {author} {\bibinfo {author} {\bibfnamefont {B.~P.}\ \bibnamefont
  {{Abbott {\it et al.} (LIGO Scientific Collaboration and Virgo
  Collaboration)}}},\ }\href {\doibase 10.1103/PhysRevLett.116.241102}
  {\bibfield  {journal} {\bibinfo  {journal} {\prl}\ }\textbf {\bibinfo
  {volume} {116}},\ \bibinfo {eid} {241102} (\bibinfo {year}
  {2016}{\natexlab{e}})},\ \Eprint {http://arxiv.org/abs/1602.03840}
  {arXiv:1602.03840 [gr-qc]} \BibitemShut {NoStop}%
\bibitem [{\citenamefont {{Racine}}(2008)}]{2008PhRvD..78d4021R}%
  \BibitemOpen
  \bibfield  {author} {\bibinfo {author} {\bibfnamefont {{\'E}.}~\bibnamefont
  {{Racine}}},\ }\href {\doibase 10.1103/PhysRevD.78.044021} {\bibfield
  {journal} {\bibinfo  {journal} {\prd}\ }\textbf {\bibinfo {volume} {78}},\
  \bibinfo {eid} {044021} (\bibinfo {year} {2008})},\ \Eprint
  {http://arxiv.org/abs/0803.1820} {arXiv:0803.1820 [gr-qc]} \BibitemShut
  {NoStop}%
\bibitem [{\citenamefont {{P{\"u}rrer}}\ \emph {et~al.}(2016)\citenamefont
  {{P{\"u}rrer}}, \citenamefont {{Hannam}},\ and\ \citenamefont
  {{Ohme}}}]{2016PhRvD..93h4042P}%
  \BibitemOpen
  \bibfield  {author} {\bibinfo {author} {\bibfnamefont {M.}~\bibnamefont
  {{P{\"u}rrer}}}, \bibinfo {author} {\bibfnamefont {M.}~\bibnamefont
  {{Hannam}}}, \ and\ \bibinfo {author} {\bibfnamefont {F.}~\bibnamefont
  {{Ohme}}},\ }\href {\doibase 10.1103/PhysRevD.93.084042} {\bibfield
  {journal} {\bibinfo  {journal} {\prd}\ }\textbf {\bibinfo {volume} {93}},\
  \bibinfo {eid} {084042} (\bibinfo {year} {2016})},\ \Eprint
  {http://arxiv.org/abs/1512.04955} {arXiv:1512.04955 [gr-qc]} \BibitemShut
  {NoStop}%
\bibitem [{\citenamefont {{Berti}}\ \emph {et~al.}(2016)\citenamefont
  {{Berti}}, \citenamefont {{Sesana}}, \citenamefont {{Barausse}},
  \citenamefont {{Cardoso}},\ and\ \citenamefont
  {{Belczynski}}}]{2016PhRvL.117j1102B}%
  \BibitemOpen
  \bibfield  {author} {\bibinfo {author} {\bibfnamefont {E.}~\bibnamefont
  {{Berti}}}, \bibinfo {author} {\bibfnamefont {A.}~\bibnamefont {{Sesana}}},
  \bibinfo {author} {\bibfnamefont {E.}~\bibnamefont {{Barausse}}}, \bibinfo
  {author} {\bibfnamefont {V.}~\bibnamefont {{Cardoso}}}, \ and\ \bibinfo
  {author} {\bibfnamefont {K.}~\bibnamefont {{Belczynski}}},\ }\href {\doibase
  10.1103/PhysRevLett.117.101102} {\bibfield  {journal} {\bibinfo  {journal}
  {\prl}\ }\textbf {\bibinfo {volume} {117}},\ \bibinfo {eid} {101102}
  (\bibinfo {year} {2016})},\ \Eprint {http://arxiv.org/abs/1605.09286}
  {arXiv:1605.09286 [gr-qc]} \BibitemShut {NoStop}%
\bibitem [{\citenamefont {{Vitale}}(2016)}]{2016PhRvD..94l1501V}%
  \BibitemOpen
  \bibfield  {author} {\bibinfo {author} {\bibfnamefont {S.}~\bibnamefont
  {{Vitale}}},\ }\href {\doibase 10.1103/PhysRevD.94.121501} {\bibfield
  {journal} {\bibinfo  {journal} {\prd}\ }\textbf {\bibinfo {volume} {94}},\
  \bibinfo {eid} {121501} (\bibinfo {year} {2016})},\ \Eprint
  {http://arxiv.org/abs/1610.06914} {arXiv:1610.06914 [gr-qc]} \BibitemShut
  {NoStop}%
\bibitem [{\citenamefont {{Vitale}}\ and\ \citenamefont
  {{Evans}}(2017)}]{2017PhRvD..95f4052V}%
  \BibitemOpen
  \bibfield  {author} {\bibinfo {author} {\bibfnamefont {S.}~\bibnamefont
  {{Vitale}}}\ and\ \bibinfo {author} {\bibfnamefont {M.}~\bibnamefont
  {{Evans}}},\ }\href {\doibase 10.1103/PhysRevD.95.064052} {\bibfield
  {journal} {\bibinfo  {journal} {\prd}\ }\textbf {\bibinfo {volume} {95}},\
  \bibinfo {eid} {064052} (\bibinfo {year} {2017})},\ \Eprint
  {http://arxiv.org/abs/1610.06917} {arXiv:1610.06917 [gr-qc]} \BibitemShut
  {NoStop}%
\bibitem [{\citenamefont {{Vitale}}\ \emph
  {et~al.}(2017{\natexlab{a}})\citenamefont {{Vitale}}, \citenamefont
  {{Lynch}}, \citenamefont {{Raymond}}, \citenamefont {{Sturani}},
  \citenamefont {{Veitch}},\ and\ \citenamefont
  {{Graff}}}]{2017PhRvD..95f4053V}%
  \BibitemOpen
  \bibfield  {author} {\bibinfo {author} {\bibfnamefont {S.}~\bibnamefont
  {{Vitale}}}, \bibinfo {author} {\bibfnamefont {R.}~\bibnamefont {{Lynch}}},
  \bibinfo {author} {\bibfnamefont {V.}~\bibnamefont {{Raymond}}}, \bibinfo
  {author} {\bibfnamefont {R.}~\bibnamefont {{Sturani}}}, \bibinfo {author}
  {\bibfnamefont {J.}~\bibnamefont {{Veitch}}}, \ and\ \bibinfo {author}
  {\bibfnamefont {P.}~\bibnamefont {{Graff}}},\ }\href {\doibase
  10.1103/PhysRevD.95.064053} {\bibfield  {journal} {\bibinfo  {journal}
  {\prd}\ }\textbf {\bibinfo {volume} {95}},\ \bibinfo {eid} {064053} (\bibinfo
  {year} {2017}{\natexlab{a}})},\ \Eprint {http://arxiv.org/abs/1611.01122}
  {arXiv:1611.01122 [gr-qc]} \BibitemShut {NoStop}%
\bibitem [{\citenamefont {{Santamar{\'{\i}}a}}\ \emph
  {et~al.}(2010)\citenamefont {{Santamar{\'{\i}}a}}, \citenamefont {{Ohme}},
  \citenamefont {{Ajith}}, \citenamefont {{Br{\"u}gmann}}, \citenamefont
  {{Dorband}}, \citenamefont {{Hannam}}, \citenamefont {{Husa}}, \citenamefont
  {{M{\"o}sta}}, \citenamefont {{Pollney}}, \citenamefont {{Reisswig}},
  \citenamefont {{Robinson}}, \citenamefont {{Seiler}},\ and\ \citenamefont
  {{Krishnan}}}]{2010PhRvD..82f4016S}%
  \BibitemOpen
  \bibfield  {author} {\bibinfo {author} {\bibfnamefont {L.}~\bibnamefont
  {{Santamar{\'{\i}}a}}}, \bibinfo {author} {\bibfnamefont {F.}~\bibnamefont
  {{Ohme}}}, \bibinfo {author} {\bibfnamefont {P.}~\bibnamefont {{Ajith}}},
  \bibinfo {author} {\bibfnamefont {B.}~\bibnamefont {{Br{\"u}gmann}}},
  \bibinfo {author} {\bibfnamefont {N.}~\bibnamefont {{Dorband}}}, \bibinfo
  {author} {\bibfnamefont {M.}~\bibnamefont {{Hannam}}}, \bibinfo {author}
  {\bibfnamefont {S.}~\bibnamefont {{Husa}}}, \bibinfo {author} {\bibfnamefont
  {P.}~\bibnamefont {{M{\"o}sta}}}, \bibinfo {author} {\bibfnamefont
  {D.}~\bibnamefont {{Pollney}}}, \bibinfo {author} {\bibfnamefont
  {C.}~\bibnamefont {{Reisswig}}}, \bibinfo {author} {\bibfnamefont {E.~L.}\
  \bibnamefont {{Robinson}}}, \bibinfo {author} {\bibfnamefont
  {J.}~\bibnamefont {{Seiler}}}, \ and\ \bibinfo {author} {\bibfnamefont
  {B.}~\bibnamefont {{Krishnan}}},\ }\href {\doibase
  10.1103/PhysRevD.82.064016} {\bibfield  {journal} {\bibinfo  {journal}
  {\prd}\ }\textbf {\bibinfo {volume} {82}},\ \bibinfo {eid} {064016} (\bibinfo
  {year} {2010})},\ \Eprint {http://arxiv.org/abs/1005.3306} {arXiv:1005.3306
  [gr-qc]} \BibitemShut {NoStop}%
\bibitem [{\citenamefont {{Abbott {\it et al.} (LIGO Scientific Collaboration
  and Virgo Collaboration)}}(2016{\natexlab{f}})}]{2016LRR....19....1A}%
  \BibitemOpen
  \bibfield  {author} {\bibinfo {author} {\bibfnamefont {B.~P.}\ \bibnamefont
  {{Abbott {\it et al.} (LIGO Scientific Collaboration and Virgo
  Collaboration)}}},\ }\href {\doibase 10.1007/lrr-2016-1} {\bibfield
  {journal} {\bibinfo  {journal} {Living Reviews in Relativity}\ }\textbf
  {\bibinfo {volume} {19}},\ \bibinfo {eid} {1} (\bibinfo {year}
  {2016}{\natexlab{f}})},\ \Eprint {http://arxiv.org/abs/1304.0670}
  {arXiv:1304.0670 [gr-qc]} \BibitemShut {NoStop}%
\bibitem [{\citenamefont {{Mc Clelland}}\ \emph {et~al.}(2015)\citenamefont
  {{Mc Clelland}}, \citenamefont {{Evans}}, \citenamefont {{Schnabel}},
  \citenamefont {Lantz}, \citenamefont {{Martin}},\ and\ \citenamefont
  {{Quetschke}}}]{dcc_instrument}%
  \BibitemOpen
  \bibfield  {author} {\bibinfo {author} {\bibfnamefont {D.}~\bibnamefont {{Mc
  Clelland}}}, \bibinfo {author} {\bibfnamefont {M.}~\bibnamefont {{Evans}}},
  \bibinfo {author} {\bibfnamefont {R.}~\bibnamefont {{Schnabel}}}, \bibinfo
  {author} {\bibfnamefont {B.}~\bibnamefont {Lantz}}, \bibinfo {author}
  {\bibfnamefont {I.}~\bibnamefont {{Martin}}}, \ and\ \bibinfo {author}
  {\bibfnamefont {V.}~\bibnamefont {{Quetschke}}},\ }\href@noop {} {\bibfield
  {journal} {\bibinfo  {journal} {LIGO Document Control Center}\ }\textbf
  {\bibinfo {volume} {\url{https://dcc.ligo.org/LIGO-T1400316/public}}}
  (\bibinfo {year} {2015})}\BibitemShut {NoStop}%
\bibitem [{\citenamefont {{Abadie {\it et al.} (LIGO Scientific Collaboration
  and Virgo Collaboration)}}(2010)}]{2010CQGra..27q3001A}%
  \BibitemOpen
  \bibfield  {author} {\bibinfo {author} {\bibfnamefont {J.}~\bibnamefont
  {{Abadie {\it et al.} (LIGO Scientific Collaboration and Virgo
  Collaboration)}}},\ }\href {\doibase 10.1088/0264-9381/27/17/173001}
  {\bibfield  {journal} {\bibinfo  {journal} {\cqg}\ }\textbf {\bibinfo
  {volume} {27}},\ \bibinfo {eid} {173001} (\bibinfo {year} {2010})},\ \Eprint
  {http://arxiv.org/abs/1003.2480} {arXiv:1003.2480 [astro-ph.HE]} \BibitemShut
  {NoStop}%
\bibitem [{\citenamefont {{Ghosh}}\ \emph {et~al.}(2016)\citenamefont
  {{Ghosh}}, \citenamefont {{Del Pozzo}},\ and\ \citenamefont
  {{Ajith}}}]{2016PhRvD..94j4070G}%
  \BibitemOpen
  \bibfield  {author} {\bibinfo {author} {\bibfnamefont {A.}~\bibnamefont
  {{Ghosh}}}, \bibinfo {author} {\bibfnamefont {W.}~\bibnamefont {{Del
  Pozzo}}}, \ and\ \bibinfo {author} {\bibfnamefont {P.}~\bibnamefont
  {{Ajith}}},\ }\href {\doibase 10.1103/PhysRevD.94.104070} {\bibfield
  {journal} {\bibinfo  {journal} {\prd}\ }\textbf {\bibinfo {volume} {94}},\
  \bibinfo {eid} {104070} (\bibinfo {year} {2016})},\ \Eprint
  {http://arxiv.org/abs/1505.05607} {arXiv:1505.05607 [gr-qc]} \BibitemShut
  {NoStop}%
\bibitem [{\citenamefont {{Veitch}}\ \emph {et~al.}(2015)\citenamefont
  {{Veitch}}, \citenamefont {{Raymond}}, \citenamefont {{Farr}}, \citenamefont
  {{Farr}}, \citenamefont {{Graff}}, \citenamefont {{Vitale}}, \citenamefont
  {{Aylott}}, \citenamefont {{Blackburn}}, \citenamefont {{Christensen}},
  \citenamefont {{Coughlin}}, \citenamefont {{Del Pozzo}}, \citenamefont
  {{Feroz}}, \citenamefont {{Gair}}, \citenamefont {{Haster}}, \citenamefont
  {{Kalogera}}, \citenamefont {{Littenberg}}, \citenamefont {{Mandel}},
  \citenamefont {{O'Shaughnessy}}, \citenamefont {{Pitkin}}, \citenamefont
  {{Rodriguez}}, \citenamefont {{R{\"o}ver}}, \citenamefont {{Sidery}},
  \citenamefont {{Smith}}, \citenamefont {{Van Der Sluys}}, \citenamefont
  {{Vecchio}}, \citenamefont {{Vousden}},\ and\ \citenamefont
  {{Wade}}}]{2015PhRvD..91d2003V}%
  \BibitemOpen
  \bibfield  {author} {\bibinfo {author} {\bibfnamefont {J.}~\bibnamefont
  {{Veitch}}}, \bibinfo {author} {\bibfnamefont {V.}~\bibnamefont {{Raymond}}},
  \bibinfo {author} {\bibfnamefont {B.}~\bibnamefont {{Farr}}}, \bibinfo
  {author} {\bibfnamefont {W.}~\bibnamefont {{Farr}}}, \bibinfo {author}
  {\bibfnamefont {P.}~\bibnamefont {{Graff}}}, \bibinfo {author} {\bibfnamefont
  {S.}~\bibnamefont {{Vitale}}}, \bibinfo {author} {\bibfnamefont
  {B.}~\bibnamefont {{Aylott}}}, \bibinfo {author} {\bibfnamefont
  {K.}~\bibnamefont {{Blackburn}}}, \bibinfo {author} {\bibfnamefont
  {N.}~\bibnamefont {{Christensen}}}, \bibinfo {author} {\bibfnamefont
  {M.}~\bibnamefont {{Coughlin}}}, \bibinfo {author} {\bibfnamefont
  {W.}~\bibnamefont {{Del Pozzo}}}, \bibinfo {author} {\bibfnamefont
  {F.}~\bibnamefont {{Feroz}}}, \bibinfo {author} {\bibfnamefont
  {J.}~\bibnamefont {{Gair}}}, \bibinfo {author} {\bibfnamefont {C.-J.}\
  \bibnamefont {{Haster}}}, \bibinfo {author} {\bibfnamefont {V.}~\bibnamefont
  {{Kalogera}}}, \bibinfo {author} {\bibfnamefont {T.}~\bibnamefont
  {{Littenberg}}}, \bibinfo {author} {\bibfnamefont {I.}~\bibnamefont
  {{Mandel}}}, \bibinfo {author} {\bibfnamefont {R.}~\bibnamefont
  {{O'Shaughnessy}}}, \bibinfo {author} {\bibfnamefont {M.}~\bibnamefont
  {{Pitkin}}}, \bibinfo {author} {\bibfnamefont {C.}~\bibnamefont
  {{Rodriguez}}}, \bibinfo {author} {\bibfnamefont {C.}~\bibnamefont
  {{R{\"o}ver}}}, \bibinfo {author} {\bibfnamefont {T.}~\bibnamefont
  {{Sidery}}}, \bibinfo {author} {\bibfnamefont {R.}~\bibnamefont {{Smith}}},
  \bibinfo {author} {\bibfnamefont {M.}~\bibnamefont {{Van Der Sluys}}},
  \bibinfo {author} {\bibfnamefont {A.}~\bibnamefont {{Vecchio}}}, \bibinfo
  {author} {\bibfnamefont {W.}~\bibnamefont {{Vousden}}}, \ and\ \bibinfo
  {author} {\bibfnamefont {L.}~\bibnamefont {{Wade}}},\ }\href {\doibase
  10.1103/PhysRevD.91.042003} {\bibfield  {journal} {\bibinfo  {journal}
  {\prd}\ }\textbf {\bibinfo {volume} {91}},\ \bibinfo {eid} {042003} (\bibinfo
  {year} {2015})},\ \Eprint {http://arxiv.org/abs/1409.7215} {arXiv:1409.7215
  [gr-qc]} \BibitemShut {NoStop}%
\bibitem [{\citenamefont {{Ajith}}\ and\ \citenamefont
  {{Bose}}(2009)}]{2009PhRvD..79h4032A}%
  \BibitemOpen
  \bibfield  {author} {\bibinfo {author} {\bibfnamefont {P.}~\bibnamefont
  {{Ajith}}}\ and\ \bibinfo {author} {\bibfnamefont {S.}~\bibnamefont
  {{Bose}}},\ }\href {\doibase 10.1103/PhysRevD.79.084032} {\bibfield
  {journal} {\bibinfo  {journal} {\prd}\ }\textbf {\bibinfo {volume} {79}},\
  \bibinfo {eid} {084032} (\bibinfo {year} {2009})},\ \Eprint
  {http://arxiv.org/abs/0901.4936} {arXiv:0901.4936 [gr-qc]} \BibitemShut
  {NoStop}%
\bibitem [{\citenamefont {{Littenberg}}\ \emph {et~al.}(2013)\citenamefont
  {{Littenberg}}, \citenamefont {{Baker}}, \citenamefont {{Buonanno}},\ and\
  \citenamefont {{Kelly}}}]{2013PhRvD..87j4003L}%
  \BibitemOpen
  \bibfield  {author} {\bibinfo {author} {\bibfnamefont {T.~B.}\ \bibnamefont
  {{Littenberg}}}, \bibinfo {author} {\bibfnamefont {J.~G.}\ \bibnamefont
  {{Baker}}}, \bibinfo {author} {\bibfnamefont {A.}~\bibnamefont {{Buonanno}}},
  \ and\ \bibinfo {author} {\bibfnamefont {B.~J.}\ \bibnamefont {{Kelly}}},\
  }\href {\doibase 10.1103/PhysRevD.87.104003} {\bibfield  {journal} {\bibinfo
  {journal} {\prd}\ }\textbf {\bibinfo {volume} {87}},\ \bibinfo {eid} {104003}
  (\bibinfo {year} {2013})},\ \Eprint {http://arxiv.org/abs/1210.0893}
  {arXiv:1210.0893 [gr-qc]} \BibitemShut {NoStop}%
\bibitem [{\citenamefont {{Cho}}\ \emph {et~al.}(2013)\citenamefont {{Cho}},
  \citenamefont {{Ochsner}}, \citenamefont {{O'Shaughnessy}}, \citenamefont
  {{Kim}},\ and\ \citenamefont {{Lee}}}]{2013PhRvD..87b4004C}%
  \BibitemOpen
  \bibfield  {author} {\bibinfo {author} {\bibfnamefont {H.-S.}\ \bibnamefont
  {{Cho}}}, \bibinfo {author} {\bibfnamefont {E.}~\bibnamefont {{Ochsner}}},
  \bibinfo {author} {\bibfnamefont {R.}~\bibnamefont {{O'Shaughnessy}}},
  \bibinfo {author} {\bibfnamefont {C.}~\bibnamefont {{Kim}}}, \ and\ \bibinfo
  {author} {\bibfnamefont {C.-H.}\ \bibnamefont {{Lee}}},\ }\href {\doibase
  10.1103/PhysRevD.87.024004} {\bibfield  {journal} {\bibinfo  {journal}
  {\prd}\ }\textbf {\bibinfo {volume} {87}},\ \bibinfo {eid} {024004} (\bibinfo
  {year} {2013})},\ \Eprint {http://arxiv.org/abs/1209.4494} {arXiv:1209.4494
  [gr-qc]} \BibitemShut {NoStop}%
\bibitem [{\citenamefont {{Graff}}\ \emph {et~al.}(2015)\citenamefont
  {{Graff}}, \citenamefont {{Buonanno}},\ and\ \citenamefont
  {{Sathyaprakash}}}]{2015PhRvD..92b2002G}%
  \BibitemOpen
  \bibfield  {author} {\bibinfo {author} {\bibfnamefont {P.~B.}\ \bibnamefont
  {{Graff}}}, \bibinfo {author} {\bibfnamefont {A.}~\bibnamefont {{Buonanno}}},
  \ and\ \bibinfo {author} {\bibfnamefont {B.~S.}\ \bibnamefont
  {{Sathyaprakash}}},\ }\href {\doibase 10.1103/PhysRevD.92.022002} {\bibfield
  {journal} {\bibinfo  {journal} {\prd}\ }\textbf {\bibinfo {volume} {92}},\
  \bibinfo {eid} {022002} (\bibinfo {year} {2015})},\ \Eprint
  {http://arxiv.org/abs/1504.04766} {arXiv:1504.04766 [gr-qc]} \BibitemShut
  {NoStop}%
\bibitem [{\citenamefont {{Poisson}}\ and\ \citenamefont
  {{Will}}(1995)}]{1995PhRvD..52..848P}%
  \BibitemOpen
  \bibfield  {author} {\bibinfo {author} {\bibfnamefont {E.}~\bibnamefont
  {{Poisson}}}\ and\ \bibinfo {author} {\bibfnamefont {C.~M.}\ \bibnamefont
  {{Will}}},\ }\href {\doibase 10.1103/PhysRevD.52.848} {\bibfield  {journal}
  {\bibinfo  {journal} {\prd}\ }\textbf {\bibinfo {volume} {52}},\ \bibinfo
  {pages} {848} (\bibinfo {year} {1995})},\ \Eprint
  {http://arxiv.org/abs/gr-qc/9502040} {gr-qc/9502040} \BibitemShut {NoStop}%
\bibitem [{\citenamefont {{Hughes}}(2002)}]{2002MNRAS.331..805H}%
  \BibitemOpen
  \bibfield  {author} {\bibinfo {author} {\bibfnamefont {S.~A.}\ \bibnamefont
  {{Hughes}}},\ }\href {\doibase 10.1046/j.1365-8711.2002.05247.x} {\bibfield
  {journal} {\bibinfo  {journal} {\mnras}\ }\textbf {\bibinfo {volume} {331}},\
  \bibinfo {pages} {805} (\bibinfo {year} {2002})},\ \Eprint
  {http://arxiv.org/abs/astro-ph/0108483} {astro-ph/0108483} \BibitemShut
  {NoStop}%
\bibitem [{\citenamefont {{Berti}}\ \emph {et~al.}(2005)\citenamefont
  {{Berti}}, \citenamefont {{Buonanno}},\ and\ \citenamefont
  {{Will}}}]{2005PhRvD..71h4025B}%
  \BibitemOpen
  \bibfield  {author} {\bibinfo {author} {\bibfnamefont {E.}~\bibnamefont
  {{Berti}}}, \bibinfo {author} {\bibfnamefont {A.}~\bibnamefont {{Buonanno}}},
  \ and\ \bibinfo {author} {\bibfnamefont {C.~M.}\ \bibnamefont {{Will}}},\
  }\href {\doibase 10.1103/PhysRevD.71.084025} {\bibfield  {journal} {\bibinfo
  {journal} {\prd}\ }\textbf {\bibinfo {volume} {71}},\ \bibinfo {eid} {084025}
  (\bibinfo {year} {2005})},\ \Eprint {http://arxiv.org/abs/gr-qc/0411129}
  {gr-qc/0411129} \BibitemShut {NoStop}%
\bibitem [{\citenamefont {{Di Valentino}}\ \emph {et~al.}(2016)\citenamefont
  {{Di Valentino}}, \citenamefont {{Melchiorri}},\ and\ \citenamefont
  {{Silk}}}]{2016PhLB..761..242D}%
  \BibitemOpen
  \bibfield  {author} {\bibinfo {author} {\bibfnamefont {E.}~\bibnamefont {{Di
  Valentino}}}, \bibinfo {author} {\bibfnamefont {A.}~\bibnamefont
  {{Melchiorri}}}, \ and\ \bibinfo {author} {\bibfnamefont {J.}~\bibnamefont
  {{Silk}}},\ }\href {\doibase 10.1016/j.physletb.2016.08.043} {\bibfield
  {journal} {\bibinfo  {journal} {Physics Letters B}\ }\textbf {\bibinfo
  {volume} {761}},\ \bibinfo {pages} {242} (\bibinfo {year} {2016})},\ \Eprint
  {http://arxiv.org/abs/1606.00634} {arXiv:1606.00634} \BibitemShut {NoStop}%
\bibitem [{\citenamefont {{Riess}}\ \emph {et~al.}(2016)\citenamefont
  {{Riess}}, \citenamefont {{Macri}}, \citenamefont {{Hoffmann}}, \citenamefont
  {{Scolnic}}, \citenamefont {{Casertano}}, \citenamefont {{Filippenko}},
  \citenamefont {{Tucker}}, \citenamefont {{Reid}}, \citenamefont {{Jones}},
  \citenamefont {{Silverman}}, \citenamefont {{Chornock}}, \citenamefont
  {{Challis}}, \citenamefont {{Yuan}}, \citenamefont {{Brown}},\ and\
  \citenamefont {{Foley}}}]{2016ApJ...826...56R}%
  \BibitemOpen
  \bibfield  {author} {\bibinfo {author} {\bibfnamefont {A.~G.}\ \bibnamefont
  {{Riess}}}, \bibinfo {author} {\bibfnamefont {L.~M.}\ \bibnamefont
  {{Macri}}}, \bibinfo {author} {\bibfnamefont {S.~L.}\ \bibnamefont
  {{Hoffmann}}}, \bibinfo {author} {\bibfnamefont {D.}~\bibnamefont
  {{Scolnic}}}, \bibinfo {author} {\bibfnamefont {S.}~\bibnamefont
  {{Casertano}}}, \bibinfo {author} {\bibfnamefont {A.~V.}\ \bibnamefont
  {{Filippenko}}}, \bibinfo {author} {\bibfnamefont {B.~E.}\ \bibnamefont
  {{Tucker}}}, \bibinfo {author} {\bibfnamefont {M.~J.}\ \bibnamefont
  {{Reid}}}, \bibinfo {author} {\bibfnamefont {D.~O.}\ \bibnamefont {{Jones}}},
  \bibinfo {author} {\bibfnamefont {J.~M.}\ \bibnamefont {{Silverman}}},
  \bibinfo {author} {\bibfnamefont {R.}~\bibnamefont {{Chornock}}}, \bibinfo
  {author} {\bibfnamefont {P.}~\bibnamefont {{Challis}}}, \bibinfo {author}
  {\bibfnamefont {W.}~\bibnamefont {{Yuan}}}, \bibinfo {author} {\bibfnamefont
  {P.~J.}\ \bibnamefont {{Brown}}}, \ and\ \bibinfo {author} {\bibfnamefont
  {R.~J.}\ \bibnamefont {{Foley}}},\ }\href {\doibase
  10.3847/0004-637X/826/1/56} {\bibfield  {journal} {\bibinfo  {journal}
  {\apj}\ }\textbf {\bibinfo {volume} {826}},\ \bibinfo {eid} {56} (\bibinfo
  {year} {2016})},\ \Eprint {http://arxiv.org/abs/1604.01424}
  {arXiv:1604.01424} \BibitemShut {NoStop}%
\bibitem [{\citenamefont {{Bernal}}\ \emph {et~al.}(2016)\citenamefont
  {{Bernal}}, \citenamefont {{Verde}},\ and\ \citenamefont
  {{Riess}}}]{2016JCAP...10..019B}%
  \BibitemOpen
  \bibfield  {author} {\bibinfo {author} {\bibfnamefont {J.~L.}\ \bibnamefont
  {{Bernal}}}, \bibinfo {author} {\bibfnamefont {L.}~\bibnamefont {{Verde}}}, \
  and\ \bibinfo {author} {\bibfnamefont {A.~G.}\ \bibnamefont {{Riess}}},\
  }\href {\doibase 10.1088/1475-7516/2016/10/019} {\bibfield  {journal}
  {\bibinfo  {journal} {\jcap}\ }\textbf {\bibinfo {volume} {10}},\ \bibinfo
  {eid} {019} (\bibinfo {year} {2016})},\ \Eprint
  {http://arxiv.org/abs/1607.05617} {arXiv:1607.05617} \BibitemShut {NoStop}%
\bibitem [{\citenamefont {{Gair}}\ \emph {et~al.}(2010)\citenamefont {{Gair}},
  \citenamefont {{Tang}},\ and\ \citenamefont
  {{Volonteri}}}]{2010PhRvD..81j4014G}%
  \BibitemOpen
  \bibfield  {author} {\bibinfo {author} {\bibfnamefont {J.~R.}\ \bibnamefont
  {{Gair}}}, \bibinfo {author} {\bibfnamefont {C.}~\bibnamefont {{Tang}}}, \
  and\ \bibinfo {author} {\bibfnamefont {M.}~\bibnamefont {{Volonteri}}},\
  }\href {\doibase 10.1103/PhysRevD.81.104014} {\bibfield  {journal} {\bibinfo
  {journal} {\prd}\ }\textbf {\bibinfo {volume} {81}},\ \bibinfo {eid} {104014}
  (\bibinfo {year} {2010})},\ \Eprint {http://arxiv.org/abs/1004.1921}
  {arXiv:1004.1921 [astro-ph.GA]} \BibitemShut {NoStop}%
\bibitem [{\citenamefont {{Sesana}}\ \emph {et~al.}(2011)\citenamefont
  {{Sesana}}, \citenamefont {{Gair}}, \citenamefont {{Berti}},\ and\
  \citenamefont {{Volonteri}}}]{2011PhRvD..83d4036S}%
  \BibitemOpen
  \bibfield  {author} {\bibinfo {author} {\bibfnamefont {A.}~\bibnamefont
  {{Sesana}}}, \bibinfo {author} {\bibfnamefont {J.}~\bibnamefont {{Gair}}},
  \bibinfo {author} {\bibfnamefont {E.}~\bibnamefont {{Berti}}}, \ and\
  \bibinfo {author} {\bibfnamefont {M.}~\bibnamefont {{Volonteri}}},\ }\href
  {\doibase 10.1103/PhysRevD.83.044036} {\bibfield  {journal} {\bibinfo
  {journal} {\prd}\ }\textbf {\bibinfo {volume} {83}},\ \bibinfo {eid} {044036}
  (\bibinfo {year} {2011})},\ \Eprint {http://arxiv.org/abs/1011.5893}
  {arXiv:1011.5893 [astro-ph.CO]} \BibitemShut {NoStop}%
\bibitem [{\citenamefont {{Stevenson}}\ \emph {et~al.}(2015)\citenamefont
  {{Stevenson}}, \citenamefont {{Ohme}},\ and\ \citenamefont
  {{Fairhurst}}}]{2015ApJ...810...58S}%
  \BibitemOpen
  \bibfield  {author} {\bibinfo {author} {\bibfnamefont {S.}~\bibnamefont
  {{Stevenson}}}, \bibinfo {author} {\bibfnamefont {F.}~\bibnamefont {{Ohme}}},
  \ and\ \bibinfo {author} {\bibfnamefont {S.}~\bibnamefont {{Fairhurst}}},\
  }\href {\doibase 10.1088/0004-637X/810/1/58} {\bibfield  {journal} {\bibinfo
  {journal} {\apj}\ }\textbf {\bibinfo {volume} {810}},\ \bibinfo {eid} {58}
  (\bibinfo {year} {2015})},\ \Eprint {http://arxiv.org/abs/1504.07802}
  {arXiv:1504.07802 [astro-ph.HE]} \BibitemShut {NoStop}%
\bibitem [{\citenamefont {{Gair}}\ \emph {et~al.}(2011)\citenamefont {{Gair}},
  \citenamefont {{Sesana}}, \citenamefont {{Berti}},\ and\ \citenamefont
  {{Volonteri}}}]{2011CQGra..28i4018G}%
  \BibitemOpen
  \bibfield  {author} {\bibinfo {author} {\bibfnamefont {J.~R.}\ \bibnamefont
  {{Gair}}}, \bibinfo {author} {\bibfnamefont {A.}~\bibnamefont {{Sesana}}},
  \bibinfo {author} {\bibfnamefont {E.}~\bibnamefont {{Berti}}}, \ and\
  \bibinfo {author} {\bibfnamefont {M.}~\bibnamefont {{Volonteri}}},\ }\href
  {\doibase 10.1088/0264-9381/28/9/094018} {\bibfield  {journal} {\bibinfo
  {journal} {\cqg}\ }\textbf {\bibinfo {volume} {28}},\ \bibinfo {eid} {094018}
  (\bibinfo {year} {2011})},\ \Eprint {http://arxiv.org/abs/1009.6172}
  {arXiv:1009.6172 [gr-qc]} \BibitemShut {NoStop}%
\bibitem [{\citenamefont {{Berti}}\ \emph {et~al.}(2011)\citenamefont
  {{Berti}}, \citenamefont {{Gair}},\ and\ \citenamefont
  {{Sesana}}}]{2011PhRvD..84j1501B}%
  \BibitemOpen
  \bibfield  {author} {\bibinfo {author} {\bibfnamefont {E.}~\bibnamefont
  {{Berti}}}, \bibinfo {author} {\bibfnamefont {J.}~\bibnamefont {{Gair}}}, \
  and\ \bibinfo {author} {\bibfnamefont {A.}~\bibnamefont {{Sesana}}},\ }\href
  {\doibase 10.1103/PhysRevD.84.101501} {\bibfield  {journal} {\bibinfo
  {journal} {\prd}\ }\textbf {\bibinfo {volume} {84}},\ \bibinfo {eid} {101501}
  (\bibinfo {year} {2011})},\ \Eprint {http://arxiv.org/abs/1107.3528}
  {arXiv:1107.3528 [gr-qc]} \BibitemShut {NoStop}%
\bibitem [{\citenamefont {{Kovetz}}\ \emph {et~al.}(2017)\citenamefont
  {{Kovetz}}, \citenamefont {{Cholis}}, \citenamefont {{Breysse}},\ and\
  \citenamefont {{Kamionkowski}}}]{2016arXiv161101157K}%
  \BibitemOpen
  \bibfield  {author} {\bibinfo {author} {\bibfnamefont {E.~D.}\ \bibnamefont
  {{Kovetz}}}, \bibinfo {author} {\bibfnamefont {I.}~\bibnamefont {{Cholis}}},
  \bibinfo {author} {\bibfnamefont {P.~C.}\ \bibnamefont {{Breysse}}}, \ and\
  \bibinfo {author} {\bibfnamefont {M.}~\bibnamefont {{Kamionkowski}}},\ }\href
  {\doibase 10.1103/PhysRevD.95.103010} {\bibfield  {journal} {\bibinfo
  {journal} {\prd}\ }\textbf {\bibinfo {volume} {95}},\ \bibinfo {pages}
  {103010} (\bibinfo {year} {2017})},\ \Eprint
  {http://arxiv.org/abs/1611.01157} {arXiv:1611.01157} \BibitemShut {NoStop}%
\bibitem [{\citenamefont {{Sesana}}\ \emph {et~al.}(2014)\citenamefont
  {{Sesana}}, \citenamefont {{Barausse}}, \citenamefont {{Dotti}},\ and\
  \citenamefont {{Rossi}}}]{2014ApJ...794..104S}%
  \BibitemOpen
  \bibfield  {author} {\bibinfo {author} {\bibfnamefont {A.}~\bibnamefont
  {{Sesana}}}, \bibinfo {author} {\bibfnamefont {E.}~\bibnamefont
  {{Barausse}}}, \bibinfo {author} {\bibfnamefont {M.}~\bibnamefont {{Dotti}}},
  \ and\ \bibinfo {author} {\bibfnamefont {E.~M.}\ \bibnamefont {{Rossi}}},\
  }\href {\doibase 10.1088/0004-637X/794/2/104} {\bibfield  {journal} {\bibinfo
   {journal} {\apj}\ }\textbf {\bibinfo {volume} {794}},\ \bibinfo {eid} {104}
  (\bibinfo {year} {2014})},\ \Eprint {http://arxiv.org/abs/1402.7088}
  {arXiv:1402.7088} \BibitemShut {NoStop}%
\bibitem [{\citenamefont {{Vitale}}\ \emph
  {et~al.}(2017{\natexlab{b}})\citenamefont {{Vitale}}, \citenamefont
  {{Lynch}}, \citenamefont {{Sturani}},\ and\ \citenamefont
  {{Graff}}}]{2017CQGra..34cLT01V}%
  \BibitemOpen
  \bibfield  {author} {\bibinfo {author} {\bibfnamefont {S.}~\bibnamefont
  {{Vitale}}}, \bibinfo {author} {\bibfnamefont {R.}~\bibnamefont {{Lynch}}},
  \bibinfo {author} {\bibfnamefont {R.}~\bibnamefont {{Sturani}}}, \ and\
  \bibinfo {author} {\bibfnamefont {P.}~\bibnamefont {{Graff}}},\ }\href
  {\doibase 10.1088/1361-6382/aa552e} {\bibfield  {journal} {\bibinfo
  {journal} {\cqg}\ }\textbf {\bibinfo {volume} {34}},\ \bibinfo {eid} {03LT01}
  (\bibinfo {year} {2017}{\natexlab{b}})},\ \Eprint
  {http://arxiv.org/abs/1503.04307} {arXiv:1503.04307 [gr-qc]} \BibitemShut
  {NoStop}%
\bibitem [{\citenamefont {{Nishizawa}}\ \emph {et~al.}(2016)\citenamefont
  {{Nishizawa}}, \citenamefont {{Berti}}, \citenamefont {{Klein}},\ and\
  \citenamefont {{Sesana}}}]{2016PhRvD..94f4020N}%
  \BibitemOpen
  \bibfield  {author} {\bibinfo {author} {\bibfnamefont {A.}~\bibnamefont
  {{Nishizawa}}}, \bibinfo {author} {\bibfnamefont {E.}~\bibnamefont
  {{Berti}}}, \bibinfo {author} {\bibfnamefont {A.}~\bibnamefont {{Klein}}}, \
  and\ \bibinfo {author} {\bibfnamefont {A.}~\bibnamefont {{Sesana}}},\ }\href
  {\doibase 10.1103/PhysRevD.94.064020} {\bibfield  {journal} {\bibinfo
  {journal} {\prd}\ }\textbf {\bibinfo {volume} {94}},\ \bibinfo {eid} {064020}
  (\bibinfo {year} {2016})},\ \Eprint {http://arxiv.org/abs/1605.01341}
  {arXiv:1605.01341 [gr-qc]} \BibitemShut {NoStop}%
\bibitem [{\citenamefont {{Breivik}}\ \emph {et~al.}(2016)\citenamefont
  {{Breivik}}, \citenamefont {{Rodriguez}}, \citenamefont {{Larson}},
  \citenamefont {{Kalogera}},\ and\ \citenamefont
  {{Rasio}}}]{2016ApJ...830L..18B}%
  \BibitemOpen
  \bibfield  {author} {\bibinfo {author} {\bibfnamefont {K.}~\bibnamefont
  {{Breivik}}}, \bibinfo {author} {\bibfnamefont {C.~L.}\ \bibnamefont
  {{Rodriguez}}}, \bibinfo {author} {\bibfnamefont {S.~L.}\ \bibnamefont
  {{Larson}}}, \bibinfo {author} {\bibfnamefont {V.}~\bibnamefont
  {{Kalogera}}}, \ and\ \bibinfo {author} {\bibfnamefont {F.~A.}\ \bibnamefont
  {{Rasio}}},\ }\href {\doibase 10.3847/2041-8205/830/1/L18} {\bibfield
  {journal} {\bibinfo  {journal} {\apjl}\ }\textbf {\bibinfo {volume} {830}},\
  \bibinfo {eid} {L18} (\bibinfo {year} {2016})},\ \Eprint
  {http://arxiv.org/abs/1606.09558} {arXiv:1606.09558} \BibitemShut {NoStop}%
\bibitem [{\citenamefont {{Nishizawa}}\ \emph {et~al.}(2017)\citenamefont
  {{Nishizawa}}, \citenamefont {{Sesana}}, \citenamefont {{Berti}},\ and\
  \citenamefont {{Klein}}}]{2017MNRAS.465.4375N}%
  \BibitemOpen
  \bibfield  {author} {\bibinfo {author} {\bibfnamefont {A.}~\bibnamefont
  {{Nishizawa}}}, \bibinfo {author} {\bibfnamefont {A.}~\bibnamefont
  {{Sesana}}}, \bibinfo {author} {\bibfnamefont {E.}~\bibnamefont {{Berti}}}, \
  and\ \bibinfo {author} {\bibfnamefont {A.}~\bibnamefont {{Klein}}},\ }\href
  {\doibase 10.1093/mnras/stw2993} {\bibfield  {journal} {\bibinfo  {journal}
  {\mnras}\ }\textbf {\bibinfo {volume} {465}},\ \bibinfo {pages} {4375}
  (\bibinfo {year} {2017})},\ \Eprint {http://arxiv.org/abs/1606.09295}
  {arXiv:1606.09295 [astro-ph.HE]} \BibitemShut {NoStop}%
\bibitem [{\citenamefont {{Haster}}\ \emph {et~al.}(2016)\citenamefont
  {{Haster}}, \citenamefont {{Wang}}, \citenamefont {{Berry}}, \citenamefont
  {{Stevenson}}, \citenamefont {{Veitch}},\ and\ \citenamefont
  {{Mandel}}}]{2016MNRAS.457.4499H}%
  \BibitemOpen
  \bibfield  {author} {\bibinfo {author} {\bibfnamefont {C.-J.}\ \bibnamefont
  {{Haster}}}, \bibinfo {author} {\bibfnamefont {Z.}~\bibnamefont {{Wang}}},
  \bibinfo {author} {\bibfnamefont {C.~P.~L.}\ \bibnamefont {{Berry}}},
  \bibinfo {author} {\bibfnamefont {S.}~\bibnamefont {{Stevenson}}}, \bibinfo
  {author} {\bibfnamefont {J.}~\bibnamefont {{Veitch}}}, \ and\ \bibinfo
  {author} {\bibfnamefont {I.}~\bibnamefont {{Mandel}}},\ }\href {\doibase
  10.1093/mnras/stw233} {\bibfield  {journal} {\bibinfo  {journal} {\mnras}\
  }\textbf {\bibinfo {volume} {457}},\ \bibinfo {pages} {4499} (\bibinfo {year}
  {2016})},\ \Eprint {http://arxiv.org/abs/1511.01431} {arXiv:1511.01431
  [astro-ph.HE]} \BibitemShut {NoStop}%
\bibitem [{\citenamefont {{Fishbach}}\ \emph {et~al.}(2017)\citenamefont
  {{Fishbach}}, \citenamefont {{Holz}},\ and\ \citenamefont
  {{Farr}}}]{2017arXiv170306869F}%
  \BibitemOpen
  \bibfield  {author} {\bibinfo {author} {\bibfnamefont {M.}~\bibnamefont
  {{Fishbach}}}, \bibinfo {author} {\bibfnamefont {D.}~\bibnamefont {{Holz}}},
  \ and\ \bibinfo {author} {\bibfnamefont {B.}~\bibnamefont {{Farr}}},\
  }\href@noop {} {\bibfield  {journal} {\bibinfo  {journal} {ArXiv e-prints}\ }
  (\bibinfo {year} {2017})},\ \Eprint {http://arxiv.org/abs/1703.06869}
  {arXiv:1703.06869 [astro-ph.HE]} \BibitemShut {NoStop}%
\bibitem [{\citenamefont {{Robitaille {\it et al.} (Astropy
  Collaboration)}}(2013)}]{2013A&A...558A..33A}%
  \BibitemOpen
  \bibfield  {author} {\bibinfo {author} {\bibfnamefont {T.~P.}\ \bibnamefont
  {{Robitaille {\it et al.} (Astropy Collaboration)}}},\ }\href {\doibase
  10.1051/0004-6361/201322068} {\bibfield  {journal} {\bibinfo  {journal}
  {\aap}\ }\textbf {\bibinfo {volume} {558}},\ \bibinfo {eid} {A33} (\bibinfo
  {year} {2013})},\ \Eprint {http://arxiv.org/abs/1307.6212} {arXiv:1307.6212
  [astro-ph.IM]} \BibitemShut {NoStop}%
\bibitem [{\citenamefont {{Hunter}}(2007)}]{2007CSE.....9...90H}%
  \BibitemOpen
  \bibfield  {author} {\bibinfo {author} {\bibfnamefont {J.~D.}\ \bibnamefont
  {{Hunter}}},\ }\href {\doibase 10.1109/MCSE.2007.55} {\bibfield  {journal}
  {\bibinfo  {journal} {Computing in Science and Engineering}\ }\textbf
  {\bibinfo {volume} {9}},\ \bibinfo {pages} {90} (\bibinfo {year}
  {2007})}\BibitemShut {NoStop}%
\end{thebibliography}%

\end{document}